\documentclass[aps,nofootinbib,amsmath,prd,notitlepage,showpacs,superscriptaddress]{revtex4-2}

\usepackage[american]{babel}
\usepackage{amsfonts}
\usepackage{amsmath}
\usepackage{amssymb}
\usepackage{epsfig}
\usepackage{latexsym}
\usepackage{paralist}
\usepackage{fancyhdr}
\usepackage{graphicx}
\usepackage{graphicx,txfonts}

\usepackage{etex}
\usepackage{float}
\usepackage{slashed}
\usepackage{xcolor}
\usepackage{soul}
\usepackage{dcolumn} 
\usepackage{bm}
\usepackage{tipa}
\usepackage{dsfont}
\usepackage{caption}
\usepackage{CJKutf8}

\newcommand{\re}{\text{\begin{CJK}{UTF8}{min}せ\end{CJK}}}
\newcommand{\ri}{\text{\begin{CJK}{UTF8}{min}ん\end{CJK}}}
\newcommand{\ru}{\text{\begin{CJK}{UTF8}{min}る\end{CJK}}}
\newcommand{\no}{\text{\begin{CJK}{UTF8}{min}の\end{CJK}}}

\allowdisplaybreaks
\usepackage{titlesec}
\titlespacing*{\section}{10pt}{10pt}{15pt}
\begin{document}

	\title{Radiative corrections to the $R$ and $R^2$ invariants from torsion fluctuations on maximally symmetric spaces 
	}

	\author{Riccardo Martini}
\email{riccardo.martini@pi.infn.it}
\affiliation{
	INFN - Sezione di Pisa, Largo Bruno Pontecorvo 3, 56127 Pisa, Italy}

	\author{Gregorio Paci}
\email{gregorio.paci@phd.unipi.it}
\affiliation{
	Universit\`a di Pisa, Largo Bruno Pontecorvo 3, 56127 Pisa, Italy}
\affiliation{
	INFN - Sezione di Pisa, Largo Bruno Pontecorvo 3, 56127 Pisa, Italy}

	\author{Dario Sauro}
\email{dario.sauro@phd.unipi.it}
\affiliation{
	Universit\`a di Pisa, Largo Bruno Pontecorvo 3, 56127 Pisa, Italy}
\affiliation{
	INFN - Sezione di Pisa, Largo Bruno Pontecorvo 3, 56127 Pisa, Italy}

\begin{abstract}

We derive the runnings of the $R$ and $R^2$ operators that stem from integrating out quantum torsion fluctuations on a maximally symmetric Euclidean background, while treating the metric as a classical field. Our analysis is performed in a manifestly covariant way, exploiting both the recently-introduced spin-parity decomposition of torsion perturbations and the heat kernel technique. The Lagrangian we start with is the most general one for 1-loop computations on maximally symmetric backgrounds involving kinetic terms and couplings to the scalar curvature that is compatible with a gauge-like symmetry for the torsion. The latter removes the twice-longitudinal vector mode from the spectrum, and it yields operators of maximum rank four. We also examine the conditions required to avoid ghost instabilities and ensure the validity of our assumption to neglect metric quantum fluctuations, demonstrating the compatibility between these two assumptions. Then, we use our findings in the context of Starobinsky's inflation to calculate the contributions from the torsion tensor to the $\beta$-function of the $R^2$ term. 
While this result is quantitatively reliable only at the $0$-th order in the slow-roll parameters or during the very early stages of inflation -- due to the background choice -- it qualitatively illustrates how to incorporate quantum effects of torsion in the path integral formalism.
\end{abstract}
	\maketitle

\section{Introduction}

The Gravitational interaction is rooted in Einstein's General theory of Relativity (GR) \cite{Einstein:1916vd}. Spacetime is described as a (pseudo-) Riemannian manifold, and the only dynamical field pertaining to gravity is the metric tensor. Although this framework works very well at planetary scales, it presents conceptual and phenomenological issues at the Planck scale for both the classical \cite{Penrose:1964wq, Hawking:1970zqf} and quantum theory \cite{tHooft:1974toh, Christensen:1979iy, Goroff:1985th}, as well as at the cosmological scales \cite{Hawking:1973uf}. The latter could be due to the presence of some yet unobserved matter fields, while the former seems to signal a breakdown of GR.

The general perspective towards the high-energy regime problems of GR is that this theory has to be thought of as an effective field theory, valid only below the Planck scale \cite{Donoghue:1994dn}. Indeed, recently an interesting point of view has been put forward, i.e., that the metric tensor acquires a non-singular vacuum expectation value (VEV) around the Planck scale, thus being the order parameter of some phase transitions associated with a symmetry breaking \cite{Percacci:2023rbo}. Adopting this viewpoint, one is naturally led to the assumption that other gravitational degrees of freedom have to be taken into account around the Planck scale. In string theory these are given by the whole spin tower of string excitations, while in supergravity one has to consider the super-partner of the graviton, i.e.,\ the gravitino \cite{VanNieuwenhuizen:1981ae}. A somewhat different description of the ultraviolet (UV) behavior of gravity is the one of Metric-Affine theories of Gravity (MAGs) \cite{Hehl:1976kj,Hehl:1994ue,Gronwald:1995em,BeltranJimenez:2019acz,BeltranJimenez:2019esp,BeltranJimenez:2020sih,BeltranJimenez:2020sqf,Heisenberg:2023wgk,Baldazzi:2021kaf,Sauro:2022chz,Sauro:2022hoh,Shapiro:2001rz,Wheeler:2023qyy}, where the geometry is not assumed \emph{a priori} to be Riemannian and the affine-connection is taken to be an independent field, i.e., it is not bound to be a functional of the metric tensor. In this paper we shall analyze some quantum aspects of a particular class of metric-affine theories.

In MAGs, aside from the usual geometric concept of curvature, spacetime is described also by torsion and non-metricity \cite{Gronwald:1995em}. The first one manifests itself with the non-closure of infinitesimal parallelograms, whereas the second one gives rise to the non-conservation of the length of vectors when they are parallel transported. If the spacetime manifold is endowed with a non-singular metric tensor, we can split the affine-connection into the sum of the standard Levi-Civita contribution plus two tensors. The first one is the contortion and it depends on the torsion, while the second is the distortion and it is a function of the non-metricity. Both these tensors also depend on the metric, which is used to raise and lower the indices of torsion and non-metricity. The compelling feature of metric-affine theories is that they naturally reproduce GR in the infrared red (IR) limit since, much below the Planck scale, the degrees of freedom due to torsion and non-metricity are non-propagating, and these fields are set to zero by their field equations \cite{Palatini:1919, Dadhich:2012htv, Iosifidis:2021tvx}.
                                                                         
MAGs can be formulated as gauge theories of the full general linear group \cite{Lord:1978qz}, bringing the description of gravity closer to other fundamental interactions. Moreover, this formulation highlights a striking difference between torsion and non-metricity. Indeed, the former is bound to the Lorentz subgroup, while the latter is connected to its orthogonal complement in $GL(4)$. Moreover, theories bearing non-metricity usually propagate a dangerous spin-$3$ mode \cite{Percacci:2020bzf,Baldazzi:2021kaf}, which is inconsistent with soft theorems \cite{Weinberg:1965nx}. Furthermore, the minimal coupling of spin-$\frac{1}{2}$ fermions to the post-Riemannian structure only depends on the completely antisymmetric part of the torsion. Thus, the behavior of usual matter fields on a post-Riemannian background geometry do not depend on the non-metricity at the classical level. For these reasons we are inclined to discard MAGs with non-metricity, analyzing only the dependence on torsion degrees of freedom.

A subclass of theories with non-trivial torsion is given by the so-called Poincar\`{e} Gauge Theories (PGT) \cite{Hehl:1976kj,Gronwald:1995em,Wheeler:2023qyy}, in which the action functional is assumed to depend on either the torsion or the full curvature tensor in a quadratic way \cite{Sezgin:1981xs}. These theories are way simpler than the most general ones, and are entirely described by six real parameters \cite{Sezgin:1979zf,Sezgin:1981xs}. Despite their formal simplicity, they have the huge drawback of being non-renormalizable. This is due to the fact that radiative corrections induced by integrating out quantum fluctuations have logarithmic divergences that depend on the torsion in a non-quadratic way, see \cite{Melichev:2023lwj} for a very recent discussion on this topic.

The quantum behavior of general theories with torsion is not known, aside from the parametric form of the flat space propagators of the spin-parity eigenmodes \cite{Baldazzi:2021kaf} (see also \cite{Aoki:2019snr}). For this reason, we have recently shown \cite{Martini:2023rnv} how to carry out a covariant spin-parity decomposition of the torsion fluctuations on a Riemannian background. Such a decomposition has then been exploited to formally define the functional integration measure over such torsional fluctuations. We remark that the functional measure over the spin-parity eigenstates is diagonal only on maximally symmetric spaces, whereas in the case of metric fluctuations it is sufficient to assume the background geometry to be an Einstein space \cite{Percacci:2017fkn}. Thus, our choice here is to require the background Riemannian geometry to be maximally symmetric, implying that our results shall provide quantum corrections to inflation due to the torsion. Indeed, in this article we shall exploit these formal developments to integrate out torsion fluctuations and to compute the beta function of the $R^2$, Einstein-Hilbert and cosmological constant terms on maximally symmetric spaces, therefore paving the way for a more thorough analysis of the quantum features of the torsion. Thus, we provide the contributions to the beta-functions of the Starobinsky Lagrangian \cite{Starobinsky:1980te} at the 0th order in the inflation parameters due to Gaussian torsion fluctuations. Indeed, in our perspective the torsion behaves much like a matter field, that can be integrated out on curved backgrounds, disregarding the metric fluctuations in a first approximation. Of course, our reasoning requires that the coupling constants parameterizing torsion interactions are such that they give rise to quantum effects below the Planck scale, and that the masses of such torsion excitations are less than the Planck mass. This is justified since, from a MAG perspective, the torsion is independent of the metric and it can enter the theory at lower scales. Nevertheless, we have to keep in mind that torsion also includes spin-parity $2^+$ and $0^+$ modes that would naturally mix with the graviton. Therefore, a complete picture would require the inclusion of metric fluctuations, which we hope to complete in the future. See \cite{Aoki:2020zqm,Fomin:2023tsn, Mondal:2023cxx, Cai:2015emx} for previous applications of torsion to cosmological scenarios.
                                                                                                                                                                               
The technical advantage of employing a spin-parity decomposition like \cite{Martini:2023rnv} or \cite{York:1974psa} is that the differential operators are considerably simpler. Indeed, one of the main difficulties in exploiting the heat kernel technique is the presence of non-minimal terms of highest rank. In \cite{Melichev:2023lwj} the way around is provided by parameterizing the theory in terms of the co-frame and the spin-connection. This allows the authors to get rid of one particular non-minimal term by a suitable choice of the gauge-fixing for local Lorentz transformations. On the other hand, in the most general MAGs other non-minimal terms arise, which cannot be dealt with by using such a procedure. Thus, our choice is to parameterize the quantum torsion fluctuations in terms of the spin-parity eigenstates, that are described by transverse tensors. While the transversality automatically removes the non-minimal terms of highest rank, using a spin-parity decomposition yields higher derivative operators. However, if such operators are orthogonal to the second-order ones, one can directly apply the results of \cite{Barvinsky:1985an}, which are valid for differential operators whose highest rank part is minimal and invertible. This recipe applies for all spin-parity sectors but the $1^-$: in this case the integration would require a differential re-definition of the torsion vector, which would yield a sixth-order operator. Thus, we choose to get rid of the $1^-$ mode that concurs to the tracefree hook antisymmetric part of the torsion by requiring the action to be invariant under a gauge-like transformation, which may be thought of as a very special kind of Curtright transformation \cite{Curtright:1980yk}.

The paper is structured as follows. We start in section \ref{sect:MAGs_review} with a general review of MAGs and some results relevant from a field-theoretic perspective, e.g., the aforementioned spin-parity decomposition of the torsion tensor. In this section, we also derive the bounds on the Lagrangian parameters necessary to ensure the absence of ghost modes and the validity of our assumption to exclude the metric field from loops. This analysis also shows the compatibility between these two requirements. In section \ref{sect:Lagrangian}, we introduce the Lagrangian, which is made out of the kinetic terms for the torsion, as well as the three couplings to the scalar curvature. The technical tools necessary to integrate out the torsion fluctuations are provided in sections \ref{sect:Schwinger_DeWitt_technique} and \ref{sect:spectrum-of-diff-constrained}. The first one is a brief but self-contained introduction to the Schwinger-DeWitt technique and its generalization, which we shall heavily employ in the rest of the paper. In the second one, we show how to properly take into account the presence of zero-modes, a complication that arises when dealing with differentially constrained fields on compact manifolds. In this respect, we give a simple classification based on their origin being dynamical, i.e., linked to the choice of the action, or kinematical, i.e., rooted in the spin-parity decomposition. After having presented the required techniques, in section\ \ref{sect:1-loop-body} we outline the functional integration of torsion fluctuations. We choose to present in full detail only the computations for the $1^+$ sector, which are the most complicated ones. Then, we provide the second Seeley-DeWitt coefficient that results from our general calculation, and we employ it to study the renormalization of the Starobinsky Lagrangian, including the cosmological constant. The appendices contain most of the computations, which would have overburdened the flow of the paper. In particular, in App.\ \ref{sect:appendix-kin-terms} we outline how to rewrite the action in a handy way, in App.\ \ref{sect:appendix-hessians} we give the complete expressions of all the non-trivial components of the Hessian, while in App.\ \ref{sect:appendix-1loop} we collect all the partial results for the functional integration in each spin-parity sector. Finally, we conclude the paper analyzing our results and proposing some future prospects in section \ref{sect:conclusions}.

\section{Field-theoretic perspective towards MAGs}\label{sect:MAGs_review}

Metric affine-theories of gravity start from the assumption that both the affine-connection $\Gamma^\rho{}_{\nu\mu}$ and the metric tensor $g_{\mu\nu}$ of the spacetime manifold are independent field variables. In the original formulation \cite{Hehl:1976kj,Gronwald:1995em,Sezgin:1979zf,Sezgin:1981xs}, the field variables of MAGs were the co-frame, the affine-connection and the metric of the local frame. Thus, the curvature tensors that characterize the geometry are the torsion of the co-frame, the curvature of the affine-connection, and the non-metricity of the local frame metric. Since observables are gauge-invariant, we can choose to work either with holonomic or non-holonomic indices, and we stick to the former for computational convenience.

Adopting the convention that the last index of the connection is the directional one, the covariant derivative of a vector reads
\begin{equation}\label{def-connection}
\nabla_\mu v^\rho = \partial_\mu v^\rho + \Gamma^\rho{}_{\sigma\mu} v^\sigma \, .
\end{equation}
Assuming that the metric is non-singular, we define the tensor $\Phi^\rho{}_{\nu\mu}$ in terms of the connection $\Gamma^\rho{}_{\nu\mu}$ as
\begin{equation}\label{def-Phi-tensor}
\Gamma^\rho{}_{\nu\mu} = \mathring{\Gamma}^\rho{}_{\nu\mu} + \Phi^\rho{}_{\nu\mu}\, , 
\end{equation}
where $\mathring{\Gamma}^\rho{}_{\mu\nu}$ is the Levi-Civita connection. As we have said above, the geometry of MAGs is characterized by curvature, torsion and non-metricity. In the holonomic formulation, the first is perceived through the rotation of vectors parallel-transported around loops. Instead, the second manifests itself in the non-closure of infinitesimal parallelograms. Lastly, the non-metricity can be measured through the variation of the norm of vectors when they are parallel transported. The formal definitions of these tensors are
\begin{subequations}\label{def-Riem}
	{\small \begin{align}
	R^\lambda{}_{\rho\mu\nu} \equiv & \, \partial_\mu \Gamma^\lambda{}_{\rho\nu} - \partial_\nu \Gamma^\lambda{}_{\rho\mu} + \Gamma^\lambda{}_{\kappa\mu} \Gamma^\kappa{}_{\rho\nu} - \Gamma^\lambda{}_{\kappa\nu} \Gamma^\kappa{}_{\rho\mu} \, ,\\\label{def-Tor}
	T^\lambda{}_{\mu\nu} \equiv & \, \Gamma^\lambda{}_{\nu\mu} - \Gamma^\lambda{}_{\mu\nu} \, ,\\\label{def-Nonmetr}
	Q_{\mu\nu\lambda} \equiv & \, - \nabla_\lambda g_{\mu\nu} \, .
	\end{align} }
\end{subequations}
The tensor $\Phi^\rho{}_{\nu\mu}$ that parameterizes the deviation from GR can be decomposed uniquely as
\begin{equation}\label{aff-conn-split}
\Phi^\rho{}_{\nu\mu} = 
K^\rho{}_{\nu\mu} + N^\rho{}_{\nu\mu} \, ,
\end{equation}
where the contortion $K^\rho{}_{\nu\mu}$ is independent of the non-metricity, while the distortion $N^\rho{}_{\nu\mu}$ does not depend on the torsion. With our conventions we have
\begin{subequations}
	{\small \begin{align}\label{def-Contortion}
	K^\rho{}_{\nu\mu} & = \frac{1}{2} \left( T_\nu{}^\rho{}_\mu + T_\mu{}^\rho{}_\nu - T^\rho{}_{\nu\mu} \right) \, ;\\\label{def-Distortion}
	N^\rho{}_{\nu\mu} & = \frac{1}{2} \left( Q_\nu{}^\rho{}_\mu + Q_\mu{}^\rho{}_\nu - Q_{\mu\nu}{}^\rho \right) \, .
	\end{align} }
\end{subequations}
The contortion is antisymmetric in the first two indices, thus taking value in the Lorentz subalgebra of $GL(d,\mathbb{R})$, while the distortion is symmetric in its last two indices. It is important to remark that also the non-metricity can contribute to the part of the connection that takes values in the Lorentz subalgebra. To highlight this, let us decompose the torsion and the non-metricity as follows
\begin{subequations}
{\small \begin{align}
 T^\rho{}_{\mu\nu} = & H^\rho{}_{\mu\nu} + t^\rho{}_{\mu\nu} \, ;\\
 Q_{\mu\nu\rho} = & C_{\mu\nu\rho} + D_{\mu\nu\rho} \, ,
\end{align} }
\end{subequations}
where $H$ is completely antisymmetric, $t$ is hook antisymmetric, $C$ is completely symmetric and $D$ is hook symmetric, i.e.,
\begin{subequations}
 {\small \begin{align}
  H_{[\mu\nu\rho]} = & H_{\mu\nu\rho}, , \qquad t^\rho{}_{[\mu\nu]} = t^\rho{}_{\mu\nu} \, , \qquad t_{[\mu\nu\rho]} = 0 \, ; \\
  C_{(\mu\nu\rho)} = & C_{\mu\nu\rho} \, , \qquad D_{(\mu\nu)\rho} = D_{\mu\nu\rho} \qquad D_{(\mu\nu\rho)} = 0 \, .
 \end{align} }
\end{subequations}
Thus, the contribution of the non-metricity to the part of the connection that takes values in the Lorentz subalgebra is given solely in terms of the hook symmetric part
\begin{equation}
 \Phi_{[\rho\nu]\mu} = K_{\rho\nu\mu} + D_{\mu\rho\nu} + \frac{1}{2} D_{\nu\rho\mu} \, .
\end{equation}
In a completely analogous way, only the hook antisymmetric part of the torsion contributes to the symmetric part of the connection
\begin{equation}
 \Phi_{\rho(\nu\mu)} = N_{\rho\nu\mu} + t_{\mu\rho\nu} - \frac{1}{2} t_{\rho\mu\nu} \, .
\end{equation}
The similarity of the hook symmetric and hook antisymmetric tensors is striking, e.g., they have the same number of components, i.e.,\ $\frac{d(d^2-1)}{3}$. These tensors also have the same spin-parity content, see \cite{Baldazzi:2021kaf}. Moreover, the conformal actions of their trace-free parts have ghosts and tachyons in the same spin-parity sectors \cite{Paci:2023twc}. 

The torsion and non-metricity have one and two non-trivial traces, respectively. Here we define them as traces of the $GL(d)$-decomposed tensors, i.e.,\
\begin{equation}
 \tau_\mu = t^\rho{}_{\mu\rho} \, , \qquad A_\mu = C^\rho{}_{\rho\mu} \, , \qquad B_\mu = D^\rho{}_{\rho\mu} \, . 
\end{equation}
With these definitions, and denoting trace-free tensors by drawing a line over them, we have the following algebraic decomposition of the torsion and non-metricity
\begin{subequations}
 {\small \begin{align}
  T^\rho{}_{\mu\nu} = & H^\rho{}_{\mu\nu} + \frac{1}{d-1} \left( \delta^\rho{}_\nu \tau_\mu - \delta^\rho{}_\mu \tau_\nu \right) + \kappa{}^\rho{}_{\mu\nu} \, ; \\
  Q_{\mu\nu\rho} = & \frac{1}{d+2} \left( g_{\mu\nu} A_\rho + g_{\rho\nu} A_\mu + g_{\mu\rho} A_\nu \right) + \overline{C}_{\mu\nu\rho} + \frac{1}{2(d-1)} \left( 2 g_{\mu\nu} B_\rho - g_{\mu\rho} B_\nu - g_{\nu\rho} B_\mu \right) + \overline{D}_{\mu\nu\rho} \, .
 \end{align} }
\end{subequations}
Since $D_{\mu\nu\rho}$ is the only part of the non-metricity that contributes to the part of the connection that takes values in the Lorentz algebra, the vector $B_\mu$ is the only one that can be coupled to spin-$\frac{1}{2}$ fermions, see, e.g., \cite{Rigouzzo:2023sbb}.


In the Einstein's view of MAGs, torsion can be thought of as a particular type of matter field on top of a Riemannian geometry. We adopt this field theory interpretation for torsion for the rest of the paper, and thus the covariant derivatives and curvatures will always be Riemannian. Accordingly, to lighten the notation, we drop the empty circle and use $\nabla_\mu$ for the Levi-Civita connection.

\subsection{Spin-parity decomposition of the torsion}\label{subsec:decomposition}

In a recent paper \cite{Martini:2023rnv} we proved that torsion fluctuations on top of a Riemannian background can be decomposed in modes that interpolate spin-parity eigenstates in the flat space limit. Such a decomposition reads
{\small \begin{align}\label{eq:spin-parity-decomposition}
\begin{split}
T^\rho{}_{\mu\nu} = & \frac{1}{d-1} \left( \delta^\rho{}_\nu (\tau_\mu + \partial_\mu \varphi) - \delta^\rho{}_\mu (\tau_\nu + \partial_\nu \varphi) \right) + \frac{1}{(d-3)! 3!} \varepsilon^{\sigma_1 \cdots \sigma_{d-3}\rho}{}_{\mu\nu} \nabla_{\sigma_1} \pi_{\sigma_2 \cdots \sigma_{d-3}} + \nabla^\rho \Pi_{\mu\nu} + \nabla_\mu \Pi_\nu{}^\rho + \nabla_\nu \Pi^\rho{}_\mu \\
& + \kappa^\rho{}_{\mu\nu} + 2 \nabla^\rho A_{\mu\nu} + \nabla_\mu A^\rho{}_\nu - \nabla_\nu A^\rho{}_\mu + \nabla_\mu S^\rho{}_\nu - \nabla_\nu S^\rho{}_\mu \\
& + \nabla^\rho \nabla_\mu \zeta_\nu - \nabla^\rho \nabla_\nu \zeta_\mu - \frac{1}{d-1} \left[ \delta^\rho{}_\nu \left( R_{\lambda\mu} \zeta^\lambda - \square \zeta_\mu \right) - \delta^\rho{}_\mu \left( R_{\lambda\nu} \zeta^\lambda - \square \zeta_\nu \right) \right]  \, ,
\end{split}
\end{align} }
where all the tensors that appear on the r.h.s.\ are transverse. 
The derivation of this result was obtained as a straight generalization of the work of York on metric perturbations \cite{York:1974psa} (see also \cite{Deser:1967zzb, York:1973ia}).

We would like to point out that another decomposition of the torsion fluctuations was proposed in \cite{Aoki:2019snr}.\footnote{The authors would like to thank Roberto Percacci for pointing out such a work.} For completeness, in the following we compare the two results.  A first difference is that in \cite{Aoki:2019snr} the background geometry is fixed to be four dimensional Minkowski, thus, the tensors are transverse w.r.t.\ partial and not covariant derivatives. We restrict to the same framework for this section. Focusing on the tracefree hook antisymmetric part, the decomposition in \cite{Aoki:2019snr} reads
{\small \begin{align}
\kappa^\rho{}_{\mu \nu} = 2\partial_{[\mu}\re_{\nu]}{}^\rho+2\left(\frac{\partial^\rho\partial_{[\mu}}{\square}-\frac{1}{3}\delta^\rho{}_{\mu}\right)\ri_{\nu]}
		+\varepsilon_{\mu\nu}{}^{\alpha\beta}\left[\partial_\alpha \no_{\beta}{}^{\rho}+\left(\frac{\partial_\beta\partial^{\rho}}{\square}-\frac{1}{3}\delta_{\beta}{}^{\rho}\right)\ru_\alpha\right]\,,
\end{align} }
where $\ri_\mu$ is a transverse $1$-form, $\ru_\mu$ is a transverse pseudo-vector, and $\re_{\mu\nu}$ and $\no_{\mu\nu}$ are symmetric, transverse and traceless.
%
%
%
It is easy to see that the two decompositions are related by the following change of field variables
\begin{eqnarray}
\begin{split}
 \kappa^\rho{}_{\mu\nu} = - \varepsilon_{\mu\nu}{}^{\alpha\beta} \partial_\alpha \no{}_\beta{}^\rho \, , \qquad && S_{\mu\nu} = - \re{}_{\mu\nu} \, ,\\
 \zeta_\mu = - \frac{1}{\square} \ri{}_\mu \, , \qquad 
 && A_{\mu\nu} = \frac{1}{3} \varepsilon_{\mu\nu}{}^{\alpha\beta} \frac{\partial_\alpha}{\square} \ru{}_\beta \, .
\end{split}
\end{eqnarray} 
Hence, $\re{}_{\mu\nu}$ and $\no{}_{\mu\nu}$ have positive and negative parity, respectively. 
The field reparametrization involve derivatives, thus non-trivial functional Jacobians arise when we take into account quantum fluctuations. However, we have to remark that, non-local field redefinitions tend to alter the physical interpretation of the monomials in a derivative expansion of the action. Moreover, locality is usually required to properly define perturbative renormalizability. Given the locality of the decomposition, we start from Eq.\ \eqref{eq:spin-parity-decomposition} for singling out $1$-loop divergences.

\section{The Lagrangian on maximally symmetric spaces}\label{sect:Lagrangian}


We are interested in working out the effects of the back-reaction of torsion fluctuations on the Riemannian background. In particular, we want to compute the contribution to the beta-function of the $R^2$ term in the higher derivative Lagrangian on a maximally symmetric background \cite{camporesi1992spinor}. Such a term is of interest in a cosmological context, since one of the best ways to describe the inflationary epoch is through the Starobinsky Lagrangian \cite{Starobinsky:1980te}. Concretely, we will start from an action functional that involves the kinetic terms for the torsion, as well as its couplings to the scalar curvature. From the computational viewpoint, we will be working with the spin-parity eigenstates of the torsion. These are parameterized by transverse fields, as we have proved in \cite{Martini:2023rnv}. Dealing with transverse fields hugely simplifies the differential operators, though, as we shall discuss in section \ref{sect:spectrum-of-diff-constrained}, it introduces some technical complications when we have to evaluate functional traces. As mentioned earlier, we will neglect metric fluctuations in the following analysis. An intuitive justification for this approximation arises from the perspective of effective field theory. Quantum effects of the metric field are generally expected to become significant only near the Planck scale. In contrast, torsion fluctuations can become relevant at much lower energy scales. For instance, in \cite{Shapiro:2001rz}, it is shown that the mass $m_{Tor}$ of the torsion field must satisfies $m_{\psi} \ll m_{Tor} \ll m_{Planck}$ when requiring unitarity in theories coupling a spinor field of mass $m_\psi$ to torsion. This is separation is plausible because in MAGs the torsion and the metric are independent degrees of freedom, allowing them to operate at different energy thresholds. In the following sections, we will examine more precisely the conditions under which this approximation holds, while also addressing its compatibility with the stability of the theory.

For these reasons, let us start by writing down the most general Lagrangian quadratic in the torsion involving kinetic terms and couplings to the scalar curvature. Since the metric is treated as a background field, the limit of maximally symmetric spaces can be taken from the onset, thus, we need not take into account the couplings of the torsion to the Ricci and Weyl tensors. The most general Lagrangian satisfying the aforementioned requirements reads
{\small \begin{align}\label{Lagriangian-max-sym}
\mathcal{L} = & \, a_1 (\nabla_\mu T^{\nu\rho\sigma}) \nabla^\mu T_{\nu\rho\sigma} + a_2 (\nabla_\mu T_{\rho\nu\sigma}) \nabla^\mu T^{\sigma\nu\rho} + a_3 (\nabla_\mu T^{\alpha\mu\beta}) \nabla_\nu T_\alpha{}^\nu{}_\beta + a_4 (\nabla_\mu T^{\alpha\mu\beta}) \nabla_\nu T_\beta{}^\nu{}_\alpha + a_5 (\nabla_\mu T^\mu{}_{\alpha\beta}) \nabla_\nu T^{\nu\alpha\beta}\\\nonumber
& \,\,\,  + a_6 (\nabla_\mu T^\mu{}_{\alpha\beta}) \nabla_\nu T^{\alpha\nu\beta} + a_7 (\nabla_\mu T^{\mu\alpha\beta}) \nabla_\alpha \tau_\beta + a_8 (\nabla_\mu \tau_\nu) \nabla^\mu \tau^\nu + a_9 (\nabla_\mu \tau^\mu) \nabla_\nu \tau^\nu \\\nonumber
& \,\,\, + b_1 R T^{\mu\nu\rho} T_{\mu\nu\rho} + b_2 R T^{\mu\nu\rho} T_{\rho\nu\mu} + b_3 R \tau_\mu \tau^\mu\, .
\end{align}}
The absence of other independent kinetic terms can be proved by exploiting the symmetries of the torsion tensor and integration by parts, see \cite{Baldazzi:2021kaf} for a thorough discussion on those terms which are relevant in the flat space limit. Moreover, since the background Riemannian structure is maximally symmetric, the last line of Eq.\ \eqref{Lagriangian-max-sym} provides effective mass terms for the torsion. Indeed, these are explicitly accounted for by redefining the coupling constants $b_i\to b_i+\frac{m_i^2}{R}$. Clearly, this would not possible on more general Riemannian backgrounds.

In order to simplify the form of the previous Lagrangian and to highlight the role of each term, let us decompose the torsion into its hook antisymmetric and completely antisymmetric parts
\begin{equation}
T^\rho{}_{\mu\nu} = t^{\,\rho}{}_{\mu\nu} + H^\rho{}_{\mu\nu} \, ,
\end{equation}
i.e.,\ $t_{[\mu\nu\rho]}=0$ and $H_{[\mu\nu\rho]} = H_{\mu\nu\rho}$. We find it convenient to define the shorthand for the gradients and divergences of the $t$ and $H$, which will be the building blocks of the kinetic terms for the torsion in the new basis we are about to introduce
{\small \begin{align}\label{def1-grad-div}
\begin{split}
& (grad \, t)_{\alpha}{}^\rho{}_{\mu\nu} \equiv \nabla_\alpha t^{\,\rho}{}_{\mu\nu} \, , \qquad (grad \, H)_{\alpha}{}^\rho{}_{\mu\nu} \equiv \nabla_\alpha H^{\,\rho}{}_{\mu\nu} \, , \qquad (grad \, \tau)_{\alpha\mu} \equiv \nabla_\alpha \tau_\mu \, , \\
& (div_1 t)_{\mu\nu} \equiv \nabla_\alpha t^{\, \alpha}{}_{\mu\nu} \, , \qquad (div_2 t)_{\alpha\beta} \equiv \frac{1}{2} \nabla_\rho \left( t_\alpha{}^\rho{}_\beta + t_\beta{}^\rho{}_\alpha \right) \, , \qquad (div_1 H )_{\mu\nu} \equiv \nabla_\alpha H^\alpha{}_{\mu\nu} \, , \qquad (div \, \tau) \equiv \nabla_\alpha \tau^\alpha \, .
\end{split}
\end{align} }
Such a basis is given by
{\small \begin{align}\label{eq:Lagr-kin}
\begin{split}
\mathcal{L}_{{\tiny {\rm kin}}} = & \, \alpha_1 (grad \, t)^2 + \alpha_2 (grad \, H)^2 + \alpha_3 (grad \, \tau)^2 + \alpha_4 (div_1 t)^2 + \alpha_5 (div_2 t)^2  \\
& \, + \alpha_6 (div_1  H)^2 + \alpha_7 (div \, \tau)^2 + \alpha_8 (div_1 t) (div_1 H) + \alpha_9 (div_1 t) (grad \, \tau) \, ,
\end{split}
\end{align} }
and the mapping between the two bases reads
{\small \begin{align}\label{eq:alphasToAs}
\begin{split}
& \alpha_1 = a_1 + \frac{1}{2} a_2 \, , \qquad \alpha_2 = a_1 - a_2 \, , \qquad \alpha_3 = a_8 \, , \qquad \alpha_4 = \frac{a_3 - a_4}{4} + a_5 + \frac{1}{2} a_6 \, , \qquad \alpha_5 = a_3 + a_4 \, , \\
& \alpha_6 = a_3 - a_4 + a_5 - a_6 \, , \qquad \alpha_7 = a_9 \, , \qquad \alpha_8 = a_4 - a_3 + 2 a_5 - \frac{1}{2} a_6 \, , \qquad \alpha_9 = a_7 \, .
\end{split}
\end{align} }
At this point, it is also convenient to further split the hook antisymmetric part $t^\rho{}_{\mu\nu}$ into a pure trace and a trace-free components as
\begin{equation}
t^\rho{}_{\mu\nu} = \kappa{}^\rho{}_{\mu\nu} + \frac{1}{d-1} \left( \delta^\rho{}_\nu \tau_\mu - \delta^\rho{}_\mu \tau_\nu \right) \, ,
\end{equation}
where $\kappa{}^\mu{}_{\nu\mu}=0$. We plug this algebraic decomposition and into the kinetic Lagrangian Eq.\ \eqref{eq:Lagr-kin} and into the interaction terms with the scalar curvature exploiting the results of the Appendix \ref{sect:appendix-kin-terms}, obtaining
{\small \begin{align}\label{eq:Lagrangian}
\mathcal{L} = & \, \alpha_1 (grad \, \kappa)^2 + \alpha_2 (grad \, H)^2 + \left(\alpha_3 + \frac{2 \alpha_1}{d-1} + 2\alpha_4 + \frac{\alpha_5}{2(d-1)^2} 
	- \frac{\alpha_9}{d-1} \right) (grad \, \tau)^2 + \alpha_4 (div_1 \kappa)^2 + \alpha_5 (div_2 \kappa)^2  \\ \nonumber
& + \alpha_6 (div_1  H)^2 + \left(\alpha_7 - 2 \alpha_4 + \frac{(d-2) \alpha_5}{(d-1)^2} 
	+ \frac{\alpha_9}{d-1} \right) (div \, \tau)^2 + \alpha_8 (div_1 \kappa) (div_1 H) + \alpha_9 (div_1 \kappa) (grad \, \tau) \\\nonumber
&	 + \left( b_1 + \frac{b_2}{2} \right) R \kappa^2 + \left[ \frac{2}{d-1} \left( b_1 + \frac{b_2}{2} \right) + b_3 + \frac{2}{d(d-1)^2} \alpha_4 - \frac{\alpha_5}{2d(d-1)^2} - \frac{\alpha_9}{2d(d-1)} \right] R \tau^2 + (b_1 -b_2) R H^2 \, .
\end{align} }                                                                                                                                                                                                                                                                                                             
Let us notice that there are two kinetic terms that provide mixings between different algebraic components of the torsion. Since the completely antisymmetric and pure vector parts of the torsion have a different spin-parity content, these two do not get mixed. Instead, the hook antisymmetric tracefree part carries both a $1^+$ and a $1^-$ modes, that give rise to the aforementioned mixings. Indeed, $\alpha_8$ parameterizes the mixing with the completely antisymmetric part, while $\alpha_9$ does the same job with the pure trace one.

\subsection{Conditions for the stability of the theory}

A few comments are in order. One may wonder whether a theory of this kind develops instabilities, such as ghostly and tachyonic degrees of freedom. Indeed, as discussed in \cite{BeltranJimenez:2020sqf, Heisenberg:2023wgk,Baldazzi:2021kaf}, this is a relatively common feature in MAGs. Since in our case the background geometry is maximally symmetric, the analysis of instabilities simplifies and boils down to check the behavior of the kinetic and potential energy for large momenta in each spin-parity sector (see \cite{Neville:1978bk,Neville:1979rb,Sezgin:1981xs,Sezgin:1979zf,Aoki:2019snr,Percacci:2020ddy,Baldazzi:2021kaf,Marzo:2021iok,Barker:2024ydb,Marzo:2024pyn} for more detailed analyses on analogous systems). One may solve the appropriate system of inequalities and discover that viable solutions devoid of ghosts do exist.

From the point of view of Ostrogradsky's instability, since many of the spin-parity eigenstates of the torsion propagate according to higher derivative dynamics, the Hamiltonian analysis would exhibit kinetic instabilities showing up as ghosts. However, two facts conspire against the robustness of such a statement. First, the higher derivatives are induced by a spin-parity decomposition of the torsion, thus being of kinematical and not dynamical origin\footnote{See \cite{Percacci:2020ddy} for an explicit example where the projectors on the spin-parity representations do not contribute to the physical propagators, but rather generate only spurious singularities.}.
Second, the present action is relevant when intended for $1$-loop computations on top of a non-dynamical purely metric background, since we truncated it to second order in the torsion. Indeed, our results are insensitive to the addition non-Gaussian terms in the torsion, for our torsion background is trivial. Thus, the most general Lagrangian for the torsion yields the same counterterms as those originated by Eq.\ \eqref{eq:Lagrangian} when the torsion background is set to zero, the Riemannian background is maximally symmetric and the metric is non-dynamical. For a conclusive study of the instabilities from a Hamiltonian point of view one would need to consider the entire dynamics of the torsion and the metric, since interactions between different fields can resolve instabilities without the introduction of ghosts \cite{deRham:2016wji}\footnote{Moreover, we point out that a recent revival of the Lee-Wick approach to higher derivative theories and its application to gravity is suggesting to reconsider the ghost problem from a different perspective, as unitarity might be restored at the quantum level \cite{Anselmi:2017yux,Anselmi:2017lia,Anselmi:2018ibi}}. While such a Hamiltonian analysis lies beyond the scope of the present paper, one can employ the results presented in \cite{Baldazzi:2021kaf}. The latter displays the flat-space kinetic terms of torsion and metric fluctuations for any metric-affine theory, from which one can read off the conditions for stability of the appropriate generalization of our Lagrangian Eq. \eqref{eq:Lagrangian}, which will involve the higher derivative action for the metric \cite{Stelle:1977ry}.

Here we show how to exploit the results of \cite{Baldazzi:2021kaf} to present the classes of models that are devoid of ghosts. When we employ a spin-parity decomposition we inevitably increase the order of the operators, thus introducing new unphysical kinematical poles. The same feature also appears when one uses spin-projectors on top of a flat-space background \cite{Neville:1978bk,Neville:1979rb}. By leveraging on the results of \cite{Baldazzi:2021kaf} we are guaranteed of avoiding this problem, as these unphysical poles cancel out when sandwiched between spin-parity eigenstates.

The coefficients of the sandwiched kinetic terms can be read off from the Appendix E in \cite{Baldazzi:2021kaf}, while the condition for the absence of ghosts is given in terms of the residue of the propagator evaluated at the pole, see section $4$ in \cite{Percacci:2020ddy}. The correct sign of this residue depends on the signature of the metric tensor, i.e., mostly plus or minus. A remark is needed when dealing with the $1^-$ sector, since we impose a symmetry that halves the degrees of freedom of transverse vectors. The coefficients presented in \cite{Baldazzi:2021kaf} correspond to performing a transverse decomposition first, and then expanding the result into the sum of traceless and traceful contributions\footnote{We also point out that we have found a typo in the $a(1^-)_{33}$ coefficient in \cite{Baldazzi:2021kaf}, as the correct result is given by letting $b_3^{TT} \rightarrow 3 b_3^{TT}$ and $m_3^{TT} \rightarrow 3 m_3^{TT}$.}. Since our decomposition \eqref{eq:spin-parity-decomposition} is transverse traceless from the onset, the stability of this sector is directly read off from the Hessians \eqref{eq:hessian-vectors} with the parameter choice given by \eqref{eq:canonical-norm-gradients} and \eqref{eq:coeff-from-sym-req}.

A final problem that typically arises in generic MAGs is related to strong couplings \cite{BeltranJimenez:2020sqf, Heisenberg:2023wgk, Delhom:2022vae}. Let us recall that the strong coupling problem is actually the consequence of a vanishing kinetic term when this is (naively) canonically normalized. Since the class of models that are devoid of ghosts that we are presenting below is such that all kinetic terms are non-vanishing, we conclude that our model is insensitive on the strong coupling issue.

As we have explained, the analysis of the stability of our model focuses on the absence of ghosts and is carried out employing the results of \cite{Baldazzi:2021kaf}, following the same strategies that are explained in \cite{Neville:1978bk,Neville:1979rb,Sezgin:1979zf,Sezgin:1981xs,Percacci:2020ddy}. Adopting the following shorthand $x\equiv 2b_1+b_2$, we find that there are six classes of solutions that are devoid of ghosts. The following conditions are imposed by requiring the absence of ghosts in the $1^+$ sector
	\begin{subequations}\label{eq:conditions-ghosts-1+}
		\begin{align}
		1. \qquad & m^2_2\leq 0 \,\,\, \land \,\,\,  m^2_1>-\frac{m^2_2}{2} \,\,\,  \land \,\,\,  \alpha_6<-\frac{3}{2} \, ;\\
		2. \qquad &  m^2_2\leq 0 \,\,\, \land \,\,\,  m^2_1>-\frac{m^2_2}{2} \,\,\, \land \alpha_6\geq -\frac{3}{2} \,\,\, \land \,\,\, 
		\alpha_8<-\frac{1}{2} \sqrt{-6 \alpha_6+24 \alpha_6 x+36 x-9}\, ;\\
		3. \qquad & m^2_2\leq 0 \,\,\, \land  \,\,\,  m^2_1>-\frac{m^2_2}{2} \,\,\, \land \alpha_6\geq -\frac{3}{2} \,\,\, \land \,\,\, 
		\alpha_8>\frac{1}{2} \sqrt{-6 \alpha_6+24 \alpha_6 x+36 x-9} \, ;\\
		4. \qquad & m^2_2>0 \,\,\, \land \,\,\,  m^2_1>m^2_2 \,\,\, \land \,\,\,  \alpha_6<-\frac{3}{2} \, ;\\
		5. \qquad & m^2_2>0 \,\,\, \land \,\,\,  m^2_1>m^2_2 \,\,\, \land \,\,\,  \alpha_6\geq -\frac{3}{2} \,\,\, \land \,\,\,  \alpha_8<-\frac{1}{2} \sqrt{-6 \alpha_6+24
			\alpha_6 x+36 x-9} \, ;\\
		6. \qquad & m^2_2>0 \,\,\, \land \,\,\,  m^2_1>m^2_2 \,\,\, \land \,\,\,  \alpha_6\geq -\frac{3}{2} \,\,\, \land \,\,\,  \alpha_8>\frac{1}{2} \sqrt{-6 \alpha_6+24 \alpha_6 x+36
			x-9} \, .
		\end{align}
	\end{subequations}
We reject the first three cases since they involve negative masses in the bare Lagrangian for the torsion. The analysis of the $1^-$ sector only yields the following condition on the third mass parameter
\begin{equation}
 m_3^2 >0 \, .
\end{equation}
Furthermore, by analyzing the $2^+$ and $0^+$ sectors we derive two further conditions imposing the model to be ghost-free, respectively
\begin{equation}\label{eq:conditions-ghosts-2+0+}
x>\frac{5}{12} \qquad \land \qquad \alpha_7 < \frac{1}{6} \left( 28 x - 17 \right) \, .
\end{equation}
The latter have to be satisfied independently from the previous ones, since they stem from different spin-parity sectors.



Now we list the conditions that must be fulfilled by the parameters of our Lagrangian \eqref{eq:Lagrangian} to ignore gravitons in loops. These allow us to confine ourselves to the setup of quantum field theory in curved space, rather than performing a quantum gravity computation. We find
\begin{align}\label{eq:conditions-masses}
\begin{split}
&m^2_1\ll m_p^2\,,\quad \quad \quad \quad \quad m^2_2\ll m_p^2\,,\quad \quad \quad \quad \quad m^2_3\ll m_p^2\,,\quad \quad \quad \quad \quad \frac{2m^2_1+m^2_2}{\left|12x-5\right|}\ll m_p^2\,,\\
& \left|\frac{2m^2_1+m^2_2+3m^2_3}{17-28x+6\alpha_7}\right|\ll m_p^2\,,\quad  \quad \left| \frac{4 \alpha_6 m^2_1+72 m^2_1 x-12 m^2_1+2 \alpha_6
	m^2_2-72 m^2_2 x+21 m^2_2 \pm \sqrt{y}}{2 \left(-6 \alpha_6-4 \alpha_8^2+12 (2 \alpha_6+3)
	x-9\right)} \right| \ll m_p^2   \, ,
\end{split}
\end{align}
where the last condition is written in terms of
{\small
\begin{align}
 y = (4 m^2_1 (\alpha_6+18 x-3)+m^2_2 (2 \alpha_6-72 x+21))^2-24 \left(2
 m^4_1-m^2_1 m^2_2-m^4_2\right) \left(-6 \alpha_6-4 \alpha_8^2+12 (2
 \alpha_6+3) x-9\right) \, .
\end{align}}
We have derived the first three conditions in \eqref{eq:conditions-masses} from the masses of the $2^-$, $1^-$ and $0^-$ states, see \cite{Baldazzi:2021kaf} and App.\ \ref{sect:appendix-hessians}. On the other hand, the fourth and fifth conditions come from the results of \cite{Baldazzi:2021kaf} involving the $2^+$ and $0^+$ sectors, respectively. The last condition pertains to the $1^+$ sector. As we have said above, in \cite{Baldazzi:2021kaf} one can find the sandwiched flat-space expression of the kinetic terms in each spin-parity sector for any class of metric-affine theories. In general, these expression are given by $n \times n$ matrices, where $n$ is the number of modes carrying that specific spin-parity quantum numbers. In the case of the $1^+$ sector we have a $2\times2$ matrix, and the masses are read off from finding the two zeros of the determinant of this matrix.

Having stated the conditions for which our approximation of neglecting gravitons in loops is well-posed, we discuss their compatibility with the requirement of absence of ghosts. The fourth and fifth inequalities in \eqref{eq:conditions-masses} tell us that the conditions in \eqref{eq:conditions-ghosts-2+0+} cannot be saturated. More concretely, by defining
\begin{eqnarray}
 z_1 = x - \frac{5}{12} \, , \qquad && z_2 = \alpha_7 - \frac{1}{6} \left( 28 x - 17 \right) \, ,
\end{eqnarray}
we must have that
\begin{eqnarray}\label{eq:bound1}
 \left| z_1 \right| > 10^{-1} \, ,  \qquad && \left|z_2\right| > 10^{-1} \, .
\end{eqnarray}
Now we focus on the last condition in \eqref{eq:conditions-masses}. Here the situation is a bit more complicated, since parameters $\alpha_6$, $\alpha_8$ and $x$ appear also in the numerator. This fact leads us to require that these couplings cannot be large, i.e.,
\begin{equation}\label{eq:bound2}
 \alpha_6 \approx 1 \, , \qquad \alpha_8 \approx 1 \, , \qquad x \approx 1 \, .
\end{equation}
Moreover, we also impose that the denominator in the last condition in \eqref{eq:conditions-masses} does not approach zero. Again, this is nothing but asking that the inequalities involving $\alpha_8$ in \eqref{eq:conditions-ghosts-1+} cannot be close to being saturated. Therefore, we define
\begin{equation}
 z_3 = 4 \alpha_8^2 -6 \alpha_6+12 (2 \alpha_6+3)
 x-9 
\end{equation}
and we demand that
\begin{equation}\label{eq:bound3}
 \left| z_3 \right| > 10^{-1} \, .
\end{equation}
In conclusion, when the bounds \eqref{eq:bound1}, \eqref{eq:bound2}, and \eqref{eq:bound3} are satisfied we are in a class of theories which are devoid of ghost and for which our approximation of discarding the metric field into loops is viable.

The Lagrangian Eq.\ \eqref{eq:Lagrangian} will be our starting point. Since we will be performing an integration at $1$-loop of the fluctuations, quadratic terms in the torsion modes is all we need. By inspection of the previous equation and Eq.\ \eqref{eq:spin-parity-decomposition} we readily see that the transverse modes ($2^-$, $1^-$ and $0^-$) will have second-order operators, while longitudinal ones are described as higher-derivative fields. Most notably, the twice-longitudinal mode parameterized by $\zeta_\mu$ will display a sixth-order operator. We shall come to this point in section\ \ref{sect:1-loop-body}. Having introduced the model, in the next two sections we present the technical tools that are necessary to the analysis.

\section{A primer on the Schwinger-DeWitt technique}\label{sect:Schwinger_DeWitt_technique}

In this section we provide a brief but pedagogical introduction to the Schwinger-DeWitt technique and its generalizations, mainly following the presentation in \cite{Barvinsky:1985an}. For the rest of this work we focus on the Euclidean formulation of gravity and, therefore, we only consider a Riemannian background.

The Schwinger-DeWitt technique offers a very general method for calculating Feynman diagrams in a covariant fashion \cite{Schwinger:1951nm,DeWitt:1964mxt}, see \cite{Vassilevich:2003xt} for an extensive review. Moreover, it can be combined with any regularization scheme, although in the following we will use dimensional regularization, in which only logarithmic divergences appear. Let us start by considering a multicomponent field $\varphi(x) \equiv \varphi^A(x)$, where the index $A$ carries both the Lorentz and gauge group representations \cite{Percacci:2017fkn}. Then, the kinetic operator of the theory, or equivalently the Hessian of the classical action $S[\varphi]$, is the matrix valued differential operator
\begin{equation}\label{eq:kinetic_operator}
\hat{O}
=
\frac{\delta^2 S[\varphi]}{\delta \varphi(x)\delta \varphi(y) } \, , 
\end{equation}
where hat stands for $\hat{O}=O^A{}_B$. The most important quantity in the Schwinger-DeWitt technique is the heat kernel $\hat{K}(s|x,y)=K^A{}_B(s|x,y)$. 
The heat kernel is defined as the solution of a diffusion process generated by $\hat{O}$ along the proper time $s$, and can be thought as a Wick rotated Schr\"{o}dinger-like equation. Formally, the process is given in terms of the following initial values problem 
\begin{equation}\label{eq:heat_equation}
\frac{\partial}{\partial s} \hat{K}(s|x,y)
+
\hat{O} \hat{K}(s|x,y)
=
0 \, , \qquad
\hat{K}(0|x,y)
=
\hat{\mathds{1}} \delta(x,y)
\end{equation}
where 
$ \hat{\mathds{1}} \equiv \delta^A{}_B$ represents the identity on the fields representation space.                      
It is easy to see that a formal solution of \eqref{eq:heat_equation} is given in terms of the following matrix elements
\begin{equation}
\hat{K}(s|x,y)
=
\langle x |   e^{-s \hat{O}} | y \rangle \, , 
\end{equation}
where we implicitly defined the operator $\hat{K}(s)=e^{-s \hat{O}}$. $K(s|x, y)$ is the kernel of the integral representation defining the evolution of fields in the proper time $s$
{\small \begin{align}
\varphi(s_1|x) = \int_0^{s_1}{\rm d}s\;K(s|x, y)\,\varphi(y)\,,
\end{align} }
much like the Hamiltonian generates the unitary time evolution of states in the Scr\"{o}dinger picture of quantum mechanics.

For us, the importance of the heat kernel relies in the following Schwinger proper time representation of the propagator $\hat{G}$ in terms of $\hat{K}$ \cite{Barvinsky:1985an}
\begin{equation}\label{eq:relation_G_HK}
\hat{G}
=
\hat{O}^{-1}
=
\int_0^\infty ds \,\hat{K}(s) \, ,
\end{equation}                                                                                                                                                                         
which is useful for multi-loop computations in coordinate space, as we are about to discuss. Notice that, in this language, the propagator is an operator too, while $G(x, y)=\langle x|\hat{G}|y \rangle$ is the standard Green function for the differential operator $\hat{O}$. Loosely speaking, equation \eqref{eq:relation_G_HK} describes the $2$-point function $G(x, y)$ as the total probability for a random walk to start at $x$ and end in $y$. 
This representation establishes a direct relation between the heat kernel and the one-loop corrections to the effective action, as we are going to show.

As it is well-known, the expansion in $\hbar$ of the effective action reads  
\begin{equation}
\Gamma[\varphi]=S[\varphi] + \hbar \Gamma_{\text{1-loop}}+ \hbar^2 \Gamma_{\text{2-loop}} + \dots \, ,
\end{equation}
where
\begin{equation}
\Gamma_{\text{1-loop}}
=
\frac{1}{2} \ln \det \hat{O}
=
\frac{1}{2} \text{Tr} \ln \hat{O} \, ,
\end{equation}
which diagramatically is $1/2$ times a single loop. 
If we assume to have a complete set of eigenfunctions $\varphi_n$ and eigenvalues $\lambda_n$ for $\hat{O}$, it is not difficult to see that $\Gamma_{\text{1-loop}}$ is related to the $\zeta$-function $\zeta_{\hat{O}}(r)=\sum_n (\lambda_n)^{-r}$ by \cite{Percacci:2017fkn}
\begin{equation}\label{eq:HK_zetafunction}
\Gamma_{\text{1-loop}}[\varphi]= - \frac{1}{2} \frac{d}{dr} \zeta_{\hat{O}}(r) |_{r=0} \, .
\end{equation}
The $\zeta$-function can be re-expressed as
\begin{equation}
\zeta_{\hat{O}}(r)
=
\sum_n (\lambda_n)^{-r}
=
\frac{1}{\Gamma(r)}\int_0^\infty \left(\sum_n e^{-\lambda_n s}\right) s^{r-1} ds
=
\frac{1}{\Gamma(r)}\int_0^\infty  s^{r-1}   \left(\text{Tr} \hat{K}(s) \right) ds,
\end{equation}
where we used the fact that $\Gamma(r) = \int_0^\infty e^{-s} s^{r-1} ds=\lambda^r \int_0^\infty   e^{-\lambda s} s^{r-1} ds$. Then, from equation \eqref{eq:HK_zetafunction}, we get
\begin{equation}\label{eq:relation_EA_HK}
\frac{1}{2} \,  \text{Tr} \ln \hat{O}
=
- \, \frac{1}{2} \int_0^\infty \frac{ds}{s} \,   \text{Tr} \, \hat{K}(s) \, .
\end{equation}
In the previous equation, $\text{Tr} $ indicates the functional trace involving the usual trace over matrix indices (denoted with $\text{tr}$) and an integral of the coincidence limit of $\hat{K}(s|x,y)$ over the background manifold, that is $\text{Tr} \, \hat{K}(s) = \int dx \, \text{tr} \, \hat{K}(s|x,x)$. Up to now, the formalism is completely general, i.e., it is valid for any kind of operator, but it is not very useful for practical calculations. Its efficiency in computing singularities in the effective action, $\beta$-functions and all sort of anomalies is due the so called early time ($s \to 0$) asymptotic expansion of the heat kernel \cite{Schwinger:1951nm,DeWitt:1964mxt}
\begin{equation}\label{ansatzSeeleyDeWittExpFinal}
\hat{K}(s|x,y)
=
\frac{\Delta^{{1/2}}(x,y)}{ (4\pi s)^{d/2} }\,\exp\left\{-\frac{\sigma(x,y)}{2s}\right\} \, \sum_{n=0}^{\infty} s^n \hat{a}_n(x,y) \, . 
\end{equation}
In the previous equation, $\sigma(x,y)$ is the Synge world function, defined by the identity $\frac{1}{2}\,\nabla_\mu\sigma\nabla^\mu\sigma=\sigma$, while $\Delta(x,y)=|\det \partial_\mu^x\partial_\nu^y \sigma(x,y) | $ is the Van Vleck-Morette determinant \cite{Poisson:2003nc}. However, it is important to stress that equation \eqref{ansatzSeeleyDeWittExpFinal} holds only for the so-called minimal operators 
\begin{equation}\label{eq:minimal_II_ord}
\hat{O} 
=
- \hat{\mathds{1}}\square
-
\hat{E} \, ,
\end{equation}
as it can be verified by solving the heat equation on flat space and ``covariantizing''. In general, the covariant derivatives in \eqref{eq:minimal_II_ord} contain both a gauge and the tangent-space connection. Notice that for $\hat{O}$ of this form, $s$ has dimensions of inverse square length. Hence, the lower end of the integral \eqref{eq:relation_EA_HK} is associated with UV physics, which is then contained in the (coincidence limit of) heat kernel coefficients $\hat{a}_n$ appearing \eqref{ansatzSeeleyDeWittExpFinal}. For example, to compute UV divergences of the effective action and $\beta$-functions, we need to compute these coefficients. This is achieved thanks to the following recursive equation \cite{DeWitt:1964mxt}
{\small \begin{align}
\label{eq:recursiveCoeff}
n \hat{a}_n + \nabla^\mu\sigma\nabla_\mu \hat{a}_n + \Delta^{-1/2}\hat{O}\left(\Delta^{1/2}\hat{a}_{n-1}\right)=0\,,
\end{align} }
obtained upon inserting \eqref{ansatzSeeleyDeWittExpFinal} in the heat equation \eqref{eq:heat_equation}. The advantage of such an approach is that Eq.~\eqref{eq:recursiveCoeff} makes particularly easy the employment of computer algebra to compute coincidence limits of the coefficients $\hat{a}_n$ and their derivatives.
A relevant aspect of these coefficients is that they depend \emph{only} on the operator considered. For example, for the operator \eqref{eq:minimal_II_ord}, they depend only on the endomorphism $\hat{E}$ and on the set of curvatures 
\begin{equation}\label{eq:generalized_curvatures}
\begin{split}
&[\nabla_{\mu},\nabla_{\nu}]V^\alpha=R^{\alpha}{}_{\beta\mu\nu} V^\beta \, ,\\
&[\nabla_{\mu},\nabla_{\nu}]\,\varphi=\hat{\mathcal{R}}_{\mu\nu} \, \varphi \, ,
\end{split}
\end{equation}
where $V^\alpha$ is a vector field, so that the first is the Riemann tensor, while $\hat{\mathcal{R}}_{\mu\nu} \equiv \mathcal{R}^A{}_{B\mu\nu}$ is the curvature of the full vector bundle where $\varphi$ lives. For instance, when the latter is a gauge field in flat space the curvature of the vector bundle is simply the field strength. Since we will work in four dimensions, we are interested in finding $\hat{a}_2(x,x)$. This is due to the fact that the coefficient associated with logarithmic divergences is, in general, $\hat{a}_{d/2}$, as it is also clear for dimensional reasons. It can be shown that, for $d=4$, the coincidence limit of $\hat{a}_2$ reads \cite{Percacci:2017fkn}
\begin{equation}\label{eq:a2-minimal-second-order}
    \hat a_2(x,x)
    =
    \frac{ \hat{\mathds{1}} }{180}\, \left(R^{\,2}_{\alpha\beta\mu\nu} - R^{\,2}_{\mu\nu}+ \frac{5}{2}\, R^2 + 6\,\Box R \right) 
    + \,\frac{1}{2} \hat E^2\,
    + \,\frac{1}{6} R \hat E \,
    + \,\frac{1}{12} \,\hat{\mathcal{R}}^{\,2}_{\mu\nu}
    + \,\frac{1}{6} \Box \hat E \,\, ,
\end{equation}
with the conventions of equations \eqref{eq:generalized_curvatures}. Therefore, we have
{\small \begin{align}\label{eq:EAdiv_minimal-second-order}
-\left.\frac{1}{2} {\rm Tr} \ln \hat{O}\right|^{\rm div} = - \frac{\mu^{-\epsilon}}{16 \pi^2 \epsilon} \int \sqrt{g} \, {\rm tr} \, \hat a_2(x,x),
\end{align} }
where we are employing dimensional regularization to compute Feynman diagrams in $d=4-\epsilon$. In the special case where the background Riemannian geometry is given by a maximally symmetric space, Eq.\ \eqref{eq:a2-minimal-second-order} becomes
\begin{equation}
{\rm tr} \hat{a}_2 = \frac{29}{2160} R^2 {\rm tr} \hat{\mathds{1}} + \frac{1}{2} {\rm tr} \hat{E}^2 + \frac{R}{6} {\rm tr} \hat{E} + \frac{1}{12} {\rm tr} \hat{\mathcal{R}}_{\mu\nu} \hat{\mathcal{R}}^{\mu\nu} \, .
\end{equation}
This equation will be employed for dealing with second-order operators throughout this paper.

It is useful to stop for a moment and have a small remark on the IR behavior of the expansion. One may wonder whether the introduction of a physical mass term would affect the previous formula. Indeed, a mass would act as a natural IR regulator that should not be removed in dimensional regularization. As a consequence, we may expect that the mass produces logarithmic divergences determined by the coefficients $\hat{a}_0$ and $\hat{a}_1$. However, such an effect can be easily taken into account as a shift of the endomorphism of the operator. Let us consider the two differential operators $\hat{\cal O} = -\hat{\square} - \hat{E}+m^2\hat{\mathds{1}}$ and $\hat{\cal O}' = -\hat{\square} - \hat{E}$, with heat kernel coefficients $\hat{a}_n$ and $\hat{a}'_n$, respectively. Isolating the contribution of the mass from $\hat{\cal O}$ and writing the Schwinger representation for ${\rm Tr} \log \hat{\cal O}$ as
\begin{align}
{\rm Tr}\log\hat{\cal O} = -\int_0^\infty\frac{{\rm d}s}{s}\; e^{-s m^2}{\rm Tr}\, e^{-s\hat{\cal O}'}\,,
\end{align}
we notice that the leading divergences (in $d=4$) come in the following form
\begin{align}\label{eq:divergenceWithMass}
\left.{\rm Tr}\,\log\hat{\cal O}\right|_{\rm div} = -\frac{1}{(4\pi)^2}\sum_{n=0}^2{\rm tr}\,\hat{a}'_n\,\Gamma\left(n-2+\frac{\epsilon}{2}\right)\sim
		-\frac{2}{16\pi^2\epsilon}\left[\frac{1}{2}{\rm tr}\,\hat{a}'_0\,m^4-{\rm tr}\,\hat{a}'_1\,m^2+{\rm tr}\,\hat{a}'_2\right]\,.
\end{align}
Plugging the following form of the primed coefficients
\begin{align}
{\rm tr}\,\hat{a}'_0 = {\rm tr}\,{\mathds 1}, \qquad {\rm tr}\,\hat{a}'_1 = \frac{R}{6}{\rm tr}\,{\mathds 1}+{\rm tr}\,\hat{E}\,,
\end{align}
into \eqref{eq:divergenceWithMass}, it is easy to check that the introduction of the mass reproduces exactly the form of the coefficient $\hat{a}_2$ with an endomorphism of the form $\hat{\cal E}= \hat{E}-m^2\hat{\mathds 1}$
\begin{align}
{\rm tr}\,\hat{a}_2 = \frac{1}{2}{\rm tr}\,\hat{a}'_0\,m^4-{\rm tr}\,\hat{a}'_1\,m^2+{\rm tr}\,\hat{a}'_2\,.
\end{align}
Such a statement is a general feature of the heat kernel formalism, and we will use it to drop explicit mass terms for the torsion field since, on maximally symmetric spaces, these contributions can easily be extracted upon a redefinition of the coupling constants for the operators of the form $RT^2$. We shall come back to this point later.

It is possible to extend this approach to more general operators. The basic ideas are either to find the appropriate generalization of \eqref{ansatzSeeleyDeWittExpFinal}, see for example \cite{Barvinsky:2021ijq}, or to perform perturbation theory with respect to the nonminimal terms \cite{Barvinsky:1985an}. We choose this second approach. In order to keep track of perturbative orders, we introduce a background length $l$. In the approach of \cite{Barvinsky:1985an}, the power of $l$ keeps track of the dimensions of the coefficients multiplying derivatives in the expansion of $\hat{O}$. In the following we will need only the case of minimal operators, but we will consider higher order derivatives: $\hat{O}=\hat{\square}^k + \hat{\Pi}(\nabla)$, for some integer $k$. In this case, $\hat{\Pi}(\nabla)$ has at least one less derivative then $\hat{\square}^k$, so that its coefficients are of order at least $O[1/l]$, a fact which considerably simplifies perturbation theory as performed in \cite{Barvinsky:1985an}. We now briefly recall how to obtain the divergent part of the effective action for these operators. We are interested in computing $\frac{1}{2} \,  \text{Tr} \ln (\hat{\square}^k + \hat{\Pi}(\nabla))$. From standard manipulations it is easy to see that
\begin{equation}
\frac{1}{2} \,  \text{Tr} \ln ( \hat{\mathds{1}}{\square}^k + \hat{\Pi}(\nabla))
=
\frac{k}{2} \, \text{Tr} \ln  (\hat{\mathds{1}}{\square} )- \frac{1}{2} \, \text{Tr} \, \sum_{p=1}^4\,\frac{1}{p}  \left( - \hat{\Pi} \frac{\hat{1}}{\square^k} \right)^p + O[1/l^5] \, ,
\end{equation}
where we stopped at $p=4$ because, as before, if we are to find logarithmic divergences in four dimensions, we only need to reach the order of curvature squared. If we now commute the operators $\hat{1}/\square^k$ to the right we see that all we have to compute are functional traces of quantities of the form
\begin{equation}
\left.\nabla_{\mu_1} \cdots \nabla_{\mu_r} \frac{ \hat{1} }{\square^n} \delta(x,y)\right|_{y=x} \, ,
\end{equation}
whose divergent part can be obtained from the expansion \eqref{ansatzSeeleyDeWittExpFinal} since $\frac{  \hat{1}  }{\square^n}=\frac{1}{(n-1)!}\left.\left[\left(  \frac{d}{dm^2}     \right)^{n-1}    \frac{ \hat{1}}{\square-m^2}      \right]\right|_{m=0}$. Therefore, we have the proper time representation
\begin{equation}
\frac{\hat{1}}{\square^n}
=
\frac{1}{(n-1)!}\int_0^\infty ds \, s^{n-1} e^{-s  ( \hat{\mathds{1}}{\square})}
\end{equation}
where we used \eqref{eq:relation_G_HK} for the inverse of the second order minimal operator. With this procedure, it is possible to obtain the logarithmic divergences for fourth order operators of the form
\begin{equation}\label{eq:fourth-order-operator}
\hat{O} =  \hat{\mathds{1}}{\square}^2 + \hat{\Omega}^{\mu\nu\rho} \nabla_\mu \nabla_\nu \nabla_\rho + \hat{D}^{\mu\nu} \nabla_\mu \nabla_\nu + \hat{H}^\mu \nabla_\mu + \hat{P} =   \hat{\mathds{1}}{\square}^2 + \hat{\Pi}(\nabla) \, ,
\end{equation}
as it has been done in \cite{Barvinsky:1985an}.
We report this result only in the special case $\hat{\Omega}^{\mu\nu\rho}=0$ and $\hat{H}^\mu=0$
{\small \begin{align}\label{eq:a2-minimal-fourth-order}
-\left.\frac{1}{2} {\rm Tr} \ln \hat{O} \right|^{\rm div} = - \frac{1}{16 \pi^2 \epsilon} \int \sqrt{g} \, {\rm tr} & \left[ \frac{ \hat{\mathds{1}}}{90} \left( R_{\mu\nu\rho\sigma} R^{\mu\nu\rho\sigma} - R_{\mu\nu} R^{\mu\nu} \right) + \frac{1}{6} \hat{\mathcal{R}}_{\mu\nu} \hat{\mathcal{R}}^{\mu\nu} + \frac{1}{36} R^2 \hat{\mathds{1}} - \hat{P} - \frac{1}{6} \hat{D}^{\mu\nu} R_{\mu\nu} \right.\\\nonumber
& \,\,\, \left.  + \frac{1}{12} \hat{D} R + \frac{1}{48} \hat{D}^2 + \frac{1}{24} \hat{D}_{\mu\nu} \hat{D}^{\mu\nu} \right] \, ,
\end{align} }
which is the one relevant for us. In the previous language, the coefficient $\hat{a}_2(x,x)$ is the integrand of this equation.

The fourth-order operators whose spectrum we will study in the subsequent sections are even simpler than the parameterization used in Eq.\ \eqref{eq:fourth-order-operator}. Indeed, they all have the general form
\begin{equation}
 \hat{O}= \square^2 + \sigma_1 R \square + \sigma_2 R^2 \, ,
\end{equation}
and the trace of the second Seeley-DeWitt coefficient on maximally symmetric spaces in $d=4$ can be written as
\begin{equation}\label{eq:a2-minimal-fourth-order-max-sym}
 {\rm tr } \, a_2 = \left[ \frac{29}{1080} + \frac{\sigma_1}{6} - \sigma_2 + \frac{\sigma_1^2}{2} \right] R^2 {\rm tr} \mathds{1} + \frac{1}{6} {\rm tr} \left( {\cal R}_{\mu\nu} {\cal R}^{\mu\nu} \right) \, ,
\end{equation}
which we will frequently use in the following.

Before delving into our analysis, it might be helpful to outline some general considerations on our approach. The $1$-loop effective action considered here, once renormalized, physically encapsulates the quantum back-reaction effects induced by matter fields on the underlying Riemannian geometry. For instance, this effective action could be varied w.r.t. the metric to investigate the graviton propagator dressed with torsion fluctuations. As we have explained at the end of section\ \ref{sect:MAGs_review}, torsion can be formally treated as a type of tensorial matter field. Consequently, this framework suffices to compute the impact of torsion.

However, while computing the finite part of the effective action would hold significant theoretical interest, we will concentrate on the more modest task of calculating the beta functions for $R$ and $R^2$ that are induced by torsion. We proceed in this way because the techniques required to derive a fully explicit, finite effective action suitable for further computations are considerably more complex than the heat kernel method employed in our analysis. Nonetheless, it is noteworthy that the heat kernel approach allows us to formally express the finite part of the effective action as the \emph{asymptotic} series \eqref{ansatzSeeleyDeWittExpFinal} with $n \geq 3$. Indeed, this is what remains after subtracting the coefficients corresponding to the divergences for $n=0, 1, 2$, i.e., after renormalization. The relation between this series and the $1$-loop effective action can be found in Eq.~\eqref{eq:relation_EA_HK}. Moreover, not only is Eq.~\eqref{ansatzSeeleyDeWittExpFinal} an asymptotic series, but the heat kernel coefficients also become increasingly complex, and their explicit form for $n > 3$ remains generally unknown.

A well-established alternative for addressing these issues is the covariant perturbation theory developed by Barvinsky and Vilkovisky in \cite{Barvinsky:1987uw-covariant-perturbation}.  In essence, this approach effectively resums series such as Eq.~\eqref{ansatzSeeleyDeWittExpFinal}, and typically yields a non-local effective action. However, this method has yet to be applied to the operators pertinent to our study and is valid only on asymptotically flat spacetimes. For these reasons, our analysis is confined to extracting the one-loop divergences, which nonetheless gives important results, including the beta functions computed here.

Thus, our focus will be on using the heat kernel technique to compute the logarithmically divergent part of the effective action according to Eq.~\eqref{eq:EAdiv_minimal-second-order}, which is sufficient for the calculation of the beta functions. As we have just saw, in this framework it is necessary to take the second variation of the starting action with respect to the quantum fields only. Since we do not include gravitons in loops, variations with respect to the metric are not required.

To further elucidate our approach, we present an analogy with the Higgs model. In this context, as a first approximation, one may compute the beta-function of the Higgs sector due to the fluctuations of the Standard-Model fields that are lighter than the Higgs boson, i.e., excluding the top quark. The Higgs field can be regarded as a constant background, a justification grounded in its mass being greater than that of the other fields. In this analogy, the metric corresponds to the Higgs field, while torsion plays the role of the Standard-Model matter and gauge fields. Importantly, we assume $m_{Tor} \ll m_{Planck}$, rendering our approximation more robust than the one we have just outlined in the Higgs model.

Having introduced the heat kernel formalism, we will now focus on the issue of zero-modes, which arises as we evaluate functional traces on compact manifolds.

\section{Spectral geometry of differentially constrained fields}\label{sect:spectrum-of-diff-constrained}
                                                                                                                                                                                                                                                                                                                                                                                                                                         
An important issue in defining perturbation theory is the construction of a parametrix, i.e., a propagator defined through the removal of a finite number of zero-modes. The conditions of ellipticity and hermiticity of a kinetic operator are usually employed to guarantee that it has finite kernel and co-kernel, therefore admitting a well defined parametrix for a given appropriate choice of boundary conditions \cite{Witten:2018lgb}. The importance of subtracting the zero-modes can be easily highlighted by considering a $1$-loop computation of the partition function within the background field method. For bosonic fields $\varphi$ on a background $\bar{\varphi}$, the result involves the computation of the inverse square root of the functional determinant of a kinetic operator $\cal O$
\begin{align*}
Z[J; \bar{\varphi}]\sim \rm{Det}\,{\cal O}^{-\frac{1}{2}}\exp\left\{-S[\bar{\varphi}]+\frac{1}{2}J_xG_{xy}J_y\right\}\times\text{ higher loops}\,,
\end{align*}
that is divergent if the determinant is computed on a spectrum including any zero-mode.

Special care is needed when removing of the kernel of a differential operator acting on differentially constrained fields, which usually arise through a decomposition into transverse and longitudinal modes. For example, one may wish to isolate the longitudinal modes with a decomposition of the form
{\small \begin{align}
\label{eq:genericDecomposition}
\varphi = \sum_{i} \hat{\cal C}_{i} \phi_i\,.
\end{align} }
In \eqref{eq:genericDecomposition}, the index $i$ runs over the spin-parity content $\phi_i \equiv \phi_i^A$ of $\varphi \equiv \varphi^A$, and $\hat{\cal C}_{i}  \equiv{{\cal C}_i^A}_B$ are the differential operators that map each longitudinal mode to a tensor with the symmetries of $\varphi$ (one may have in mind a traceless symmetric $2$-tensor and the conformal Killing form). Given a kinetic operator of the form \eqref{eq:kinetic_operator}
we wish to solve the equation of motion
{\small \begin{align}
\label{eq:dynamicalZeroModes}
\hat{{\cal O}} \, \varphi=0\,.
\end{align} }
By empolying the decomposition \eqref{eq:genericDecomposition}, we decouple the computation into the following set of differential equation
{\small \begin{align}
\label{eq:decomposedZeroModes}
\hat{{\cal O}}  \,   \hat{\cal C}_{i} \, \phi_i    =0\,.
\end{align} }
Looking at \eqref{eq:decomposedZeroModes}, we notice how the decomposition introduces a new set of solutions to the equations of motion, namely those modes belonging to the kernel of the longitudinal operators
{\small \begin{align}
\label{eq:kinematicalZeroModesEq}
\hat{\cal C}_{i} \, \phi_i    =0\,.
\end{align} }
When computing the traces and determinants over the longitudinal parts, solutions of \eqref{eq:kinematicalZeroModesEq} are responsible for a miscounting of the zero-modes as they appear in the logarithmic divergences of the effective action, but were not meant to be included in the spectrum from the start, as they do not contribute to $\varphi$. For this reason we will dub these solutions \emph{kinematical zero-modes} and we will need to remove them by hand from the decomposed spectrum.

Once we single-out the kinematical zero-modes, we want to identify the actual solutions of \eqref{eq:dynamicalZeroModes}, i.e., the solutions up to kinematical zero-modes. Since these are solely due to the choice of an action, we will dub them \emph{dynamical zero-modes}. To disentangle them from the kinematical ones, we can imagine decomposing the action of the longitudinal operators $\cal C$ on a set of eigenfunctions through the spectral theorem, automatically removing the null eigenvalues
{\small \begin{align}
\hat{{\cal O}}  \,   \hat{\cal C}_{i} \, \phi_i  
=
\hat{{\cal O}}     \int {\rm d}\mu_{\cal C} \, \lambda\, \hat{        \mathbb{P}}_i^{ \lambda} \, \phi_i
= 
\int {\rm d}\mu_{\cal C}\;\lambda\, \hat{        \mathbb{P}}_i^{ \lambda} \, \hat{   \tilde{ {\mathcal O} }  }_i \, \phi_i
=
 \hat{\cal C}_{i}  \hat{   \tilde{ {\cal O} }  }_i      \phi_i   \,.
\end{align} }
In the previous equation, the spectral measure ${\rm d}\mu_{\cal C}$ is defined over the non vanishing spectrum of $\hat{{\cal C}}_i$, therefore factoring out the kinematical zero-modes. The commutation procedure shows us that the dynamical zero-modes of the kinetic operator $\hat{\cal O}$, once we decompose our field content, can be made in one-to-one correspondence with the kernel of the operators $\hat{\tilde{{\cal O}}}_i$ obtained after commuting with the longitudinal operators $\hat{{\cal C}}_i$, up to possible overlap between the kernel of $\hat{{\cal C}}_i$ and that of $\hat{\tilde{{\cal O}}}_i$.

As a final remark, note that the highest-spin content of the field $\varphi_A$, i.e., the fully transverse mode, will project through a trivial operator ${\cal C}^1 = \delta^{TT}$ corresponding to the identity on the transverse (and possibly traceless) modes. Therefore, the operator $\hat{\tilde{{\cal O}}}_1$ will be equal in form to $\hat{\cal O}$ and its kernel will be a proper subset of the kernel of $\hat{\cal O}$. Obviously, since the amount of the dynamical zero modes of \eqref{eq:dynamicalZeroModes}
and \eqref{eq:decomposedZeroModes} must be the same, the kernel of $\hat{\tilde{{\cal O}}}_1$ will be in general smaller than that of $\hat{\cal O}$.	

Throughout the body of the paper we shall frequently exploit the fact that, on a maximally symmetric space and in absence of other background fields beside the metric, the spectrum of a given differential operator is the disjoint union of the spectra on their spin-parity eigenstates, modulo zero-modes. Indeed, in the presence of any such non-trivial background tensor field the contributions of the spin-parity components will not be orthogonal to each other anymore. Given the relevance of this result for our work, we will briefly prove it here. 

Let us first recall how we find the functional Jacobian for the decomposition \eqref{eq:genericDecomposition}. The functional measure is defined by the normalization of the Gaussian integral \cite{Percacci:2017fkn,Mazur:1990ak,Mottola:1995sj}, i.e.,
{\small \begin{align}
1 = \int D\varphi\,\exp\left\{-\frac{1}{2}\int\sqrt{g}\,\varphi^A\varphi^A\right\}
		=\int \left( {\tiny{\prod_i}} D\phi_i\right)\;\det J \exp\left\{-\frac{1}{2}\sum_{i, j}\int\sqrt{g}\,\phi_i\hat{\cal C}_{i}^\dagger\hat{\cal C}_{j}\phi_j \right\}\,.
\end{align} }
The operators $\hat{\cal C}_i$ map the transverse tensors that parameterize the longitudinal modes to a tensor with the symmetries of $\varphi$, and on maximally symmetric spaces they are orthogonal to each other, thus
{\small \begin{align}
\label{eq:general-jacobian-decomposed}
\det J  = \prod_i \det \left(\hat{\cal C}_i^\dagger\hat{\cal C}_i\right)^{1/2}\,.
\end{align} }
Let us now consider the determinant of a second order differential operator $\hat{\cal O}$
{\small \begin{align}
\det(\hat{\cal O})^{-1/2} = \int D\varphi\,\exp\left\{-\frac{1}{2}\int\sqrt{g}\,\varphi^A\hat{\cal O}^{AB}\varphi^B\right\}
		=\int \left( {\tiny{\prod_i}} D\phi_i\right)\;\det J \exp\left\{-\frac{1}{2}\sum_{i, j}\int\sqrt{g}\,\phi_i\hat{\cal C}_{i}^\dagger\hat{\cal O}\hat{\cal C}_{j}\phi_j \right\}\,.
\end{align} }
The crucial point is that on maximally symmetric spaces we can always write $\hat{\cal O} =  \hat{\mathds{1}}( -\square +\alpha R)$ for some constant $\alpha$. Then, commuting $\hat{\cal O}$ with $\hat{\cal C}_i^\dagger$ simply shifts the constant $\alpha\to \tilde{\alpha}_i$. Therefore, for $\hat{\tilde{{\cal O}}}_i = \hat{\mathds{1}}( -\square +\tilde{\alpha}_iR)$ we can write
{\small \begin{align}
\det(\hat{\cal O})^{-1/2} =\int \left( {\tiny{\prod_i}} D\phi_i\right)\;\det J 
		\exp\left\{-\frac{1}{2}\sum_{i, j}\int\sqrt{g}\,\phi_i\hat{\tilde{{\cal O}}}_i\hat{\cal C}_{i}^\dagger\hat{\cal C}_{j}\phi_j \right\}\,.
\end{align} }
Due to the orthogonality of each spin-parity sector on maximally symmetric spaces, we have that the spectrum of $\hat{\cal O}$ is the disjoint union of the spectra on each spin-parity mode, i.e., 
{\small \begin{align}
\det(\hat{\cal O})^{-1/2} =\det J \left.\prod_i \det\left(\hat{\cal C}_i^\dagger\hat{\cal C}_i\right)^{-1/2}\det\hat{\tilde{{\cal O}}}_i^{-1/2}\right|_{\phi_i} 
		=\left.\prod_i \det\hat{\tilde{{\cal O}}}_i^{-1/2}\right|_{\phi_i} \,,
\end{align} }
where the last equality is due to \eqref{eq:general-jacobian-decomposed}.

In the following we shall see how this whole picture gets more complicated when we choose to work on compact maximally symmetric spaces, i.e., $d$-spheres. Indeed, since on compact spaces the spectrum is discrete, zero-modes contribute, contrary to what happens when computing functional determinants on non-compact spaces.
For notational simplicity we will also drop the hat notation, since the fields on which the operators act will always be specified.

\subsection{Kinematical zero-medes}

We start our analysis of kinematical zero-modes by looking at $S_{\perp}{}_{\mu\nu}$. This tensor enters the decomposition Eq.\ \eqref{eq:spin-parity-decomposition} via the first Curtright form, whose kernel is defined by
\begin{align*}
\nabla_{\mu}S_{\perp}{}^{\rho}{}_{\nu}-\nabla_{\nu}S_{\perp}{}^{\rho}{}_{\mu}=0\,.
\end{align*}
Since the background metric is positive definite, we know that global solutions of the previous equation are in one-to-one correspondence with solutions of
{\small \begin{align}
\int {\rm d}^dx\sqrt{g}\,\left(\nabla_{\mu}S_{\perp}{}^{\rho}{}_{\nu}-\nabla_{\nu}S_{\perp}{}^{\rho}{}_{\mu}\right)\left(\nabla^{\mu}S_{\perp}{}_{\rho}{}^{\nu}-\nabla^{\nu}S_{\perp}{}_{\rho}{}^{\mu}\right)=2\int{\rm d}^dx\sqrt{g}\,S_{\perp}{}_\mu{}^\nu\left(-\square+\frac{R}{d(d-1)}\right)S_{\perp}{}^\mu{}_\nu=0\,,
\end{align} }
where we have integrated by parts and specialized to maximally symmetric spaces, also exploiting the transversality of $S_{\perp}{}^\mu{}_\nu$. Solving the previous equation is a much simpler task, since it can be straightforwardly done through the spectrum of the Laplace operator, which is known on maximally symmetric spaces. Indeed, on the sphere the spectrum for spin-$2$ symmetric transverse traceless fields is given by the following set of eigenvalues $\lambda_S(l)$ with degeneracies $D_S(l)$ \cite{Codello:2008vh}
{\small \begin{align}
\lambda_S(l)=\frac{l(l+d-1)-2}{d(d-1)}R\,,\qquad D_S(l)=\frac{(d+1)(d-2)(l+d)(l-1)(2l+d-1)(l+d-3)!}{2(d-1)!(l+1)!}\,,\qquad l=2\,,\;3\,,\dots
\end{align} }
A simple check shows that the equation $l(l+d-1)-2=-1$ has no solution for $l$ in the integers larger that $1$, so there is no configuration of $S_{\perp}{}^\mu{}_\nu$ that contributes to the kinematical zero-modes on the sphere.

We can proceed similarly for the spin-parity $1^+$ mode encoded in $A_{\perp}{}_{\mu\nu}$. Our aim is to solve
\begin{align*}
2\nabla^\rho A_{\perp}{}_{\mu\nu}+\nabla_{\mu}A_{\perp}{}^\rho{}_\nu-\nabla_{\nu}A_{\perp}{}^\rho{}_\mu=0\,.
\end{align*}
Once again, we can square and integrate the previous equation, obtaining
\begin{align*}
6\int{\rm d}^d x \sqrt{g}\,A_{\perp}{}^{\mu\nu}\left(-\square-\frac{d-2}{d(d-1)}R\right)A_{\perp}{}_{\mu\nu}=0\,.
\end{align*}
The spectrum of the Bochner Laplacian on $p$-forms on the $d$-sphere can be found from the results of \cite{Elizalde:1996nb} and reads:
{\small \begin{align}
\label{eq:eigenvaluesPForm}
\lambda_p(l)=\frac{l(l+d-1)-p}{d(d-1)}R\,,\qquad D_p(l)=\frac{(2l+d-1)(l+d-1)!}{p!(d-p-1)!(l-1)!(l+p)(l+d-p-1)}\,,\qquad l=1\,,\,2\,,\dots
\end{align} }
By plugging $p=2$ in \eqref{eq:eigenvaluesPForm}, we find one value of $l$ that contributes to kinematical zero-modes, i.e.,\ $l=1$. The multiplicity that results for this zero mode is $D_2(1)=10$ when $d=4$.

Finally, we focus on the twice-longitudinal modes. The contribution due to the kinematical zero-modes come from the solutions of
\begin{align*}
\nabla^\rho\nabla_\mu\zeta_{\perp}{}_\nu-\nabla^\rho\nabla_\nu\zeta_{\perp}{}_\mu
+\frac{1}{d-1}\left[\delta^\rho{}_\nu\square\zeta_{\perp}{}_\mu - \delta^\rho{}_\mu\square\zeta_{\perp}{}_\nu
+\frac{R}{d}\left(\delta^\rho{}_\mu\zeta_{\perp}{}_\nu-\delta^\rho{}_\nu\zeta_{\perp}{}_\mu\right)\right]=0\,.
\end{align*}
Using the same analysis employed for $S_{\perp}{}^\mu{}_\nu$ and $A_{\perp}{}_{\mu\nu}$, we obtain that the solutions of the previous equation on the $d$-sphere are in one-to-one correspondence to solutions of
{\small \begin{align}
\left(-\square+\frac{R}{d}\right)\left(-\square-\frac{R}{d}\right)\zeta_{\perp}{}_{\mu}=0\,,
\end{align} }
which, in turns, can be simplified to 
{\small \begin{align}
\left(-\square-\frac{R}{d}\right)\zeta_{\perp}{}_{\mu}=0\,,
\end{align} }
thanks to the fact that the Laplacian $-\square$ does not have negative eigenvalues. Plugging $p=1$ in \eqref{eq:eigenvaluesPForm}, we find that for $l=1$ there is one solution whose degeneracy is
{\small \begin{align}
D_1(1) = \frac{d}{2}(d+1)\,.
\end{align} }

\subsection{Dynamical zero-modes}

We now turn the attention to the dynamical zero-modes. Since all the dynamical operators of interest for the present work are polynomials in the Bochner Laplacian, we first look at commuting our Curtright forms with $-\square$. By doing so, we identify the shift in the endomorphism that we need to consider for each operator appearing in the decomposition.

Starting from the symmetric transverse-traceless tensor $\overline{S}_{\perp}{}^\mu{}_{\nu}$ we have
{\small \begin{align}
\square \left( \nabla_\mu\overline{S}_{\perp}{}^\rho{}_\nu - \nabla_\nu \overline{S}_{\perp}{}^\rho{}_\mu
 \right) = & \nabla_\mu \left( \square + \frac{d-3}{d(d-1)} R \right) \overline{S}_{\perp}{}^\rho{}_\nu - \left( \nu \leftrightarrow \mu \right)\,.
\end{align} }
Similarly, we commute the Curtright form for the antisymmetric part of the spectrum $A_{\perp}{}_{\mu\nu}$ to get
{\small \begin{align}
\square \left( 2 \nabla^\rho A_{\perp}{}_{\mu\nu} + \nabla_\mu A_{\perp}{}^\rho{}_\nu - \nabla_\nu A_{\perp}{}^\rho{}_\mu  \right) = & 2 \nabla^\rho \left( \square + \frac{d+1}{d(d-1)} R \right) A_{\perp}{}_{\mu\nu} \\\nonumber
&+ \nabla_\mu \left( \square + \frac{d+1}{d(d-1)} R \right)  A_{\perp}{}^\rho{}_\nu - \nabla_\nu \left( \square + \frac{d+1}{d(d-1)} R \right) A_{\perp}{}^\rho{}_\mu \,.
\end{align} }
And, finally, we apply the same procedure to the $1^-$ sector and find
{\small \begin{align}
\square \left( \nabla^\rho \nabla_\mu \zeta_\perp{}_\nu + \frac{1}{d-1}  \delta^\rho{}_\nu \left( \square - \frac{R}{d} \right) \zeta_\perp{}_\mu - \left( \nu \leftrightarrow \mu \right)  \right) = &  \nabla^\rho \nabla_\mu \left( \square + \frac{2 R}{d} \right) \zeta_\perp{}_\mu \\\nonumber
& + \frac{1}{d-1} \left[ \delta^\rho{}_\nu \left( \square - \frac{R}{d} \right) \left( \square + \frac{2R}{d} \right) \zeta_\perp{}_\mu \right] - \left( \nu \leftrightarrow \mu \right)  .
\end{align} }

Once we have these results, we can analyze the spectrum of the Bochner Laplacian on transverse hook antisymmetric tracefree tensors, which parameterize the $2^-$ mode. As usual, the spectrum of $-\square$ on the non-transverse configurations of $\kappa^\rho{}_{\mu\nu}$ is the disjoint union of the spectra on the transverse and longitudinal components and we can write
{\small \begin{align}
\label{eq:kappaTrace}
\left.{\rm Tr} \, {\rm e}^{-s\left( - \square -\Upsilon R\right)} \right|_{\kappa^\rho{}_{\mu\nu}} = &
				\left.{\rm Tr} \, {\rm e}^{-s\left( - \square-\Upsilon R \right)} \right|_{\kappa_{\perp}{}^\rho{}_{\mu\nu}}  
				+ \left.{\rm Tr} \, {\rm e}^{-s\left[ - \square - \left(\frac{d-3}{d(d-1)}+\Upsilon\right) R \right]} \right|_{\overline{S}_\perp{}^\mu{}_{\nu}} 
				+ \left.{\rm Tr} \, {\rm e}^{-s\left[ - \square - \left(\frac{d+1}{d(d-1)}+\Upsilon\right) R \right]} \right|_{A_\perp{}_{\mu\nu}} \\\nonumber
				& + \left.{\rm Tr} \, {\rm e}^{-s\left[ - \square - \left(\frac{2}{d}+\Upsilon\right) R \right]} \right|_{\zeta_\perp{}_\mu} - k(d) - y(d)\,,
\end{align} }
where $\Upsilon=\Upsilon(a_i, b_j)$ contains the dependence on the coupling constants $a_i, \, b_j$ appearing in our starting Lagrangian \eqref{Lagriangian-max-sym}. Note that in the previous equation there is no evaluation on scalar modes since $\kappa^\rho{}_{\mu\nu}$ is traceless. On the r.h.s., $k(d)$ is due to the kinematical zero-modes that we have counted in the previous section. This receives no contribution from $S_{\perp}{}^\mu{}_\nu$, and it amounts to
{\small \begin{align}
\label{eq:kinematicalKappa}
k(d) = D_{p=2}(1)\,{\rm e}^{s\left(\frac{3}{d(d-1)}+\Upsilon\right)R} + D_{p=1}(1)\,{\rm e}^{s\left(\frac{1}{d}+\Upsilon\right)R} = 
			\frac{d(d+1)(d-1)}{6}\,{\rm e}^{s\left(\frac{3}{d(d-1)}+\Upsilon\right)R} + \frac{d}{2}(d+1)\,{\rm e}^{s\left(\frac{1}{d}+\Upsilon\right)R}
			\,\overset{\tiny d=4}{=}\,20\,{\rm e}^{s\left(\frac{1}{4}+\Upsilon\right)R}\,.
\end{align} }
However, the traces over the transverse fields with lower spins introduce in general further vanishing contributions coming from the dynamical zero modes $y(d)$. One might identify the dynamical zero modes directly trying to match eigenvalues of $-\square$ with the specific endomorphism that appears in each trace. However, since the traces in Eq\ \eqref{eq:kappaTrace} are performed over transverse fields, it is not clear what is the multiplicity associated with each dyanamical zero mode. In fact, those traces will have to be themselves computed through the same strategy we are following for $\kappa_{\perp}{}^\rho{}_{\mu\nu}$. Therefore, using the following identities
{\small \begin{align}
&\left(-\square + \alpha R\right)\left(\nabla_\mu\zeta_{\perp\nu}+\nabla_\nu\zeta_{\perp\mu}\right) =
		\nabla_\mu\left[-\square +\left(\alpha-\frac{d+1}{d(d-1)}\right)R\right]\zeta_{\perp\nu} + (\,\mu\leftrightarrow\nu\,)\,,\\
&\left(-\square+\alpha R\right)\left(\nabla_\mu\nabla_\nu-\frac{1}{d}g_{\mu\nu}\square\right)\sigma = 
		\left(\nabla_\mu\nabla_\nu-\frac{1}{d}g_{\mu\nu}\square\right)\left[-\square+\left(\alpha-\frac{2}{d-1}\right) R\right]\sigma\,,
\end{align} }
we can write
{\small \begin{align}
\begin{split}
\left.{\rm Tr} \, {\rm e}^{-s\left[ - \square - \left(\frac{d-3}{d(d-1)} +\Upsilon\right)R \right]} \right|_{\overline{S}^\mu{}_{\nu}} 
		=& \left.{\rm Tr} \, {\rm e}^{-s\left[ - \square - \left(\frac{d-3}{d(d-1)} +\Upsilon\right)R \right]} \right|_{\overline{S}_\perp{}^\mu{}_{\nu}}
		+ \left.{\rm Tr} \, {\rm e}^{-s\left[ - \square - \left(\frac{2}{d}+\Upsilon\right) R \right]} \right|_{\zeta_{\perp\mu}}
		+ \left.{\rm Tr} \, {\rm e}^{-s\left[ - \square - \left(\frac{3}{d}+\Upsilon\right) R \right]} \right|_{\sigma}
		-1\cdot\,{\rm e}^{\left(\frac{3}{d}+\Upsilon\right)sR}\\
&-(d+1)\,{\rm e}^{\left(\frac{2d-3}{d(d-1)}+\Upsilon\right)sR}-\frac{d(d+1)}{2}\,{\rm e}^{\left(\frac{1}{d}+\Upsilon\right)sR}-\sum_l\left[
		\delta_{\frac{\lambda_{0}(l)}{R}, \frac{3}{d}+\Upsilon}D_{p=0}(l)+\delta_{\frac{\lambda_{1}(l)}{R}, \frac{2}{d}+\Upsilon}D_{p=1}(l)\right]\,,
\end{split}
\end{align} }
where the coefficients in the second lines are the multiplicities of the zero modes computed from the spectrum of box using \eqref{eq:eigenvaluesPForm}. Note that in the above manipulations we used the fact that harmonic functions on connected manifolds without boundary are constant functions due to the maximum principle and, therefore, have multiplicity one.
Expressing the trace over transverse vectors in terms of the full vector and its longitudinal part, and taking into account the corresponding zero modes, we can explicitly check that the scalar contributions cancel\footnote{As a matter of fact, one could already guess this from the fact that $\overline{S}{}^\rho{}_\mu$ is traceless. Indeed, one can isolate its longitudinal components simply extracting $\zeta_\mu$ without dealing with extra scalar modes. In other words, the scalar fields appearing in the decomposition of $\overline{S}{}^\rho{}_\mu$ are just the longitudinal components of $\zeta_\mu$ that we now reinsert in the vector field.} and we have
{\small \begin{align}\label{eq:STrace}
\begin{split}
\left.{\rm Tr} \, {\rm e}^{-s\left[ - \square - \left(\frac{d-3}{d(d-1)} +\Upsilon\right)R \right]} \right|_{\overline{S}^\mu{}_{\nu}} 
=&\left.{\rm Tr} \, {\rm e}^{-s\left[ - \square - \left(\frac{d-3}{d(d-1)} +\Upsilon\right)R \right]} \right|_{\overline{S}_\perp{}^\mu{}_{\nu}}
		+ \left.{\rm Tr} \, {\rm e}^{-s\left[ - \square - \left(\frac{2}{d}+\Upsilon\right) R \right]} \right|_{\zeta_{\mu}}\\
&-(d+1)\,{\rm e}^{\left(\frac{2d-3}{d(d-1)}+\Upsilon\right)sR}-\frac{d(d+1)}{2}\,{\rm e}^{\left(\frac{1}{d}+\Upsilon\right)sR}
		-\sum_l\delta_{\frac{\lambda_1(l)}{R}, \frac{2R}{d}+\Upsilon}D_{p=1}(l)\,.
\end{split}
\end{align} }
%

Let us now turn our attention to the trace computed over $A_{\perp\mu\nu}$. Note that $A_{\mu\nu}$ is trivially traceless and, therefore, the twice-longitudinal modes do not contribute to its decomposition
{\small \begin{align}
A_{\mu\nu} = A_{\perp\mu\nu}+\nabla_{\mu}\xi_{\nu}-\nabla_{\nu}\xi_{\mu} = A_{\perp\mu\nu}+\nabla_{\mu}\xi_{\perp\nu}-\nabla_{\nu}\xi_{\perp\mu}\,,
\end{align} }
however, restricting to transverse $1$-forms does change the count of kinematical zero-modes in $A_{\mu\nu}$, since it affects the domain of the differential operator $\nabla_{\mu}g^\alpha_\nu-\nabla_{\nu}g^\alpha_\mu$. As usual, the best strategy to count these zero-modes would be to square the longitudinal operator acting on $\xi_\mu$, but then one has to deal with the divergence of the $1$-form shifting the spectrum. The most convenient way will be, again to decompose $\xi_\mu$ as well, and to count the kinematical zero-modes of $A_{\mu\nu}$ through the scalar part of $\xi_\mu$. In fact, by doing so we realize that the longitudinal modes extracted from $\xi_\mu$ are, actually, the only kinematical zero-modes contributing to the decomposition of the $2$-form $A_{\mu\nu}$. On the other hand, since we have already outlined the recipe for taking into account the model-dependent dynamical zero-modes, we will not write their contribution henceforth.

We use the following two formulae, which are valid on maximally symmetric spaces,
{\small \begin{align}
&\left(-\square+\alpha R\right)\left(\nabla_\mu g^\alpha_\nu-\nabla_\nu g^\alpha_\mu\right)\xi_{\perp\alpha} =
		\left(\nabla_\mu g^\alpha_\nu-\nabla_\nu g^\alpha_\mu\right)\left[-\square+\left(\alpha+\frac{d-3}{d(d-1)}\right) R\right]\xi_{\perp\alpha}\,,\\
&\left(-\square+\alpha R\right)\nabla_\mu\varphi = \nabla_\mu\left[-\square+\left(\alpha-\frac{1}{d}\right)R\right]\varphi\,,
\end{align} }
to write
{\small \begin{align}
\label{eq:ATrace}
\begin{split}
\left.{\rm Tr} \, {\rm e}^{-s\left[ - \square - \left(\frac{d+1}{d(d-1)} +\Upsilon\right)R \right]} \right|_{A_{\mu\nu}} =& 
		\left.{\rm Tr} \, {\rm e}^{-s\left[ - \square - \left(\frac{d+1}{d(d-1)} +\Upsilon\right)R \right]} \right|_{A_\perp{}_{\mu\nu}}
		+\left.{\rm Tr} \, {\rm e}^{-s\left[ - \square - \left(\frac{3}{d(d-1)}+\Upsilon\right) R \right]} \right|_{\xi_\perp{}_{\mu}}\\
=&\left.{\rm Tr} \, {\rm e}^{-s\left[ - \square - \left(\frac{d+1}{d(d-1)} +\Upsilon\right)R \right]} \right|_{A_\perp{}_{\mu\nu}}
		+\left.{\rm Tr} \, {\rm e}^{-s\left[ - \square - \left(\frac{3}{d(d-1)}+\Upsilon\right) R \right]} \right|_{\xi_{\mu}}\\
		&
		-\left.{\rm Tr} \, {\rm e}^{-s\left[ - \square - \left(\frac{d+2}{d(d-1)}+\Upsilon\right) R \right]} \right|_{\varphi}
		+1\cdot{\rm e}^{\left(\frac{d+2}{d(d-1)}+\Upsilon\right)sR}\,.
\end{split}
\end{align} }

Finally, we have
{\small \begin{align}
\label{eq:zetaTrace}
\left.{\rm Tr} \, {\rm e}^{-s\left[ - \square - \left(\frac{2}{d}+\Upsilon\right) R \right]} \right|_{\zeta_\mu}=&
		\left.{\rm Tr} \, {\rm e}^{-s\left[ - \square - \left(\frac{2}{d}+\Upsilon\right) R \right]} \right|_{\zeta_\perp{}_\mu}
		+\left.{\rm Tr} \, {\rm e}^{-s\left[ - \square - \left(\frac{3}{d}+\Upsilon\right) R \right]} \right|_{\sigma}
		-1\cdot{\rm e}^{\left(\frac{3}{d}+\Upsilon\right)s R}\,.
\end{align} }
Inserting \eqref{eq:kinematicalKappa}, \eqref{eq:STrace}, \eqref{eq:ATrace} and \eqref{eq:zetaTrace} into \eqref{eq:kappaTrace}, we find 
{\small \begin{align}
\begin{split}
\left.{\rm Tr} \, {\rm e}^{-s\left( - \square -\Upsilon R\right)} \right|_{\kappa_{\perp}{}^\rho{}_{\mu\nu}} =& 
				\left.{\rm Tr} \, {\rm e}^{-s\left( - \square -\Upsilon R\right)} \right|_{\kappa^\rho{}_{\mu\nu}}  
				-\left.{\rm Tr} \, {\rm e}^{-s\left[ - \square - \left(\frac{d-3}{d(d-1)}+\Upsilon\right) R \right]} \right|_{\overline{S}^\mu{}_{\nu}}
				-\left.{\rm Tr} \, {\rm e}^{-s\left[ - \square - \left(\frac{d+1}{d(d-1)}+\Upsilon\right) R \right]} \right|_{A_{\mu\nu}}\\
				&
				+\left.{\rm Tr} \, {\rm e}^{-s\left[ - \square - \left(\frac{3}{d(d-1)} + \Upsilon \right)R \right]} \right|_{\xi_{\mu}}-\left.{\rm Tr} \, {\rm e}^{-s\left[ - \square - \left(\frac{d+2}{d(d-1)}+\Upsilon\right) R \right]} \right|_{\varphi}
		+\left.{\rm Tr} \, {\rm e}^{-s\left[ - \square - \left(\frac{3}{d}+\Upsilon\right) R \right]} \right|_{\sigma}\\
		&
		-{\rm e}^{\left(\frac{3}{d}+\Upsilon\right) sR}+{\rm e}^{\left(\frac{d+2}{d(d-1)}+\Upsilon\right)sR}
		-(d+1)\,{\rm e}^{\left(\frac{2d-3}{d(d-1)}+\Upsilon\right)sR}+\frac{d(d+1)(d-1)}{6}\,{\rm e}^{s\left(\frac{3}{d(d-1)}+\Upsilon\right)R}\,.
\end{split}
\end{align} }

We remark that two main facts allow us to perform such an analysis explicitly: the transverse-traceless decomposition of $\kappa^\rho{}_{\mu\nu}$, that allows us to simplify the differential operators, and the maximally symmetric background, for which we can compute explicitly the eigenvalues of box on $p$-forms. One has to repeat the above analysis for all the rest of the tensors appearing in Lagrangian we study in the next section. We performed such an analysis, however the computations are extremely lengthy and depend heavily on the coupling constants. Therefore, we report here only the most necessary formulae concerning the commutations of differential operators coming from the dynamics with the ones coming from the torsion decomposition Eq.\ \eqref{eq:spin-parity-decomposition}.

When tracing the fourth-order differential operator over the $2^+$ modes, we need the following identity
{\small \begin{align}
\left( \square^2 + a R \square + b R^2 \right) \left( \nabla_\mu \xi_\nu + \nabla_\nu \xi_\mu - \frac{2}{d} g_{\mu\nu} \nabla_\lambda \xi^\lambda \right) = & \nabla_\mu \Delta_{\xi} \xi_\nu + \nabla_\nu \Delta_{\xi} \xi_\mu - g_{\mu\nu} \nabla_\lambda \Delta_{\xi} \xi^\lambda \, ,
\end{align} }
where
{\small \begin{align}
\Delta_{\xi} = \square^2 + \left( a + \frac{2(d+1)}{d(d-1)} \right) R \square + \left( \frac{(d+1)^2}{d^2(d-1)^2} + \frac{a(d+1)}{d(d-1)} + b \right) R^2 \, .
\end{align} }

The fourth order operator acting on the $1^+$ sector mixes $A_{\mu\nu}$ with $\Pi_{\mu\nu}$ and has the following form
\begin{equation}
\Delta_{1^+} = \hat{\mathds{1}} \square^2 + \hat{A} R \square + \hat{B} R^2 \, ,
\end{equation}
for generic $\hat{A}$ and $\hat{B}$ matrices
	{\small \begin{align}
	\hat{A} = & \begin{bmatrix}
	a_1 && a_2\\
	a_2 && a_3
	\end{bmatrix} \, , \qquad
	\hat{B} =  \begin{bmatrix}
	b_1 && b_2\\
	b_2 && b_3
	\end{bmatrix} \, ,
	\end{align} }
while the operator acting on the longitudinal modes after commuting with the decomposition is
\begin{equation}
\Delta_{1^-} = \hat{\mathds{1}} \square^2 + \hat{C} R \square + \hat{P} R^2 \, ,
\end{equation}
with
	{\small \begin{align}
	\hat{C} = & \begin{bmatrix}
	\frac{2(d-3)}{d(d-1)} + a_1 && a_2\\
	a_2 && \frac{2(d-3)}{d(d-1)} + a_3
	\end{bmatrix} \, , \quad
	\hat{B} =  \begin{bmatrix}
	\frac{(d-3)^3}{d^2(d-1)^2} + \frac{(d-3)}{d(d-1)} a_1 + b_1 && \frac{(d-3)}{d(d-1)} a_2 + b_2\\
	\frac{(d-3)}{d(d-1)} a_2 + b_2 && \frac{(d-3)^3}{d^2(d-1)^2} + \frac{(d-3)}{d(d-1)} a_3 + b_3
	\end{bmatrix} \, .
	\end{align} }

As a final remark, we stress that we are actually interested in the divergent part of the trace of the logarithm of differential operators, rather that the traces themselves. As explained in section \ref{sect:Schwinger_DeWitt_technique}, this implies that in the (asymptotic) expansion of the traces we only look at the coefficient of $s^0$. For the zero-modes, regardless their kinematical or dynamical origin, this means that they contribute only through their multiplicity. Hence, in $d-\epsilon$ dimensions we have

{\small \begin{align}
\begin{split}
\left.{\rm Tr}\log \, {\rm e}^{-s\left( - \square -\Upsilon R\right)} \right|_{\kappa_{\perp}{}^\rho{}_{\mu\nu}}^{\rm div} =& 
		\left.{\rm Tr}\log \, {\rm e}^{-s\left( - \square -\Upsilon R\right)} \right|_{\kappa^\rho{}_{\mu\nu}}^{\rm div}  
		-\left.{\rm Tr}\log \, {\rm e}^{-s\left[ - \square - \left(\frac{d-3}{d(d-1)}+\Upsilon\right) R \right]} \right|_{\overline{S}^\mu{}_{\nu}}^{\rm div}
		-\left.{\rm Tr}\log \, {\rm e}^{-s\left[ - \square - \left(\frac{d+1}{d(d-1)}+\Upsilon\right) R \right]} \right|_{A_{\mu\nu}}^{\rm div}\\
&+\left.{\rm Tr}\log \, {\rm e}^{-s\left[ - \square - \left(\frac{3}{d(d-1)}+\Upsilon\right) R \right]} \right|_{\xi_{\mu}}^{\rm div}
		-\left.{\rm Tr}\log \, {\rm e}^{-s\left[ - \square - \left(\frac{d+2}{d(d-1)}+\Upsilon\right) R \right]} \right|_{\varphi}^{\rm div}
		+\left.{\rm Tr}\log \, {\rm e}^{-s\left[ - \square - \left(\frac{3}{d}+\Upsilon\right) R \right]} \right|_{\sigma}^{\rm div}\\
&-\frac{1}{3\epsilon}(d+1)(d+2)(d-3)\,.
\end{split}
\end{align} }
Note, however, that most of eigenvalue equations for the zero-modes in the lower spin sectors will heavily depend on the values of the coupling constants and, for generic couplings, there will be no contribution from the dynamical zero-modes.

\section{Heat kernel integration of torsion fluctuations}\label{sect:1-loop-body}

In sections \ref{sect:Schwinger_DeWitt_technique} and \ref{sect:spectrum-of-diff-constrained} we have given some very brief introductions to the heat kernel technique and the spectral geometry of differentially constrained fields. Now we want to exploit these developments to integrate out the Gaussian torsion fluctuations stemming from the Lagrangian \eqref{eq:Lagrangian}. Since the computations are rather lengthy, we are going to deal only with the $1^+$ sector explicitly in the main text. We choose to present the results for this sector because it shows most of the complications that can arise in such kinds of computations. We defer the reader to the appendix \ref{sect:appendix-1loop} for most of the remaining results. Nevertheless, at the end of this section we shall briefly outline the technical difficulties concerning the $1^-$ sector, and how we circumvent them by restricting the relative values of some coupling constants. Before commencing we briefly analyze a different parameterization of the axial part of the torsion. In \cite{Martini:2023rnv} we have described it as a pseudo-vector (in $d=4$), whereas here we shall adopt a hybrid choice. This amounts to describing the longitudinal component with a transverse $2$-form $\Pi_{\mu\nu}$, and the transverse one with the pseudo-scalar $\pi$, in formulae
{\small \begin{align}  
H^\rho{}_{\mu\nu} = & \nabla^\rho \Pi_{\mu\nu}  + \nabla_\mu \Pi_\nu{}^\rho + \nabla_\nu \Pi^\rho{}_\mu + \frac{1}{6} \varepsilon^{\lambda\rho}{}_{\mu\nu} \nabla_\lambda \pi \\\nonumber
= &  \frac{1}{6} \varepsilon^{\lambda\rho}{}_{\mu\nu} \left( \theta_\lambda + \nabla_\lambda \pi \right) \, ,
\end{align} }
where all the tensors that appear on the right-hand-side are. Since $\theta_\mu$ enters the decomposition of the axial part of the torsion only algebraically, it does not give rise to any non-trivial Jacobian. This is not the case for $\Pi_{\mu\nu}$, whose contribution to $\det (J_{\rm diff})$ is accounted for by considering how it contributes to the normalization of the gaussian integral. Since on maximally symmetric spaces the spin-parity eigenspaces are orthogonal, the Jacobian is found by the normalization of the fluctuations over the $1^+$ sector only
{\small \begin{align}
\int {\rm det }(J_{\rm diff}) D \Pi_{\mu\nu} {\rm e}^{- \int \sqrt{g} \left( \nabla_\mu \Pi_{\nu\rho} + \nabla_\nu \Pi_{\rho\mu} + \nabla_\rho \Pi_{\mu\nu} \right) \varepsilon^{\gamma\mu\nu\rho} \varepsilon_\gamma{}^{\lambda\alpha\beta} \left( \nabla_\lambda \Pi_{\alpha\beta} + \nabla_\alpha \Pi_{\beta\lambda} + \nabla_\beta \Pi_{\lambda\alpha} \right) } = 1 \, ,
\end{align} }                           
where we have implicitly made a natural choice for the supermetric over $3$-forms, and we have exploited the fact that the metric has no quantum fluctuations. Thus, the Jacobian factor needed to balance out the contribution of $\Pi_{\mu\nu}$ to the normalization of the gaussian integral is
\begin{equation}
{\rm det }(J_{\rm diff}) = \left[ \det \left( -\square + \frac{R}{12} \right)  \right]^{\frac{1}{2}} \Bigg|_{\Pi_\perp{}_{\mu\nu}} \, .
\end{equation}                                                                                                                                                   
Notice that, since $-\square$ only has positive eigenvalues on the sphere, there are no zero-modes associated with this functional determinant. Moreover, the latter is evaluated on transverse $2$-forms as
{\small \begin{align}
{\rm Tr \, e}^{-s\left( -\square + \frac{R}{12} \right)} \Big|_{\Pi_\perp{}_{\mu\nu}} = {\rm Tr \, e}^{-s\left( -\square + \frac{R}{12} \right)} \Big|_{\Pi_{\mu\nu}} - {\rm Tr \, e}^{-s\left( -\square \right)} \Big|_{\xi_\perp{}_\mu} \, .
\end{align} }
Exploiting the results collected in \cite{Codello:2008vh} for the spectrum of transverse vectors, we find that the second Seeley-DeWitt coefficient that accounts for this Jacobian factor is given by
\begin{equation}
 {\rm tr} (a_2({\rm det }(J_{\rm diff}))) = - \frac{607}{1440} R^2\, .
\end{equation}

\subsection{Integration of the $1^+$ sector}\label{subsect:two-forms-body}

As we have just seen, the heat kernel integration over transverse $2$-forms is computed as difference between ordinary $2$-forms and transverse vectors. However, the differential operator in the $1^+$ sector is of fourth order, see App.\ \ref{sect:appendix-hessians}. Thus, here we cannot exploit the results of \cite{Codello:2008vh}, that are valid for second-order operators. Instead, we have to further express the contribution of transverse vectors in terms of ordinary vectors minus the scalar part, in formulae
{\small \begin{align}\label{eq:formal-trace-2forms}
\left. {\rm Tr \, e}^{-s \Delta_{1^+}} \right|_{\left( A_\perp, \Pi_\perp \right)} = & \left.                                                                                                                                                                                                                                                                                                                                                                                                                                                                                               
{\rm Tr \, e}^{-s \Delta_{1^+}} \right|_{\left( A, \Pi \right)} - 
\left. {\rm Tr \, e}^{-s \Delta_{1^-}} \right|_{\left( \xi_1, \xi_2 \right)} +                                                                                                
\left. {\rm Tr \, e}^{-s \Delta_{0^+}} \right|_{\left( \phi_1, \phi_2 \right)} - 2 {\rm e}^{s \lambda} \, ,
\end{align} }
where $\Delta_{1^-}=\Delta_{1^+} ( \square \rightarrow \square + \frac{R}{12})$ and $\Delta_{0^+}=\Delta_{1^+} ( \square \rightarrow \square + \frac{R}{3})$ and we have set $d=4$. The last term is due to the zero-modes, and it arises because constant scalars do not contribute to the spectrum of vectors. Thus, to obtain the final result stemming from the $1^+$-sector we have to work out the precise expressions the three Hessians, and then evaluate the traces over systems of $2$-forms, vectors and scalars. In order to lighten the presentation, here we will concentrate on the trace over $2$-forms, deferring the results for vectors and scalars to the appendix \ref{sect:appendix-1loop}.

By plugging the spin-parity decomposition \eqref{eq:spin-parity-decomposition} into the Lagrangian \eqref{Lagriangian-max-sym} we can read off the Hessian in the $1^+$ sector
\begin{equation}
\hat{H}_{\tiny 1^+} = \begin{bmatrix}
a && b\\
b && c
\end{bmatrix} \,
\square^2 \mathds{1} + \begin{bmatrix}
 e && f\\
 f && g
\end{bmatrix} \,
R \, \square \mathds{1} + \begin{bmatrix}
h && l \\
l && n
\end{bmatrix} \,
R^2 \mathds{1} \, . 
\end{equation}
The first row and column refer to the $A_{\mu\nu}$ field, while the others pertain to $\Pi_{\mu\nu}$, and $\mathds{1}=\frac{1}{2} \left( \delta^\mu{}_\alpha  \delta^\nu{}_\beta - \delta^\nu{}_\alpha  \delta^\mu{}_\beta \right)$ is the identity in the space of $2$-forms.
The actual values of the coefficients that enter the previous equation are
\begin{subequations}\label{eq:coefficients-1+-first}
 {\small \begin{align}
  a = & 12 \alpha_1 + 8 \alpha_4 \, , \qquad \qquad \qquad b = 2 \alpha_8 \, , \qquad \qquad \qquad c=2 \alpha_6 + 6 \alpha_2 \, ; \\
  e = & \frac{4}{d(d-1)} \left[ (d-2) (3 \alpha_1 + 4 \alpha_4) - 6 (2b_1 + b_2) \right] \, , \qquad f = - \frac{2(d-2)\alpha_8}{d(d-1)} ;\\
  g = & - \left( \frac{6(d+1)\alpha_2}{d(d-1)} + \frac{8(d-2)\alpha_6}{d(d-1)} + 6 (b_1-b_2) \right) \, , \qquad h = - \frac{2(d-2)}{d^2(d-1)^2} \left( 3d(d-1) (2b_1+b_2) + 6 (d+1) \alpha_1 + 4 (d-2) \alpha_4 \right) \, ; \\
  l = & - \frac{4(d-2)^2\alpha_8}{d^2(d-1)^2} \, , \qquad \qquad \qquad \qquad n = \frac{4(d-2)}{d(d-1)} \left( \frac{2(d-2) \alpha_6}{d(d-1)} - \frac{3(d-5)\alpha_2}{d(d-1)} + 3(b_1-b_2) \right) \, .
 \end{align} }
\end{subequations}
As it is customary, in order to use the known formulae \eqref{eq:fourth-order-operator}-\eqref{eq:a2-minimal-fourth-order} for fourth-order operators, we first have to canonically normalize and diagonalize the principal part of the differential operator. Implicitly assuming $a>0$ and $c>0$, we start by letting $A_{\mu\nu} \rightarrow \frac{1}{\sqrt{a}} A_{\mu\nu}$ and $\Pi_{\mu\nu} \rightarrow \frac{1}{\sqrt{c}} \Pi_{\mu\nu}$, obtaining\footnote{Since the full Hessian is block-diagonal in the spin-parity eigenstates, this ultralocal and constant change of normalization in field space can always be performed.}
\begin{equation}
\hat{H}_{\tiny 1^+} = \begin{bmatrix}
1 && \frac{b}{\sqrt{ac}}\\
\frac{b}{\sqrt{ac}} && 1
\end{bmatrix} \,
\square^2 \mathds{1} + \begin{bmatrix}
\frac{e}{a} && \frac{f}{\sqrt{ac}}\\
\frac{f}{\sqrt{ac}} && \frac{g}{c}
\end{bmatrix} \,
R \, \square \mathds{1} + \begin{bmatrix}       
\frac{h}{a} && \frac{l}{\sqrt{ac}} \\
\frac{l}{\sqrt{ac}} && \frac{n}{\sqrt{ac}}
\end{bmatrix} \,
R^2 \mathds{1} \, . 
\end{equation}
To diagonalize the Hessian we perform a rotation
\begin{subequations}
	{\small \begin{align}
	A_{\mu\nu} & \equiv \cos(\alpha) B_{\mu\nu} + \sin(\alpha) C_{\mu\nu}\\
	\Pi_{\mu\nu} & \equiv - \sin(\alpha) B_{\mu\nu} + \cos(\alpha) C_{\mu\nu} \, ,
	\end{align} }
\end{subequations}                     
whence, the principal part of the Hessian becomes
\begin{equation}
\hat{H}_{\tiny 1^+ \, \rm PP} = \begin{bmatrix}
1 - \frac{2b}{\sqrt{ac}} \sin \alpha \cos \alpha && \frac{b}{\sqrt{ac}} \left( 2 \cos^2 \alpha - 1 \right)\\
\frac{b}{\sqrt{ac}} \left( 2 \cos^2 \alpha - 1 \right) && 1 + \frac{2b}{\sqrt{ac}} \sin \alpha \cos \alpha
\end{bmatrix} \square^2 \mathds{1} \, ,
\end{equation}
where the first row and column now pertain to $B_{\mu\nu}$. Thus, we have that for $\alpha=\frac{\pi}{4}+k\frac{\pi}{2}$ with $k \in \mathbb{Z}$ the coefficient matrix of the principal part is proportional to the identity. For simplicity, we choose $\alpha=\frac{\pi}{4}$, and we also let
\begin{equation}
 B_{\mu\nu} \rightarrow \frac{\sqrt[4]{ac}}{\sqrt{\sqrt{ac}-b}} B_{\mu\nu} \, , \qquad C_{\mu\nu} \rightarrow \frac{\sqrt[4]{ac}}{\sqrt{\sqrt{ac}+b}} C_{\mu\nu} \, .
\end{equation}
Accordingly, the final form of the Hessian reads
\begin{equation}\label{eq:final-hessian-2forms}
\begin{split}
\hat{H}_{\tiny 1^+} = & \begin{bmatrix}
1 && 0\\
0 && 1
\end{bmatrix} \,
\square^2 \mathds{1} + \frac{\sqrt{ac}}{2} \begin{bmatrix}
\frac{1}{\sqrt{ac}-b}\left(\frac{e}{a} + \frac{g}{c} - \frac{2f}{\sqrt{ac}} \right) && \frac{1}{\sqrt{ac-b^2}} \left( \frac{e}{a} - \frac{g}{c} \right)\\
\frac{1}{\sqrt{ac-b^2}} \left( \frac{e}{a} - \frac{g}{c} \right) && \frac{1}{\sqrt{ac}+b}\left(\frac{e}{a} + \frac{g}{c} + \frac{2f}{\sqrt{ac}} \right)
\end{bmatrix} \,
R \, \square \mathds{1}\\                             
& + \frac{\sqrt{ac}}{2} \begin{bmatrix}
\frac{1}{\sqrt{ac}-b}\left(\frac{h}{a} + \frac{n}{c} - \frac{2l}{\sqrt{ac}} \right) && \frac{1}{\sqrt{ac-b^2}} \left( \frac{h}{a} - \frac{n}{c} \right)\\
\frac{1}{\sqrt{ac-b^2}} \left( \frac{h}{a} - \frac{n}{c} \right) && \frac{1}{\sqrt{ac}+b}\left(\frac{h}{a} + \frac{n}{c} + \frac{2l}{\sqrt{ac}} \right)\end{bmatrix} \,
R^2 \mathds{1} \, .
\end{split}
\end{equation}
This operator belongs to a special subclass of fourth-operators, see Eq.\ \eqref{eq:fourth-order-operator}, and the coefficients in front of each independent tensor structure can be parameterized as
\begin{equation}
 \hat{\Omega}^{\mu\nu\rho} = 0 \, ,\qquad
 \hat{D}^{\mu\nu} = g^{\mu\nu} R \begin{bmatrix}
 \sigma_1 && \sigma_3\\
 \sigma_3 && \sigma_5
 \end{bmatrix} \, \mathds{1} \, ,\qquad
 \hat{H}^\mu = 0 \, , \qquad
 \hat{P} = R^2 \begin{bmatrix}
 \sigma_2 && \sigma_4\\
 \sigma_4 && \sigma_6
 \end{bmatrix} \, \mathds{1} \, .
\end{equation}
Therefore, in the present case the mapping between the coefficients is
\begin{subequations}\label{eq:coefficients-1+-second}
 {\small \begin{align}
  \sigma_1 = & \frac{\sqrt{ac}}{2(\sqrt{ac}-b)}\left(\frac{e}{a} + \frac{g}{c} - \frac{2f}{\sqrt{ac}} \right)  \, , \qquad \sigma_2 = \frac{\sqrt{ac}}{2(\sqrt{ac}-b)}\left(\frac{h}{a} + \frac{n}{c} - \frac{2l}{\sqrt{ac}} \right) \, ;\\
  \sigma_3 = & \frac{\sqrt{ac}}{2\sqrt{ac-b^2}} \left( \frac{e}{a} - \frac{g}{c} \right) \, , \qquad\qquad\,\,\,\,\,\,\,\, \sigma_4 = \frac{\sqrt{ac}}{2\sqrt{ac-b^2}} \left( \frac{h}{a} - \frac{n}{c} \right) \, ;\\
  \sigma_5 = & \frac{\sqrt{ac}}{2(\sqrt{ac}+b)}\left(\frac{e}{a} + \frac{g}{c} + \frac{2f}{\sqrt{ac}} \right) \, , \qquad \sigma_6 = \frac{\sqrt{ac}}{2(\sqrt{ac}+b)}\left(\frac{h}{a} + \frac{n}{c} + \frac{2l}{\sqrt{ac}} \right) \, .
 \end{align} }
\end{subequations}
In four dimensions and on maximally symmetric spaces, the logarithmically divergent part of the effective action for such kind of operators acting on $2$-forms becomes
\begin{equation}\label{eq:a2-2forms}
\left. {\rm tr} \left( a_2 \left( \Delta_{1^+} \right) \right)\right|_{\left( A_{\mu\nu}, \Pi_{\mu\nu} \right)} =  R^2 \left[- \frac{31}{90} - 6 (\sigma_2 + \sigma_6) + \sigma_1 + \sigma_5 + 3 (\sigma_1^2 + \sigma_5^2 + 2 \sigma_3^2) \right] \, .
\end{equation}
The result for the integration over $2$-forms is obtained by plugging in the coefficients and taking $d=4$, see Eqs.\ \eqref{eq:coefficients-1+-first} and \eqref{eq:coefficients-1+-second}. The final form of the second Seeley-DeWitt coefficient for the $1^+$-sector is found by taking into account the contributions from vectors and scalars, as well as zero-modes, see Eq.\ \eqref{eq:formal-trace-2forms}. Thus, using the results of the App.\ \ref{sect:appendix-1loop}, we find
{\small \begin{align}
 \left.{\rm tr} \left( a_2 \left( \Delta_{1^+} \right) \right) \right|_{\left( A_\perp{}_{\mu\nu} , \Pi_\perp{}_{\mu\nu} \right)} = & R^2 \left[ - \frac{7}{10} - 6 \left( \sigma_2 + \sigma_6 \right) - \frac{1}{2} \left( \sigma_1 + \sigma_5 \right) + \frac{1}{2} \left( \sigma_1^2 + \sigma_5^2 + 2 \sigma_3^2 \right) \right] \, .
\end{align} }
The last step of the computation is provided by expressing the $\sigma_i$ coefficients in terms of the coupling constants that enter the Lagrangian \eqref{eq:Lagrangian}. Will not write such an expression here since it is very lengthy, but we report in the App.\ \ref{sect:appendix-1loop}.

After having outlined the recipe of functional integration over transverse fields on maximally symmetric spaces, we turn to the final result that stems form adding up the contributions of all the spin-parity sectors. 


\subsection{Logarithmic divergences}

In the previous subsection we outlined the general features of evaluating functional traces of fourth-order operators over transverse fields by focusing of the $1^+$-sector, collecting the results for the other spin-parity sectors in the appendix \ref{sect:appendix-1loop}. As we have stressed from the beginning, we choose to get rid of the vector mode parameterized by $\zeta_\mu$. This field has a sixth-order operator and has the same quantum numbers of the vector that appears in the York decomposition of metric perturbations (including the Weyl weight). Technically, it is possible to ``kill" this degree of freedom because, even though the six coefficients that appear in the $\zeta_\mu$-entries of the $1^-$-Hessian depend only on five coupling constants, only three expressions are actually linearly independent (see the App.\ \ref{sect:appendix-hessians}). The requirement that the Lagrangian be independent of this field can be translated into imposing that the action be invariant under the following transformation of the torsion
\begin{equation}\label{eq:Curtrught-sym}
 T^\rho{}_{\mu\nu} \rightarrow T^\rho{}_{\mu\nu} + \nabla^\rho \nabla_\mu \zeta_\nu - \nabla^\rho \nabla_\nu \zeta_\mu + \frac{1}{d-1} \left[ \delta^\rho{}_\nu \left( \square - \frac{R}{d} \right) \zeta_\mu - \delta^\rho{}_\mu \left( \square - \frac{R}{d} \right) \zeta_\nu \right] \, .
\end{equation}
This is a gauge-like invariance that may be thought of as the longitudinal counterpart of those proposed by Curtright in \cite{Curtright:1980yk}. We shall not delve into the geometrical consequences of this symmetry requirement, and postpone it to future works.

In order to give the final result of the second Seeley-DeWitt coefficient we will make some reasonable simplifications. First of all, since the torsion is made out of three algebraically irreducible parts, we are free to choose the normalization of three kinetic terms, one for each component. Our choice is to canonically normalize the kinetic terms involving the gradients of the three algebraically irreducible parts, i.e., $(grad \, \kappa)^2$, $(grad \, H)^2$ and $(grad\, \tau)^2$. Thus, in $d=4$ we have
\begin{equation}\label{eq:canonical-norm-gradients}
 \alpha_1 = \frac{1}{2} \, , \qquad \alpha_2 = \frac{1}{2} \, , \qquad \alpha_3 = \frac{1}{6} + \frac{\alpha_9}{3} - \frac{\alpha_5}{18} - 2 \alpha_4 \, .
\end{equation}
Moreover, as we have just discussed, we get rid of the vector mode that is parameterized by $\zeta_\mu$, which results in the following conditions
\begin{equation}\label{eq:coeff-from-sym-req}
 \alpha_4 = \frac{6}{4} (2b_1+b_2) - \frac{9}{8} \, , \qquad \alpha_5 = \frac{3}{2} - 6 (2b_1+b_2) \, , \qquad \alpha_9 = -1 \, .
\end{equation}
Having fixed these coupling constants, we find the final form of the second Seeley-DeWitt coefficient, which is reported in Eq.\ \eqref{eq:final-result-without-masses} in App.\ \ref{sect:appendix-1loop}. 

As we discussed in Sec.~\ref{sect:Lagrangian}, the renormalization of the Planck mass due to the mass terms of the torsion can be read off directly from final form of the second Seeley-DeWitt coefficient simply by performing the substitution $b_i\to b_i+\frac{m_i^2}{R}$. We have chosen to report this coefficient in App.~\ref{sect:appendix-1loop}, see Eq.~\eqref{eq:final-result-without-masses}. However, such a substitution affects the contributions of ${\rm tr}\hat{a}_2$ to the operators of the effective action in the derivative expansion, i.e., in powers of $R$ (see Eq.\ \eqref{eq:divergenceWithMass}). Once we include masses for the three independent torsion contractions, we have at order $R^2$:

\begin{small}
	{\small \begin{align}\label{eq:result-R-squared}\nonumber
		{\rm tr} & \left(a_2\right) = R^2 \left\{ -\frac{127553}{69120} + \frac{1}{1152} \Bigg[ - \frac{761}{4} + 3420 b_1 + 46096 b_1^2 + 1764 b_2 + 11536 b_2^2 + 46048 b_1 b_2  + \frac{27 m_3^2 \left(26 m_1^2+13 m_2^2+9 m_3^2\right)}{\left(2
			m_1^2+m_2^2\right)^2}  \right.\\\nonumber
		& \left.\left.           +\frac{1}{\left(-2
			\alpha_8^2+12 \alpha_8+\alpha_6 (8 \alpha_8+6)+9\right)^2} [  24 \left(96 \alpha_8^4-128
		\alpha_6 \alpha_8^3-480 \alpha_8^3-768 \alpha_6 \alpha_8^2 -504
		\alpha_8^2-384 \alpha_6^2 \alpha_8 -2880 \alpha_6 \alpha_8-1512
		\alpha_8 \right.\right.\right.\\\nonumber
		& \left.\left.\left.\left. +288 \alpha_6^2+216 \alpha_6  +1152 b_1^2 \left(2 \alpha_6^2+6
		\alpha_6+66 \alpha_8^2+108 \alpha_8+45\right)+144 b_2^2 \left(4 \alpha_6^2+12
		\alpha_6 +600 \alpha_8^2+864 \alpha_8+333\right) \right.\right.\right.\right.\\\nonumber
		& \left.\left.\left.\left. +48 b_2 \left(4 (20 \alpha_8+9) \alpha_6^2-4 \alpha_8 (55 \alpha_8+12)
		\alpha_6+3 \left(-16 \alpha_8^3+\alpha_8^2+132 \alpha_8+54\right)\right)+48 b_1 \left(8 (20 \alpha_8+9)
		\alpha_6^2 \right.\right.\right.\right.\right.\\\nonumber
		& \left.\left. +4 \left(34 \alpha_8^2+192 \alpha_8+81\right) \alpha_6 +3 \left(16
		\alpha_8^3-34 \alpha_8^2+48 \alpha_8+27\right)+12 b_2 \left(4 \alpha_6^2+12
		\alpha_6-3 \left(94 \alpha_8^2+144 \alpha_8+51\right)\right)\right)+405\right)] \\
		& \left.\left. -\frac{72 \sqrt{2}
			\sqrt{(2 \alpha_6+3) (4 \alpha_8+3)}}{\sqrt{2} \sqrt{(2 \alpha_6+3)
				(4\alpha_8+3)}-2 \alpha_8} \left(\frac{12 (2 b_1+b_2-1)}{4 \alpha_8+3}-\frac{72 b_1-72 b_2+16 \alpha_6+15}{2 \alpha_6+3}
		+\frac{4 \sqrt{2} \alpha_8}{\sqrt{(2 \alpha_6+3) (4 \alpha_8+3)}}\right) \right.\right.\\\nonumber
		& \left.\left.  -\frac{72 \sqrt{2} \sqrt{(2 \alpha_6+3) (4 \alpha_8+3)}}{\sqrt{2} \sqrt{(2 \alpha_6+3) (4 \alpha_8+3)}+2 \alpha_8}
		\left(\frac{12 (2 b_1+b_2-1)}{4 \alpha_8+3} -\frac{72 b_1-72 b_2+16 \alpha_6+15}{2 \alpha_6+3}-\frac{4 \sqrt{2} \alpha_8}{\sqrt{(2 \alpha_6+3)
				(4 \alpha_8+3)}}\right) \right.\right.\\\nonumber
		& \left.\left.  -\frac{6 \left(\alpha_6+\frac{3}{2}\right) (8 \alpha_8+6)}{\left(\sqrt{2} \sqrt{(2 \alpha_6+3) (4
				\alpha_8+3)}-2 \alpha_8\right)^2} 
		\left(-\frac{12 (2 b_1+b_2-1)}{4 \alpha_8+3} +\frac{72 b_1-72 b_2+16 \alpha_6+15}{2 \alpha_6+3}-\frac{4 \sqrt{2} \alpha_8}{\sqrt{(2 \alpha_6+3) 
				(4 \alpha_8+3)}}\right)^2 \right.\right.\\\nonumber
		& \left. \left. -\frac{6 \left(\alpha_6+\frac{3}{2}\right) (8 \alpha_8+6)}{\left(\sqrt{2} \sqrt{(2 \alpha_6+3) (4 \alpha_8+3)}+2 \alpha_8\right)^2}
		\left(-\frac{12 (2 b_1+b_2-1)}{4 \alpha_8+3}+\frac{72 b_1-72 b_2+16
			\alpha_6+15}{2 \alpha_6+3}+\frac{4 \sqrt{2} \alpha_8}{\sqrt{(2 \alpha_6+3) (4
				\alpha_8+3)}}\right)^2 \right.\right.\\\nonumber
		& \left.\left. -\frac{6 (24
			b_2 \alpha_6+64 \alpha_8 \alpha_6+24 \alpha_6+60 \alpha_8+48
			b_1 (\alpha_6+6 \alpha_8+6)-36 b_2 (8 \alpha_8+5)+9)^2}{(2 \alpha_6+3)
			(4 \alpha_8+3) \left(-2 \alpha_8^2+12 \alpha_8+\alpha_6 (8 \alpha_8+6)+9\right)} \right.\right.\\\nonumber
		& \left.\left. +\frac{3}{49} \left(\frac{366 m_3^2 \left(2 m_1^2+m_2^2\right)+27 m_3^4}{\left(2
			m_1^2+ m_2^2\right)^2}+979\right) \right]  \right\} \, .
		\end{align} }
\end{small}
The contribution to the second Seeley-DeWitt coefficient that is linear in $R$ and renormalizes Newton's constant reads
\begin{small}
	{\small \begin{align}\label{eq:result-R}\nonumber
		{\rm tr} & \left(a_2\right) = \frac{R}{1152} \left\{ 4 \left(m_1^2 (23048 b_1+11512 b_2+855)+m_2^2 (11512 b_1+5768 b_2+441)\right) \right.\\\nonumber
		& \left. +\frac{24}{\left(-2
			\alpha_8^2+12 \alpha_8+\alpha_6 (8 \alpha_8+6)+9\right)^2} \left[ 48 \left(m_1^2 \left(34 (4 \alpha_6-3) \alpha_8^2+16 (2 \alpha_6 (5 \alpha_6+24)+9) \alpha_8+9 (4 \alpha_6 (2 \alpha_6+9)+9) \right.\right.\right.\right.\\\nonumber
		& \left. \left. \left.\left. +48 \alpha_8^3+48 b_1 (2
		\alpha_6 (\alpha_6+3)+6 \alpha_8 (11 \alpha_8+18)+45)+12 b_2 (4 \alpha_6 (\alpha_6+3)-6 \alpha_8 (47 \alpha_8+72)-153)\right) \right.\right.\right.\\\nonumber
		& \left. \left.\left. +m_2^2 \left(-48 \alpha_8^3+12 b_1 (4 \alpha_6 (\alpha_6+3)-6 \alpha_8 (47 \alpha_8+72)-153)+6
		\left(6 \alpha_6^2+(4 \alpha_6 (\alpha_6+3)+333) b_2+27\right) \right.\right.\right.\right.\\\nonumber
		& \left.\left.\left.\left. +\alpha_8^2 (-220 \alpha_6+3600 b_2+3)+4 \alpha_8 (4 \alpha_6 (5 \alpha_6-3)+1296 b_2+99)\right)\right) \right] \right.\\\nonumber
		& \left. -\frac{72 \sqrt{2}
			\sqrt{(2 \alpha_6+3) (4 \alpha_8+3)}}{\sqrt{2} \sqrt{(2 \alpha_6+3)
				(4\alpha_8+3)}-2 \alpha_8} \left( \frac{12 \left(2 m_1^2+m_2^2\right)}{4 \alpha_8+3}-\frac{72 (m_1-m_2)
			(m_1+m_2)}{2 \alpha_6+3} \right) \right.\\
		& \left. -\frac{72 \sqrt{2} \sqrt{(2 \alpha_6+3) (4 \alpha_8+3)}}{\sqrt{2} \sqrt{(2 \alpha_6+3) (4 \alpha_8+3)}+2 \alpha_8} \left( \frac{12 \left(2 m_1^2+m_2^2\right)}{4 \alpha_8+3}-\frac{72 (m_1-m_2)
			(m_1+m_2)}{2 \alpha_6+3} \right) \right.\\\nonumber
		& \left. -\frac{12 \left(\alpha_6+\frac{3}{2}\right) (8 \alpha_8+6)}{\left(\sqrt{2} \sqrt{(2 \alpha_6+3) (4
				\alpha_8+3)}-2 \alpha_8\right)^2}  \left(\frac{4 \sqrt{2} \alpha_8}{\sqrt{(2 \alpha_6+3) (4 \alpha_8+3)}}+\frac{72
			(b_2-b_1)}{2 \alpha_6+3}+\frac{12 (2 b_1+b_2-1)}{4 \alpha_8+3}\right) \times\right.\\\nonumber
		& \left. \times	\left(\frac{12 \left(2 m_1^2+m_2^2\right)}{4 \alpha_8+3}-\frac{72 (m_1-m_2)
			(m_1+m_2)}{2 \alpha_6+3}\right) \right. \\\nonumber
		& \left. -\frac{12 \left(\alpha_6+\frac{3}{2}\right) (8 \alpha_8+6)}{\left(\sqrt{2} \sqrt{(2 \alpha_6+3) (4 \alpha_8+3)}+2 \alpha_8\right)^2}   \left(\frac{4 \sqrt{2} \alpha_8}{\sqrt{(2 \alpha_6+3) (4 \alpha_8+3)}}+\frac{72
			(b_1-b_2)}{2 \alpha_6+3}-\frac{12 (2 b_1+b_2-1)}{4 \alpha_8+3}\right) \times \right.\\\nonumber
		& \left. \times \left(\frac{72 (m_1-m_2) (m_1+m_2)}{2 \alpha_6+3}-\frac{12 \left(2
			m_1^2+m_2^2\right)}{4 \alpha_8+3}\right) \right.\\\nonumber
		& \left.  - \frac{ 144 (8 \alpha_6 (8 \alpha_8+3)+60 \alpha_8+48 b_1 (\alpha_6+6 \alpha_8+6)+12 b_2 (2 \alpha_6-24 \alpha_8-15)+9) \left(4 m_1^2 (\alpha_6+6
			\alpha_8+6)+m_2^2 (2 \alpha_6-3 (8 \alpha_8+5))\right) }{(2 \alpha_6+3)
			(4 \alpha_8+3) \left(-2 \alpha_8^2+12 \alpha_8+\alpha_6 (8 \alpha_8+6)+9\right)} \right\}
		\end{align} }
\end{small}
Finally, mass terms for the torsion also generate an effective cosmological constant induced by the torsion fluctuations which is parameterized by the part of the ${\rm tr} a_2$ which is independent of the constant scalar curvature
\begin{small}
	{\small \begin{align}\label{eq:result-Lambda}\nonumber
		{\rm tr} & \left(a_2\right) = \frac{1}{1152} \left\{ 16 \left(2881 m_1^4+2878 m_1^2 m_2^2+721 m_2^4\right) + \right.\\\nonumber
		& \left. + \frac{24}{\left(-2
			\alpha_8^2+12 \alpha_8+\alpha_6 (8 \alpha_8+6)+9\right)^2}  \left( 144 \left(8 m_1^4 (2 \alpha_6 (\alpha_6+3)+6 \alpha_8 (11 \alpha_8+18)+45) \right.\right.\right.\\\nonumber
		& \left.\left.\left. +4
		m_1^2 m_2^2 (4 \alpha_6 (\alpha_6+3)-6 \alpha_8 (47 \alpha_8+72)-153)+m_2^4 (4 \alpha_6 (\alpha_6+3)+24 \alpha_8 (25 \alpha_8+36)+333)\right) \right) \right.\\
		& \left. -\frac{6 \left(\alpha_6+\frac{3}{2}\right) (8 \alpha_8+6)}{\left(\sqrt{2} \sqrt{(2 \alpha_6+3) (4
				\alpha_8+3)}-2 \alpha_8\right)^2}  \left(\frac{72 (m_1-m_2) (m_1+m_2)}{2 \alpha_6+3}-\frac{12 \left(2
			m_1^2+m_2^2\right)}{4 \alpha_8+3}\right)^2  \right.\\\nonumber
		& \left. -\frac{6 \left(\alpha_6+\frac{3}{2}\right) (8 \alpha_8+6)}{\left(\sqrt{2} \sqrt{(2 \alpha_6+3) (4 \alpha_8+3)}+2 \alpha_8\right)^2} \left(\frac{72 (m_1-m_2) (m_1+m_2)}{2 \alpha_6+3}-\frac{12 \left(2
			m_1^2+m_2^2\right)}{4 \alpha_8+3}\right)^2 \right.\\\nonumber
		& \left. -\frac{864 \left(4 m_1^2 (\alpha_6+6 \alpha_8+6)+m_2^2 (2 \alpha_6-3 (8 \alpha_8+5))\right)^2}{(2 \alpha_6+3)
			(4 \alpha_8+3) \left(-2 \alpha_8^2+12 \alpha_8+\alpha_6 (8 \alpha_8+6)+9\right)}  \right\}
		\end{align} }
\end{small}
The first numerical factor on the right-hand-side of Eq.\ \eqref{eq:result-R-squared} comes from adding up the geometric contributions due to each spin-parity sector, see the App.\ \ref{sect:appendix-1loop}, while the second one arises from the fact that we fixed the actual values of some of the coupling constants. Clearly, none of them appears in the Eq.s \eqref{eq:result-R} and \eqref{eq:result-Lambda}, since there is no dependence on the couplings of the starting Lagrangian \eqref{eq:Lagrangian}.

We observe that the effect of the masses $m_i$ is to produce logarithmic divergences to all orders in the curvature expansion of the effective action. This is expected since any combination of masses of the form $m_1^{2n_1}m_2^{2n_2}m_3^{2n_3}$ contributes to operators of the form $R^{n_1+n_2+n_3}$.

\section{An Application: Renormalization of Starobinsky inflation in torsionful theories}\label{subsect:final-result}

As an example to show the relevance of our computation we briefly apply it to the Starobinsky model \cite{Starobinsky:1980te}. The Starobinsky model is particularly relevant in the context of inflationary cosmology, its main advantage being that it admits a de Sitter background, which is nevertheless unstable. The typical evolution of the universe in this context is governed by a quasi de Sitter geometry, according to which the universe goes through an inflationary de Sitter phase that stops at a finite time. In order to fine tune the inflationary phase to a lifetime that matches observations, it is customary to introduce a ``slow-roll'' parameters $\varepsilon$ and $\eta$ to describe the potential for the fluctuations of the scalar curvature \cite{Baumann_TASI_lectures}. When they vanish, the Hubble expansion rate is constant and we recover the actual de Sitter metric; in this limit inflation goes on forever. For small $\varepsilon$ and $\eta$ we have the transition to a non-inflationary epoch at late but finite time. From high energy point of view, the Starobinsky model is described by the following action
\begin{align}\label{eq:staroAction}
S_{\rm star}[g]
=
\int d^4x  \sqrt{g} \left( 
m_p^2 R 
+
\frac{1}{\alpha} R^2 
\right)  \, ,
\end{align}
where $m_p^2 = \frac{1}{16\pi G}$ is the Planck mass and the previous Lagrangian can be understood as an effective model induced by quantum fluctuations of matter fields. Indeed, one way of obtaining it is through a self consistency approach, where one investigates the quantum corrections to the energy momentum tensor induced by matter fields and studies the semiclassical Einstein equations
\begin{align}
R_{\mu\nu}-\frac{1}{2}Rg_{\mu\nu} = \frac{1}{m_p^2} \langle T_{\mu\nu}\rangle\,.
\end{align}
In our approach, we treat the torsion as a particular kind of matter field and notice that it can induce $R^2$ terms as well. Working at the level of the action, rather then at the level of the equations of motion, a similar semi-classical approach would consider the $1$-loop effect of torsion propagating on a classical Riemannian background and include their effect in the running of the couplings $\alpha$ and $m_p$.

While one should study how the torsion couples to scalar and tensor fluctuations of the metric\footnote{In particular, we expect two mixing happening between metric and torsion degrees of freedom: one between the spin-$2$ fields, i.e., between $h^{TT}_{\mu\nu}$ and $S_{\mu\nu}$, and one between the scalar inflaton and the spin-$0$ degrees of freedom in the torsion decomposition. No mixing is expected in the $1^-$ sector, since this is a pure gauge d.o.f. on the metric side.}, in this work we focus on a less ambitious task and look at its properties on a fixed de Sitter background. From a cosmological point of view, this corresponds to the $0$-th order in the $\varepsilon$ and $\eta$ expansion. While quantitatively far from a complete calculation, such an approximation still allows us to extract qualitative information, like the production of an effective cosmological constant due to torsion fluctuations. In particular, our work can be useful to define the initial conditions determining the de Sitter phase, provided that a viable model for the torsion dynamics is chosen on some phenomenological base. We leave a more refined work including dynamical fluctuations of the metric for future research.

From this result it is customary to get the beta function for the coupling of the $R^2$ term in the Starobinsky's Lagrangian. We adopt dimensional regularization and the ${MS}$-scheme, in which the counterterms are chosen to eliminate only the $\epsilon$-poles. Le us consider the Starobinsky's bare Lagrangian, Eq.\ \eqref{eq:staroAction}. The one-loop effective action reads
\begin{align}
\Gamma_{star}[g]
 =
 S_{star}[g]
 +
 \Gamma_{eff} 
 =
  S_{star}[g]
 +
 \Gamma_{fin} [g]
 +
  \Gamma_{div} [g]
 \, ,
\end{align}
where the 1-loop divergences in dimensional regularization are
\begin{align}
\Gamma_{div}
=
-\frac{\mu^{-\epsilon}}{16 \pi^2\epsilon} \int d^4 x \sqrt{g}\, 
 {\rm tr}\left(a_2\right) 
= -\frac{\mu^{-\epsilon}}{16 \pi^2\epsilon} \int d^4 x \sqrt{g} \left( 
 {\cal A}_2 R^2 + {\cal A}_1 R 
\right)  \, ,
\end{align}
Of course, $\mu$ is a mass scale introduced as usual to keep the coupling $ \frac{1}{\alpha} $ dimensionless. We split the $4d$ bare Lagrangian into its renormalized part and counterterms
{\small \begin{align}
S_{\rm star}[g]
=
 \int d^4x  \sqrt{g} \left( 
m_p^2 R
+
\frac{1}{\alpha}  R^2 
\right) 
=
S^{ren}_{\rm star}[g]
+
S^{ct}_{\rm star}[g]
=
\mu^{-\epsilon} \int d^dx  \sqrt{g} \left[
\left(
{m_p^2}_R  
+
\delta Z_{m_p^2} 
\right) R
+
\left(
\frac{1}{\alpha_R} 
+
\delta Z_{ 1/\alpha } 
\right) R^2 
\right] \, ,
\end{align} }
where the subscript ``$R$'' stands for renormalized. Therefore, the counterterms are
\begin{align}
\begin{split}
\delta Z_{m_p^2} = \frac{{\cal A}_1}{16 \pi^2 \epsilon}  \, , \qquad
\delta Z_{ 1/\alpha }   = \frac{{\cal A}_2}{16 \pi^2 \epsilon}   \, .
\end{split}
\end{align}
Accordingly, we get the following relation between bare and renormalized couplings
\begin{align}\label{renormalized_coupling}
\frac{1}{\alpha}
=&
\mu^{-\epsilon}
\left(
\frac{1}{\alpha_R}
+
\frac{1}{16\pi^2} \frac{{\cal A}_2}{\epsilon}
\right) \, ,\\
m_p^2
=&
\mu^{-\epsilon}
\left(
{m_p^2}_R  
+
\frac{1}{16\pi^2} \frac{{\cal A}_1}{\epsilon}
\right) \,.
\end{align}
By using the $\mu$-independence of $1/\alpha$ and taking the $\epsilon \to 0$ limit we get
\begin{align}
\beta_{\alpha}
\equiv 
\mu\frac{\partial\alpha_R}{\partial \mu}
= 
- \frac{{\cal A}_2}{16\pi^2}\alpha_R^2 \, .
\end{align}
Integrating this equation we get the running
\begin{align}\label{eq:integratedBeta}
\frac{\alpha_R(\mu)}{16 \pi^2}
=
\frac{\alpha_R(\mu_0)}{16 \pi^2 +{\cal A}_2 \ln(\mu/\mu_0) } \, 
\end{align}
giving the value of the coupling constant at an arbitrary scale $\mu$ in terms of its value at a reference scale $\mu_0$. Similar arguments hold for Newton's constant $G_N = \frac{16\pi}{m_p^2}$. Depending on the sign of the trace in Eq.\ \eqref{eq:result-R-squared}, we see that torsion fluctuations may enhance the asymptotic freedom of the model or work against it. A more phenomenological analysis is necessary to restrict the physically interesting Lagrangians and gain further insights on the faith of the asymptotic behavior of the Starobinsky model. However, during inflation the Hubble expansion rate is usually of order $\sqrt{\alpha_R}$. Therefore, if we wish to reproduce the phenomenology of inflation, it is clear that the coupling needs to be small close to the UV regime. The asymptotically free case of ${\cal A}_2>0$ is therefore the one we are looking for. It is interesting to notice that the coupling diverges at a finite value of the renormalization scale, i.e., for $\log(\mu/\mu_0)=-\frac{16\pi^2}{{\cal A}_2}$ in our scheme, signaling a possible second-order phase transition. While the critical value of the scale and of the coupling are scheme dependent, it would be interesting to check whether the existence of the phase transition is actually a universal result (see \cite{Gies:2018vwk,Gies:2019nij} for details on the scheme dependence of asymptotically free solutions in gauge theories). We hope that future works may shed more light on these topics.

\vspace{0.8cm}

\section{Conclusions and outlooks}\label{sect:conclusions}

In this paper we showed how one can employ both the spin-parity decomposition of torsion fluctuations, from \cite{Martini:2023rnv}, and the results of \cite{Barvinsky:1985an} to work out the trace of the second Seeley-DeWitt coefficient for torsion theories. Of course, the model we have started with is not the most general one, since we only have couplings to the Ricci scalar because of the assumption of being on a maximally symmetric background. Within this approximation, our model displays all the couplings one can construct between torsion and curvature. Moreover, given that the curvature is constant, we can account for mass terms upon the redefinition $b_i\to b_i+\frac{m_i}{R}$ (if the background is not flat). While the assumption of maximally symmetric background highly reduces the generality of our analysis, such an approximation becomes physically relevant to the inflationary epoch, i.e., to the well-known de Sitter phase. Moreover, there is another huge approximation that we perform, i.e., to discard the $1^-$ sector of the torsion. We hope that further work can shed more light on such a setup. Eventually, we stress that our result holds when we can discard metric fluctuations. This is the case when the typical energy of a process in much below the Planck mass and above the mass threshold of the torsion $m_{Tor}$. This approximation is valid in our model when conditions \eqref{eq:bound1}, \eqref{eq:bound2} and \eqref{eq:bound3} are satisfied.
	
There are two important natural generalizations of this work. The first is to take into account some non-trivial torsion background. While here we only looked at the contribution of torsion fluctuations to the flow of $R^2$, $R$ and the cosmological constant, having a background for $T^\rho{}_{\mu\nu}$ would provide us with torsion-dependent counterterms, and it would allow us to renormalize torsion operators as well.
One is then interested in going beyond the quadratic truncation and, in particular, to include marginal interactions of the torsion, to obtain insights about the class of renormalizable torsion theories. The second generalization is to include fluctuations of the metric. Indeed, the computation that we have performed in this paper is valid below the Planck scale, where quantum metric perturbations are frozen. This has been justified by assuming that the coupling constants in the torsion sector as such that the dynamics become non-trivial below the Planck scale, and that the masses of the torsion excitation lie below $m_p$. The inclusion of quantum fluctuations of the metric would be necessary in order to prove the perturbative $1$-loop renormalizability of the theory, which would be highly non-trivial due to the possible mixings of the torsion degrees of freedom with those of the graviton. Such mixings would definitely be present, since the torsion carries spin-parity $2^+$ and $0^+$ degrees of freedom. This is a task that we plan to tackle in the future.
 
An important remark is in order: Starobinsky's inflation is driven by the conformal anomaly. Intuitively, the presence of a conformal anomaly on a maximally symmetric space leads to an inflationary phase, where the role of the cosmological constant is played by the anomaly itself. From this point of view, it could be interesting to obtain the conformal action for the full torsion tensor and use our results to check if torsion fluctuations can give rise to a consistent inflationary dynamics. It would also be interesting to explicitly find some regions in the coupling space where ${\rm tr}(a_2)>0$. In fact, also in this case torsion would not allow to exit inflation. We hope to analyze these interesting physical consequences of our work in the future.   
	
At this point, we would like to make some remarks about RG flow of torsionful theories. A maximally symmetric Riemannian background may be adopted, together with a completely general torsion background, as a starting point for the study of the renormalizability and unitarity of antisymmetric MAGs. We argue that the decomposition of the torsion Eq.\ \eqref{eq:spin-parity-decomposition}, together with the York decomposition of the graviton, and the symmetry \eqref{eq:Curtrught-sym} are enough to obtain differential operators of the form studied in \cite{Barvinsky:1985an} (up to invertible field redefinitions in the $2^-$ and $1^-$ sectors). This is due to the fact that the transversality of the fields wipes out the non-minimal terms of higher rank, while the maximal symmetry of the background simplifies the tensorial structure of the coefficients in the operator. As a matter of fact, although one is not able to disentangle the Ricci scalar from the Ricci and Riemann tensors, all the non-trivial mixings between torsion and metric quantum fluctuations would already arise at this level. Furthermore, all the nine kinetic terms for the torsion, along with those terms that are quartic and cubic in the torsion, will be generated in the trace of the $a_2$. Thus, the only counterterms that will not contribute at all are those whose tensor structure is $R \nabla T$, $Ric \, T^2$ and $Weyl \, T^2$. 
	
A final formal point concerns the Curtright-like symmetries that prevent the presence of ghosts and tachyons at the linearized level \cite{Baldazzi:2021kaf, Percacci:2020ddy}. First of all it would interesting to study whether they can be interpreted as gauge symmetries. That would give a deeper meaning to the spin-parity decomposition since, as it happens for other gauge theories, the longitudinal modes would gain the role of gauge degrees of freedom and could be disregarded without fear. If one succeeds in defining a gauge-like structure for these transformations, one should then define a gauge-covariant Lie derivative, such that the torsion decomposition is preserved by diffeomorphisms along the manifold. Regardless of this interpretation, it is important to verify that the RG flow preserves these symmetries, implying constraints on the relative coefficients of the counterterms.
	
\section*{Acknowledgment}

We are thankful to O. Zanusso and R. Percacci for useful comments and discussions during the development of this draft.
	

\appendix
\section{Spin-parity decomposed kinetic terms}\label{sect:appendix-kin-terms}

In this appendix we will collect some expression that were left out of the main body of the paper. In particular, we will provide some useful formulae for the gradients and divergences of the torsion, as well as the expressions of the Hessians in each spin-parity sector. All the forthcoming results are valid only on a maximally symmetric Riemannian background.

First of all, the gradient and divergences of the hook antisymmetric part of the torsion enter the Lagrangian in different ways, and it proves useful to further single out the contributions of the traceless and pure trace parts. For the square of the gradient we have
\begin{equation}
 \int \sqrt{g} \, (grad \, t)^2 = \int \sqrt{g} \left[ (grad \,\kappa)^2 + \frac{2}{d-1} (grad \, \tau)^2 \right] \, ,
\end{equation}
while for the squares of the divergence we find
\begin{subequations}
	{\small \begin{align}
	& \int \sqrt{g} \, (div_1 t)^2 = \int \sqrt{g} \left[ (div_1 \kappa)^2 - 4 (div_1 \kappa) (grad \, \tau) + 2 (grad \, \tau)^2 - 2 (div \, \tau)^2 + \frac{2}{d} R \tau^2 \right] \, , \\
	& \int \sqrt{g} \, (div_2 t)^2 = \int \sqrt{g} \left[ (div_2 \kappa)^2 + \frac{d-2}{(d-1)^2} (div \, \tau)^2 + \frac{1}{2(d-1)^2} (grad \, \tau)^2 - \frac{2}{d-1} (div_1 \kappa) (grad \, \tau) \right] \, .
	\end{align} }
\end{subequations}
Finally, the remaining terms that involve $t^\rho{}_{\mu\nu}$ and parameterize its mixing with the other components are written in terms of the first divergence only, and read
\begin{subequations}
 {\small \begin{align}
  & \int \sqrt{g} \, (div_1 t) (grad \, \tau) = \int \sqrt{g} \left[ (div_1 \kappa ) (grad \, \tau) - \frac{1}{d-1} (grad \, \tau)^2 + \frac{1}{d-1} (div \, \tau)^2 - \frac{R}{d(d-1)} \tau^2 \right] \, , \\
  & \int \sqrt{g} \, (div_1 t) (div_1 H) = \int \sqrt{g} \, (div_1 \kappa ) (div_1 H) \, .
 \end{align} }
\end{subequations}

In the main text we have employed the spin-parity decomposition found in \cite{Martini:2023rnv} to simplify the form of the final differential operators. Inserting it into the nine terms that we have chosen as a basis for the kinetic terms highlights which spin-parity sectors contribute to each term. As one may have expected, the squares of the gradients contain all the sectors. Indeed, for the square of the gradient of the hook antisymmetric tracefree part we find
{\small \begin{align}
 & \int \sqrt{g} (grad \, \kappa )^2 = \int \sqrt{g} \left[ - \kappa^\rho{}_{\mu\nu} \square \kappa_\rho{}^{\mu\nu} + 6 A_{\mu\nu} \square^2 A^{\mu\nu} + \frac{6(2d-1)}{d(d-1)} R A_{\mu\nu} \square A^{\mu\nu} + \frac{6(d-2)(d+1)}{d^2(d-1)^2} R^2 A_{\mu\nu} A^{\mu\nu} \right.\\\nonumber
 & \left. \qquad\qquad\qquad\qquad\qquad + 2 S_{\mu\nu} \square^2 S^{\mu\nu} - \frac{6}{d(d-1)} R S_{\mu\nu} S^{\mu\nu} - \frac{2(d-3)}{d(d-1)^2} R^2 S_{\mu\nu} S^{\mu\nu} \right. \\\nonumber
 & \left. \qquad\qquad\qquad\qquad\qquad - \frac{2(d-2)}{d-1} \xi_\mu \square^3 \xi^\mu - \frac{4(d-2)}{d(d-1)} R \xi_\mu \square^2 \xi^\mu + \frac{2(d-2)}{d^2(d-1)} R^2 \xi_\mu \square \xi^\mu + \frac{4(d-2)}{d^3(d-1)} R^3 \xi_\mu \xi^\mu  \right] \, .
\end{align} }
On the other hand, the square of the gradient of the completely antisymmetric part reads
{\small \begin{align}
  & \int \sqrt{g} (grad \, H)^2 = \int \sqrt{g} \left[ 3 \Pi_{\mu\nu} \square^2 \Pi^{\mu\nu} - \frac{3(d+1)}{d(d-1)} R \Pi_{\mu\nu} \square \Pi^{\mu\nu} -\frac{6(d-5)(d-2)}{d^2(d-1)^2} R^2 \Pi_{\mu\nu} \Pi^{\mu\nu} + \frac{1}{6} \pi \square \left( \square + \frac{R}{4} \right) \pi \right] \, ,
\end{align} }
where the $0^-$ term is given in $d=4$ for simplicity. The analogous result for the torsion vector is
{\small \begin{align}
  & \int \sqrt{g} (grad \, \tau)^2 = \int \sqrt{g} \left[ -\tau_\mu \square \tau^\mu + \varphi \square \left( \square + \frac{R}{d} \right) \varphi \right] \, .
\end{align} }
The squares of the two independent divergences of $\kappa $ can be expressed as
\begin{subequations}
 {\small \begin{align}
  & \int \sqrt{g} (div_1 \kappa )^2 = \int \sqrt{g} \left[ 4 A_{\mu\nu} \square^2 A^{\mu\nu} + \frac{8(d-2)}{d(d-1)} A_{\mu\nu} R \square A^{\mu\nu} + \frac{4(d-2)^2}{d^2(d-1)^2} A_{\mu\nu} R^2 A^{\mu\nu} \right.\\\nonumber
  & \qquad\qquad\qquad\qquad\qquad \left. - \frac{2(d-2)^2}{(d-1)^2} \xi_\mu \square^3 \xi^\mu - \frac{2(d-2)^2}{d(d-1)^2} \xi_\mu R \square^2 \xi^\mu + \frac{2(d-2)^2}{d^2(d-1)^2} \xi_\mu R^2 \square \xi^\mu + \frac{2(d-2)^2}{d^3(d-1)^2} \xi_\mu R^3 \xi^\mu \right] \, , \\
  & \int \sqrt{g} (div_1 \kappa ) (grad \, \tau) = \int \sqrt{g} \left[ -\frac{d-2}{d-1} \tau^\mu \square^2 \xi_\mu + \frac{d-2}{d^2(d-1)} R^2 \xi^\mu \tau_\mu \right] \, ; \\
  & \int \sqrt{g} (div_2 \kappa )^2 = \int \sqrt{g} \left[ S_{\mu\nu} \square^2 S^{\mu\nu} - \frac{2 R}{d-1} S_{\mu\nu} \square S^{\mu\nu} + \frac{R^2}{(d-1)^2} S_{\mu\nu} S^{\mu\nu} - \frac{(d-2)^2}{(d-1)^2} \xi_\mu \square^3 \xi^\mu + \frac{(d-2)^2}{d^2(d-1)^2} R^2 \xi_\mu \square \xi^\mu \right] \, .
 \end{align} }
\end{subequations}
Moreover, the analogous expression for the divergence of the completely antisymmetric part is
\begin{equation}
 \int \sqrt{g} (div_1 H)^2 = \int \sqrt{g} \left[ \Pi_{\mu\nu} \square^2 \Pi^{\mu\nu} - \frac{4(d-2)}{d(d-1)} R \Pi_{\mu\nu} \square \Pi^{\mu\nu} + \frac{4(d-2)^2}{d^2(d-1)^2} R^2 \Pi_{\mu\nu} \Pi^{\mu\nu} \right] \, ,
\end{equation}
while the divergence of the vector part simply reads
\begin{equation}
 \int \sqrt{g} (div \, \tau)^2 = \int \sqrt{g} \varphi \square^2 \varphi \, .
\end{equation}
Finally, the only mixed term is decomposed as
\begin{equation}
 \int \sqrt{g} (div_1 \kappa ) (div_1 H) = \int \sqrt{g} \left[ 2 A_{\mu\nu} \square^2 \Pi^{\mu\nu} - \frac{2(d-2)}{d(d-1)} R A_{\mu\nu} \square \Pi^{\mu\nu} - \frac{4(d-2)^2}{d^2(d-1)^2} R^2 A_{\mu\nu} \Pi^{\mu\nu} \right] \, .
\end{equation}
The expressions of the spin-parity decomposed squares of the three irreducible algebraic components of the torsion were given implicitly in \cite{Martini:2023rnv} and will not be written here.

\section{Hessians}\label{sect:appendix-hessians}

We move on to the components of the Hessian corresponding to the Lagrangian \eqref{Lagriangian-max-sym} in each spin-parity sector. We start from the $2^-$-sector, in which we have
\begin{equation}\label{eq:hessian-kappa}
 \hat{H}_{\tiny \kappa\kappa} = - 2 \alpha_1 \square \mathds{1} + (2 b_1 + b_2) R \mathds{1} \, .
\end{equation}
In the $2^+$-sector we find
{\small \begin{align}\label{eq:hessian-S}
 \hat{H}_{\tiny SS} = & \left( 4 \alpha_1 + 2 \alpha_5 \right) \square^2 \mathds{1} - \left( \frac{12 \alpha_1}{d(d-1)} + \frac{4\alpha_5}{d-1} + 4b_1 + 2b_2 \right) R \square \mathds{1} \\\nonumber
 & + \frac{2}{d(d-1)^2} \left[ (2b_1+b_2)d(d-1) - 2 \alpha_1 (d-3) + \alpha_5 d \right] R^2 \mathds{1} \, .
\end{align} }
The $1^+$ and $1^-$ sectors are the only two that are two-dimensional. We start from the first one, for which the three non-trivial expressions are
\begin{subequations}\label{eq:hessian-1forms}
 {\small \begin{align}
  \hat{H}_{\tiny AA} = & \left( 12 \alpha_1 + 8 \alpha_4 \right) \square^2 \mathds{1} + \left[ \frac{4}{d(d-1)} \left( (6d-3) \alpha_1 + 4 (d-2) \alpha_4 \right) - 6(2b_1 + b_2) \right] R \square \mathds{1} \\\nonumber
  & - \frac{2(d-2)}{d^2(d-1)^2} \left[ 3d(d-1) (2b_1+b_2) + 6 (d+1) \alpha_1 + 4 (d-2) \alpha_4 \right] R^2 \mathds{1} \, ,\\
  \hat{H}_{\tiny \Pi \Pi} = & (2 \alpha_6 + 6 \alpha_2) \square^2 \mathds{1} - \left( \frac{6(d+1)\alpha_2}{d(d-1)} + \frac{8(d-2)\alpha_6}{d(d-1)} + 6 (b_1-b_2) \right) R \square \mathds{1} \\\nonumber
  & + \frac{4(d-2)}{d(d-1)} \left( \frac{2(d-2) \alpha_6}{d(d-1)} - \frac{3(d-5)\alpha_2}{d(d-1)} + 3(b_1-b_2) \right) R^2 \mathds{1} \, ,\\
  \hat{H}_{\tiny \Pi A} = & 2 \alpha_8 \square^2 \mathds{1} - \frac{2(d-2)\alpha_8}{d(d-1)} R \square \mathds{1} - \frac{4(d-2)^2\alpha_8}{d^2(d-1)^2} R^2 \mathds{1} \, .
 \end{align} }
\end{subequations}
On the other hand, in the $1^-$-sector we have
\begin{subequations}\label{eq:hessian-vectors}
	{\small \begin{align}
	\hat{H}_{\tiny \xi\xi} = & - \frac{d-2}{(d-1)^2} \left[ 4 (d-1) \alpha_1 + (d-2) (4 \alpha_4 + \alpha_5 ) \right] \square^3 \mathds{1} \\\nonumber
	& + \frac{d-2}{d(d-1)^2} \left[ 2d(d-1)(2b_1 + b_2) - 8 (d-1) \alpha_1 + (d-2) (\alpha_5 - 4 \alpha_4) \right] R \square^2 \mathds{1}\\\nonumber
	& + \frac{d-2}{d^2(d-1)^2} \left[  4 (d-1) \alpha_1 + (d-2) (4 \alpha_4 + \alpha_5 ) \right] R^2 \square \mathds{1}\\\nonumber
	& + \left[ 2d(d-1)(2b_1 + b_2) - 8 (d-1) \alpha_1 + (d-2) (\alpha_5 - 4 \alpha_4) \right] R^3 \mathds{1} \, ,\\
	\hat{H}_{\xi \tau} = & \frac{d-2}{(d-1)^2} \left[ \alpha_5 + 4 \alpha_4 - (d-1) \alpha_9 \right] \square^2 \mathds{1} \\\nonumber
	& + \frac{d-2}{d^2 (d-1)^2} \left[ (d-1) \alpha_9 - \alpha_5 - 4 \alpha_4 \right] R^2 \mathds{1} \, ,\\
	\hat{H}_{\tau\tau} = & - \frac{1}{(d-1)^2} \left[ 4 (d-1) \alpha_1 + 2 (d-1)^2 \alpha_3 + 4 \alpha_4 + \alpha_5 - 2 (d-1) \alpha_9 \right] \square \mathds{1} \\\nonumber
	& + \frac{1}{d(d-1)^2} \left[ 2 d (d-1) \left( 2 b_1 + b_2 + (d-1) b_3 \right) + 4 \alpha_4 - \alpha_5 - (d-1) \alpha_9 \right] R \mathds{1} \, .
	\end{align} }
\end{subequations}
The scalar part of the Hessian is given by the $0^+$ and $0^-$ sectors. The first one reads
{\small \begin{align}\label{eq:hessian-scalar}
 \hat{H}_{\tiny \varphi\varphi} = & \frac{2}{d-1} \left[ 2 \alpha_1 + (d-1) \alpha_3 + \alpha_5 + (d-1) \alpha_7 \right] \square^2\\\nonumber
 & + \frac{2}{d(d-1)^2} \left[ 2 (d-1) \alpha_1 + (d-1)^2 \alpha_3 + \alpha_5 - (d-1) \left( 2b_1 + b_2 + (d-1) b_3 \right) \right] R \square \, ,
\end{align} }
while for the second one we find
\begin{equation}\label{eq:hessian-pseudoscalar}
 \hat{H}_{\tiny \pi\pi} = 4 \alpha_2 \square^2 + \frac{1}{3} \left( \frac{\alpha_2}{4} - b_1+b_2 \right) R \square \, . \\
\end{equation}

\section{Coincidence limit of the Seeley-DeWitt coefficients}\label{sect:appendix-1loop}

In this section we will collect the partial results that concur to the final form of the second Seeley-DeWitt coefficient given in Eq.\ \eqref{eq:result-R-squared}. We first provide the results for second-order operators in \ref{subsect:second-order}, and then focus on fourth-order ones in \ref{subsect:fourth-order}.

One of the ingredients that enters both calculations is the trace of the generalized curvature, see section\ \ref{sect:Schwinger_DeWitt_technique}. On maximally symmetric spaces, such a trace only depends on the scalar curvature and on the background dimensions. For tracefree hook antisymmetric tensors we find
\begin{equation}\label{eq:trace-curv-squared-TFhook-as}
 {\rm tr} \left( ({\cal R}_{\mu\nu})_{\tiny \kappa} ({\cal R}^{\mu\nu})_{\tiny \kappa} \right) = - \frac{2 (d+2)(d-2) R^2}{d(d-1)} \, .
\end{equation} 
The analogous result for $2$-forms is
\begin{equation}\label{eq:trace-curv-squared-2form}
 {\rm tr} \left( ({\cal R}_{\mu\nu})_{\tiny A} ({\cal R}^{\mu\nu})_{\tiny A} \right) = - \frac{2 (d-2) R^2}{d(d-1)} \, .
\end{equation}
On the other hand, the contribution due to a symmetric tracefree field is
\begin{equation}\label{eq:trace-curv-TFsym}
 {\rm tr} \left( ({\cal R}_{\mu\nu})_{\tiny \overline{S}} ({\cal R}^{\mu\nu})_{ \overline{S}} \right) = - \frac{2(d+2)R^2}{d(d-1)} \, .
\end{equation}
Finally, for vectors we have
\begin{equation}\label{eq:trace-curv-vector}
 {\rm tr} \left( ({\cal R}_{\mu\nu})_{\tiny \tau} ({\cal R}^{\mu\nu})_{\tiny \tau} \right) = - \frac{2R^2}{d(d-1)} \, .
\end{equation}
Notice that at this order in the curvature the results would be the same also if we had taken into account traces over transverse tensors. This is because the difference would affect terms with more than two derivatives, like, e.g., the third Seeley-DeWitt coefficient.

\subsection{Second-order operators}\label{subsect:second-order}

\subsubsection{$2^-$-sector}

The second-order differential operator acting on the hook antisymmetric tracefree tensor $\kappa_{\perp}{}^\rho{}_{\mu\nu}$ that we read off from Eq.\ \eqref{eq:hessian-kappa} is
\begin{equation}\label{eq:diff-op-kappa}
\Delta_{2^-} = - (\square + \Upsilon R) \mathds{1} \, .
\end{equation}
Here $\Upsilon=-\frac{2b_1+b_2}{2 \alpha_1}$ and ${\rm Tr}(\mathds{1})=16$ since we are dealing with a tracefree hook antisymmetric tensor in $d=4$. Notice that this number pertains to not-transverse fields, c.f. the table in \cite{Martini:2023rnv}. Comparing with Eq.\ \eqref{eq:kappaTrace}, the trace over transverse $\kappa$ is expressed as
{\small \begin{align}
\begin{split}
\left.{\rm Tr \, e}^{-s \Delta_{2^-}} \right|_{ \kappa_\perp{}^\rho{}_{\mu\nu} } = & 
\left.{\rm Tr \, e}^{-s \Delta_{2^-}} \right|_{ \kappa^\rho{}_{\mu\nu} } - 
\left.{\rm Tr \, e}^{-s \Delta_{2^+}} \right|_{\overline{S}_\perp{}_{\mu\nu}} - \left.{\rm Tr \, e}^{-s \Delta_{1^+}} \right|_{A_{\mu\nu}} - \left.{\rm Tr \, e}^{-s \Delta_{1^-}} \right|_{\zeta_\perp{}_\mu}\\
& +
\left.{\rm Tr \, e}^{-s \tilde{\Delta}_{1^-}} \right|_{\xi_\perp{}_\mu}  + 
10 \, {\rm e}^{-s \Delta_{1^+} (-\square \rightarrow \frac{R}{6})} + 10 \, {\rm e}^{-s \Delta_{1^-} (-\square \rightarrow \frac{R}{4})} \, .
\end{split}
\end{align} }
The terms that enter the first line with a minus sign are those that appear in the decomposition of the torsion tensor Eq.\ \eqref{eq:spin-parity-decomposition}, except for the fact that $A_{\mu\nu}$ is not transverse. We have employed the results given in \cite{Codello:2008vh} to evaluate the traces over $\overline{S}_\perp{}_{\mu\nu}$ and $\zeta_{\perp}{}_\mu$. The trace over $\xi_\perp{}_\mu$ counterbalances the presence of the longitudinal modes in $A_{\mu\nu}$. The zero modes are due to the kinematical zero modes of the Curtright forms, i.e.\, $(\overline{C} \, A_\perp)^\rho{}_{\mu\nu}$ and the $(\overline{C} \, \zeta_\perp)^\rho{}_{\mu\nu}$, respectively (see section\ \ref{sect:spectrum-of-diff-constrained}).
In $d=4$ the operators on the longitudinal modes are given in terms of the starting one using the results of section\ \ref{sect:spectrum-of-diff-constrained} as
{\small \begin{align}
\begin{split}
& \Delta_{2^+} = \Delta_{2^-} \left( -\square \rightarrow -\square - \frac{R}{12} \right) \, , \qquad \Delta_{1^+} = \Delta_{2^-} \left( -\square \rightarrow -\square - \frac{5 R}{6} \right) \, ; \\
& \Delta_{1^-} = \Delta_{2^-} \left( -\square \rightarrow -\square - \frac{R}{2} \right) \, , \qquad \tilde{\Delta}_{1^-} = \Delta_{2^-} \left( -\square \rightarrow -\square - \frac{13 R}{12} \right) \, .
\end{split}
\end{align} }
The second Seeley-DeWitt coefficient that comes from evaluating the contribution of \eqref{eq:diff-op-kappa} on non-transverse configurations of $\kappa$ is
\begin{equation}
\left. {\rm tr} \left( a_2 (\Delta_{2^-}) \right)\right|_{\kappa^\rho{}_{\mu\nu}} = R^2 \left[ \frac{13}{270} + \frac{8}{3} \Upsilon + 8 \Upsilon^2  \right]. 
\end{equation}
Likewise, the transverse traceless symmetric tensor $\overline{S}_\perp{}_{\mu\nu}$ contributes with
\begin{equation}
\left. {\rm tr} \left( a_2 (\Delta_{2^+}) \right)\right|_{\overline{S}_\perp{}_{\mu\nu}} = R^2 \left[ - \frac{31}{432} - \frac{5\Upsilon}{6} + \frac{5\Upsilon^2}{2} \right] \, .
\end{equation}
On the other hand, the non-transverse $2$-form $A_{\mu\nu}$ gives rise to the following part
\begin{equation}
\left. {\rm tr} \left( a_2 (\Delta_{1^+}) \right) \right|_{A_{\mu\nu}} = R^2 \left[ \frac{1069}{360} + 6 \Upsilon + 3 \Upsilon^2 \right] \, .
\end{equation}
Finally, the contributions of the two transverse vectors $\zeta_\perp{}_\mu$ and $\xi_\perp{}_\mu$ is given by
\begin{subequations}
 {\small \begin{align}
  \left. {\rm tr} \left( a_2 (\Delta_{1^-}) \right)\right|_{\zeta_\perp{}_\mu} = & R^2 \left[ \frac{713}{1440} + \frac{7 \Upsilon}{4} + \frac{3 \Upsilon^2}{2} \right] \, . \\
  \left. {\rm tr} \left( a_2 (\tilde{\Delta}_{1^-}) \right) \right|_{\xi_\perp{}_\mu} = & R^2 \left[ \frac{1459}{720} + \frac{7 \Upsilon}{2} + \frac{3 \Upsilon^2}{2} \right] \, .
 \end{align} }
\end{subequations}
Adding up all these contributions, as well as those due to the zero-modes, yields the second Seeley-DeWitt corresponding to the transverse traceless hook antisymmetric part of the torsion
{\small \begin{align}\label{eq:a2-kappa-perp}
\left. {\rm tr} \left( a_2 \left( \Delta_{2^-} \right) \right)\right|_{ \kappa_\perp{}^\rho{}_{\mu\nu} } = \left[ -\frac{419}{864} + \frac{5 (2 b_1+b_2)^2}{2 \alpha_1^2}+\frac{3 (2 b_1 + b_2)}{4 \alpha_1} \right] R^2 \, .
\end{align} }

\subsubsection{$1^-$-sector}

Now we turn to the other second-order operator, i.e., the one acting on the $1^-$ sector, see Eq.\ \eqref{eq:hessian-vectors}, where we recall that we have chosen the coupling constant such as the action functional is independent of the two-times longitudinal mode parameterized by $\zeta_\perp{}_\mu$. The differential operator is
\begin{equation}
 \Delta_{1^-} = - \square \mathds{1} - r R \mathds{1} \, ,
\end{equation}
where $\mathds{1}$ is the identity on vectors and
\begin{equation}
 r = -\frac{4 (12 \alpha_1+18 \alpha_3+4 \alpha_4+\alpha_5-6 \alpha_9)}{4 \alpha_4-\alpha_5-3 \alpha_9+24 (2 b_1+b_2+3 b_3)} \, .
\end{equation}
Exploiting the results of \cite{Codello:2008vh}, we find
\begin{equation}\label{eq:a2-tau-perp}
 \left. {\rm tr} \left( a_2 \left( \Delta_{1^-} \right)\right)\right|_{\tau_\perp{}_\mu} =  \left[-\frac{7}{1440} + \frac{3 (4 \alpha_4-\alpha_5-3 \alpha_9+24 (2 b_1+b_2+3 b_3))^2}{32 (12 \alpha_1+18 \alpha_3+4 \alpha_4+\alpha_5-6 \alpha_9)^2}-\frac{4 \alpha_4-\alpha_5-3 \alpha_9+24 (2 b_1+b_2+3 b_3)}{16 (12 \alpha_1+18 \alpha_3+4 \alpha_4+\alpha_5-6 \alpha_9)} \right] R^2 \, .
\end{equation}

\subsection{Fourth-order operators}\label{subsect:fourth-order}

\subsubsection{$1^+$-sector}

For taking the functional trace over the two-dimensional sector of transverse $2$-forms we employ Eq.\ \eqref{eq:formal-trace-2forms}. The trace over non-transverse $2$-forms was given in Eq.\ \eqref{eq:a2-2forms}, now we focus on the longitudinal modes. The differential operator acting on vectors is found in terms of that acting on $2$-forms \eqref{eq:final-hessian-2forms} by letting $\square \rightarrow \square + \frac{R}{12}$, whereas the one that acts on scalars is related to the starting one by $\square \rightarrow \square + \frac{R}{3}$.
Thus, we parameterize these two operators according to the general form of fourth-order differential operators \eqref{eq:fourth-order-operator}. For the vector we have
\begin{equation}
 \hat{\Omega}^{\mu\nu\rho} = 0 \, , \qquad \hat{D}^{\mu\nu} = \begin{bmatrix}
 \theta_1 && \theta_2\\
 \theta_2 && \theta_3
 \end{bmatrix} g^{\mu\nu} R \, \mathds{1} \, , \qquad
 \hat{H}^\mu = 0 \, , \qquad
 \hat{P} = \begin{bmatrix}
 \eta_1 && \eta_2\\
 \eta_2 && \eta_3
 \end{bmatrix} R^2 \, \mathds{1} \, ,
\end{equation}
where, using $\square \rightarrow \square + \frac{R}{12}$, we find
{\small \begin{align}
\begin{split}
 & \theta_1 = \sigma_1 + \frac{1}{6} \, , \qquad \theta_2 = \sigma_3 + \frac{1}{6} \, , \qquad \theta_3 = \sigma_5 + \frac{1}{6} \, ;\\
 & \eta_1 = \sigma_2 + \frac{\sigma_1}{12} + \frac{1}{144} \, , \qquad \eta_3 = \sigma_6 + \frac{\sigma_5}{12} + \frac{1}{144} \, ,
\end{split}
\end{align} }
and we have not written $\eta_2$ since it does not enter the expression of the $a_2$.  
By specializing to maximally symmetric spaces as in Eq.\ \eqref{eq:a2-minimal-fourth-order-max-sym} and employing the result of the trace of the curvature squared acting on vectors Eq.\ \eqref{eq:trace-curv-vector} we get the contribution due to the vectors as
{\small \begin{align}
\left. \left( a_2 \left( \Delta_{1^-} \right) \right) \right|_{\left( \xi_1{}_\mu, \xi_2{}_\mu \right)} = & R^2 \left[ \frac{43}{270} - 4 \left( \eta_1 + \eta_3 \right) + \frac{2}{3} \left( \theta_1 + \theta_3 \right) + 2 \left( \theta_1^2 + \theta_3^2 + 2 \theta_2^2 \right) \right] \\\nonumber
= & R^2 \left[ \frac{74}{135} + 2 \sigma_1^2+\sigma_1-4 \sigma_2+4 \sigma_3^2+\frac{4 \sigma_3}{3}+2
\sigma_5^2+\sigma_5-4 \sigma_6 \right]\, .
\end{align} }

Now we turn to the contribution due to the scalar sector, see Eq. \eqref{eq:formal-trace-2forms}. In complete analogy, the parameterization for such a sector is
\begin{equation}
\hat{\Omega}^{\mu\nu\rho} = 0 \, , \qquad \hat{D}^{\mu\nu} = \begin{bmatrix}
\chi_1 && \chi_2\\
\chi_2 && \chi_3
\end{bmatrix} g^{\mu\nu} R \, \, , \qquad
\hat{H}^\mu = 0 \, , \qquad
\hat{P} = \begin{bmatrix}
\psi_1 && \psi_2\\
\psi_2 && \psi_3
\end{bmatrix} R^2 \, \, .
\end{equation}
In this case, the coefficients entering the previous equation are related to those appearing in the operator acting on $2$-forms by
{\small \begin{align}
\begin{split}
& \chi_1 = \sigma_1 + \frac{2}{3} \, , \qquad \chi_2 = \sigma_3 + \frac{2}{3} \, , \qquad \chi_3 = \sigma_5 + \frac{2}{3} \, ;\\
& \psi_1 = \sigma_2 + \frac{\sigma_1}{3} + \frac{1}{9} \, , \qquad \psi_3 = \sigma_6 + \frac{\sigma_5}{3} + \frac{1}{9} \, ,
\end{split}
\end{align} }
where also here we have not taken into account $\psi_2$ for it does not enter the trace of the $a_2$.
Thus, the resulting trace of the second Seeley-DeWitt coefficient for the scalar sector is
{\small \begin{align}
\left. {\rm tr} \left( a_2 \left( \Delta_{0^+} \right) \right) \right|_{\left( \phi_1, \phi_2 \right)} = & R^2 \left[ \frac{29}{540} - \left( \psi_1 + \psi_3 \right) + \frac{1}{6} \left( \chi_1 + \chi_3 \right) + \frac{1}{2} \left( \chi_1^2 + \chi_3^2 + 2 \chi_2^2 \right) \right]\\\nonumber
= & R^2 \left[ \frac{149}{540} + \frac{\sigma_1^2}{2}-\frac{\sigma_1}{2}-4 \sigma_2+\sigma_3^2+\frac{4 \sigma_3}{3}+\frac{\sigma_5^2}{2}-\frac{\sigma_5}{2}-4 \sigma_6 \right] \, .
\end{align} }
Adding up the contributions due to the tensors, vectors and scalars, and taking into account the zero-modes yields the final expression of the second Seeley-DeWitt coefficient stemming from the $1^+$-sector
{\small
{\small \begin{align}\nonumber
& \left. {\rm tr} \left( a_2 \left( \Delta_{1^+} \right) \right) \right|_{(A_\perp{}_{\mu\nu},\Pi_\perp{}_{\mu\nu})} = R^2 \left[ -\frac{7}{10} -\frac{(3 \alpha_2+\alpha_6) (12 \alpha_1+8 \alpha_8) \left(\frac{-12 b_1-6
		b_2+2 (3 \alpha_1+4 \alpha_4)}{3 \alpha_1+2 \alpha_8}+\frac{36 b_1-36
		b_2+15 \alpha_2+8 \alpha_6}{3 \alpha_2+\alpha_6}\right)^2}{96 \left(8 (3
	\alpha_2+\alpha_6) (3 \alpha_1+2 \alpha_8)-4 \alpha_8^2\right)} \right.\\\nonumber
& \,\,\, \left. + \frac{1}{12 (6 \alpha_1 (3 \alpha_2+\alpha_6)+(12 \alpha_2+4 \alpha_6-\alpha_8) \alpha_8)^2} \left(6
	\alpha_8^4-60 \alpha_2 \alpha_8^3 -8 \alpha_6 \alpha_8^3-108 \alpha_2^2 \alpha_8^2-234 \alpha_1 \alpha_2 \alpha_8^2-96 \alpha_2
	\alpha_4 \alpha_8^2 -78 \alpha_1 \alpha_6 \alpha_8^2-144 \alpha_2
	\alpha_6 \alpha_8^2 \right.\right.\\\nonumber
	& \,\,\, \left.\left. -56 \alpha_4 \alpha_6 \alpha_8^2+1188 \alpha_1
	\alpha_2^2 \alpha_8+168 \alpha_1 \alpha_6^2 \alpha_8+96 \alpha_4
	\alpha_6^2 \alpha_8 +864 \alpha_2^2 \alpha_4 \alpha_8+576 \alpha_1
	\alpha_2 \alpha_6 \alpha_8 +576 \alpha_2 \alpha_4 \alpha_6
	\alpha_8+2349 \alpha_1^2 \alpha_2^2 +576 \alpha_2^2 \alpha_4^2  \right.\right.\\\nonumber
	& \,\,\, \left.\left. +288
	\alpha_1^2 \alpha_6^2+64 \alpha_4^2 \alpha_6^2+240 \alpha_1 \alpha_4 \alpha_6^2+2160 \alpha_1 \alpha_2^2 \alpha_4+384 \alpha_2
	\alpha_4^2 \alpha_6 +1404 \alpha_1^2 \alpha_2 \alpha_6+1440 \alpha_1 \alpha_2 \alpha_4 \alpha_6 +144 b_1^2 \left(81 \alpha_1^2 +108
	\alpha_8 \alpha_1+9 \alpha_2^2 \right.\right.\right.\\\nonumber
	& \,\,\, \left.\left.\left. +\alpha_6^2+39 \alpha_8^2+6 \alpha_2 \alpha_6\right)+36 b_2^2 \left(324 \alpha_1^2+432 \alpha_8 \alpha_1+9
	\alpha_2^2+\alpha_6^2  +138 \alpha_8^2+6 \alpha_2 \alpha_6\right)+6
	b_2 \left(108 (3 \alpha_2-2 \alpha_6) \alpha_1^2  +12 \left(9 \alpha_2^2+6
	(\alpha_6+6 \alpha_8) \alpha_2 \right.\right.\right.\right.\\
	& \,\,\, \left.\left.\left.\left.  +\alpha_6 (\alpha_6-24 \alpha_8)\right) \alpha_1-144 \alpha_2^2 (\alpha_4-\alpha_8) +\alpha_2 \left(135
	\alpha_8^2+96 \alpha_6 \alpha_8-96 \alpha_4 \alpha_6\right)-8 \left(3
	\alpha_8^3-6 \alpha_4 \alpha_8^2+12 \alpha_6 \alpha_8^2-2 \alpha_6^2 \alpha_8+2 \alpha_4 \alpha_6^2\right)\right) \right.\right.\\\nonumber
	& \,\,\, \left.\left.  -12 b_1 \left(-12 \alpha_8^3+81 \alpha_2 \alpha_8^2+24 \alpha_4 \alpha_8^2-48 \alpha_6
	\alpha_8^2 -144 \alpha_2^2 \alpha_8-16 \alpha_6^2 \alpha_8-96 \alpha_2 \alpha_6 \alpha_8+16 \alpha_4 \alpha_6^2+144 \alpha_2^2 \alpha_4+54 \alpha_1^2 (3 \alpha_2-2 \alpha_6)  \right.\right.\right.\\\nonumber
	& \,\,\, \left.\left.\left.  +96 \alpha_2 \alpha_4
	\alpha_6+6 b_2 \left(324 \alpha_1^2+432 \alpha_8 \alpha_1-18 \alpha_2^2-2 \alpha_6^2+147 \alpha_8^2 -12 \alpha_2 \alpha_6\right)-12 \alpha_1
	\left(9 \alpha_2^2+6 (\alpha_6-3 \alpha_8) \alpha_2+\alpha_6
	(\alpha_6+12 \alpha_8)\right)\right) \right) \right.\\\nonumber
	& \,\,\, \left.  -\frac{\sqrt{(3 \alpha_2+\alpha_6) (3 \alpha_1+2 \alpha_8)}}{4 \sqrt{2}
		\left(\sqrt{(6 \alpha_2+2 \alpha_6) (12 \alpha_1+8 \alpha_8)}-2 \alpha_8\right)}    \left(\frac{-12 b_1-6 b_2+2 (3
		\alpha_1+4 \alpha_4)}{3 \alpha_1+2 \alpha_8}   +\frac{2 \sqrt{2} \alpha_8}{\sqrt{(3 \alpha_2+\alpha_6) (3 \alpha_1+2 \alpha_8)}}+\frac{-36 b_1+36
		b_2-15 \alpha_2-8 \alpha_6}{3 \alpha_2+\alpha_6}\right) \right.\\\nonumber
	& \,\,\, \left. -\frac{\sqrt{(3 \alpha_2+\alpha_6) (3 \alpha_1+2 \alpha_8)}}{4 \sqrt{2} \left(2 \alpha_8+\sqrt{(6 \alpha_2+2 \alpha_6) (12
				\alpha_1+8 \alpha_8)}\right)}
	\left(\frac{-12 b_1-6 b_2+2 (3 \alpha_1+4 \alpha_4)}{3 \alpha_1+2 \alpha_8} -\frac{2 \sqrt{2} \alpha_8}{\sqrt{(3 \alpha_2+\alpha_6) (3 \alpha_1+2
			\alpha_8)}}+\frac{-36 b_1+36 b_2-15 \alpha_2-8 \alpha_6}{3 \alpha_2+\alpha_6}\right) \right.\\\nonumber
		& \,\,\, \left.      -\frac{(3 \alpha_2+\alpha_6) (12 \alpha_1+8
	\alpha_8)}{192 \left(\sqrt{(6 \alpha_2+2 \alpha_6) (12
		\alpha_1+8 \alpha_8)}-2 \alpha_8\right)^2}    \left(\frac{-12 b_1-6 b_2+2 (3 \alpha_1+4 \alpha_4)}{3 \alpha_1+2 \alpha_8} +\frac{2 \sqrt{2} \alpha_8}{\sqrt{(3 \alpha_2+\alpha_6) (3
			\alpha_1+2 \alpha_8)}}+\frac{-36 b_1+36 b_2-15 \alpha_2-8 \alpha_6}{3
		\alpha_2+\alpha_6}\right)^2 \right.\\\nonumber
	& \,\,\, \left. -\frac{(3 \alpha_2+\alpha_6) (12
	\alpha_1+8 \alpha_8)}{192 \left(2 \alpha_8+\sqrt{(6
		\alpha_2+2 \alpha_6) (12 \alpha_1+8 \alpha_8)}\right)^2}   \left(\frac{-12 b_1-6 b_2+2 (3 \alpha_1+4 \alpha_4)}{3 \alpha_1+2 \alpha_8} -\frac{2 \sqrt{2} \alpha_8}{\sqrt{(3 \alpha_2+\alpha_6) (3 \alpha_1+2 \alpha_8)}}+\frac{-36 b_1+36 b_2-15 \alpha_2-8 \alpha_6}{3 \alpha_2+\alpha_6}\right)^2 \right]
\end{align} }
}

\subsubsection{$2^+$-sector}

The fourth-order operator acting on $\overline{S}_{\mu\nu}$ is parameterized as
\begin{equation}
\Delta_{2^+} = \left( \square^2 + \rho_1 R \square + \rho_2 R^2 \right) \mathds{1} \, ,
\end{equation}
where $\mathds{1}$ is the identity on symmetric trace-free $2$-tensors. Comparing the previous equation with Eq.\ \eqref{eq:hessian-S} and choosing $d=4$, we find
\begin{eqnarray}
 \rho_1 = - \frac{\alpha_1 + \frac{4}{3} \alpha_5 + 2 \left( 2b_1 + b_2 \right)}{2 \left( 2 \alpha_1 + \alpha_5 \right)} \, \qquad && \rho_2 = \frac{ 2\alpha_5 - \alpha_1  6\left( 2b_1 + b_2 \right)}{18 \left( 2 \alpha_1 + \alpha_5 \right)} \, .
\end{eqnarray}
The trace of this fourth-order operator over transverse traceless configurations is found by using
{\small \begin{align}\label{eq:formal-trace-2+-tentative}
 \left.{\rm Tr \, e}^{-s \Delta_{2^+}} \right|_{\overline{S}_\perp{}_{\mu\nu}} = & \left.{\rm Tr \, e}^{-s \Delta_{2^+}} \right|_{\overline{S}_{\mu\nu}} - \left.{\rm Tr \, e}^{-s \Delta_{1^-}} \right|_{\xi_\mu} - \left.{\rm Tr \, e}^{-s \Delta_{0^+}} \right|_{\sigma}\\\nonumber
 & + \left.{\rm Tr \, e}^{-s \tilde{\Delta}_{0^+}} \right|_{\phi} + {\rm e}^{\lambda_1 \, s} + 5 \, {\rm e}^{\lambda_2 \, s} + 10 {\rm e}^{\lambda_3 \, s}
\end{align} }
where the kinematical zero modes are due to the embedding of the sphere in $\mathbb{R}^5$ and to the Killing vectors and only the multiplicities of the zero-modes contributes to the second Seeley-DeWitt coefficient. The specific form of the operators acting on the longitudinal modes can be written in terms of the starting one as
\begin{equation}
 \Delta_{1^-} = \Delta_{2^+} \left( \square \rightarrow \square + \frac{5}{12} R \right) \, , \qquad \Delta_{0^+} = \Delta_{2^+} \left( \square \rightarrow \square + \frac{2}{3} R \right) \, , \qquad \tilde{\Delta_{0^+}} = \Delta_{2^+} \left( \square \rightarrow \square + \frac{2}{3} R \right) \, .
\end{equation}
Notice that, since the operators acting on the scalar sector are identical, Eq.\ \eqref{eq:formal-trace-2+-tentative} actually simplifies to
\begin{equation}\label{eq:formal-trace-2+}
  \left.{\rm Tr  e}^{-s \Delta_{2^+}} \right|_{\overline{S}_\perp{}_{\mu\nu}} = \left.{\rm Tr  e}^{-s \Delta_{2^+}} \right|_{\overline{S}_{\mu\nu}} - \left.{\rm Tr  e}^{-s \Delta_{1^-}} \right|_{\xi_\mu} + 5 \, {\rm e}^{\lambda_1 \, s} + 10 {\rm e}^{\lambda_2 \, s} \, ,
\end{equation}
as it was proved in section\ \ref{sect:spectrum-of-diff-constrained}. The contribution to the $a_2$ due to the symmetric traceless tensor is
\begin{equation}
\left. {\rm tr} \left( a_2 \left( \Delta_{2^+} \right) \right) \right|_{\overline{S}_{\mu\nu}} = R^2 \left[ \frac{3}{40} -9 \rho_2 + \frac{3}{2} \rho_1 + \frac{9}{2} \rho_1^2 \right] \, .
\end{equation}
On the other hand, the form of the operator acting on the vector that parameterizes the longitudinal mode is
\begin{equation}
 \Delta_{1^-} = \left[ \square^2 + \left( \rho_1 + \frac{5}{6} \right) R \square + \left( \rho_2 + \frac{5}{12} \rho_1 + \frac{25}{144}  \right) R^2 \right] \mathds{1} \, ,
\end{equation}
where now $\mathds{1}$ is the identity on vectors. The second Seeley-DeWitt coefficient stemming from this operator is
\begin{equation}
 \left. {\rm tr} \left( a_2 \left( \Delta_{1^-} \right) \right) \right|_{\xi_{\mu}} = R^2 \left[ \frac{359}{270} + \frac{7}{3} \rho_1  - 4 \rho_2 + 2 \rho_1^2 \right] \, .
\end{equation}
Therefore, adding up the contributions as in Eq.\ \eqref{eq:formal-trace-2+}, we obtain the final form of the trace of the second Seeley-DeWitt coefficient due to the $2^+$-sector
{\small \begin{align}
 \left.{\rm tr} \left( a_2 \left( \Delta_{2^+} \right) \right) \right|_{\overline{S}_\perp{}_{\mu\nu}} = & \left[ - \frac{17}{27} + \frac{1}{72 (2\alpha_1 + \alpha_5)^2} \left( -71 \alpha_1^2-326 \alpha_1 \alpha_5-64 \alpha_5^2+1872 b_1^2 +  \right.\right.\\\nonumber
 & \left.\left. +24 b_1 (-51
 \alpha_1+7 \alpha_5+78 b_2)+468 b_2^2+b_2 (84 \alpha_5-612 \alpha_1) \right) \right] R^2 \, .
\end{align} }

\subsubsection{$0^+$-sector}

From the Hessian \eqref{eq:hessian-scalar} we find that the fourth-order operator acting on the scalar sector is parameterized as
\begin{equation}
 \hat{\Omega}^{\mu\nu\rho} = 0 \, , \qquad \hat{D}^{\mu\nu} = \omega_1 g^{\mu\nu} R  \, , \qquad \hat{H}^\mu = 0 \, , \qquad \hat{P} = 0 \, .
\end{equation}
By using Eq.\ \eqref{eq:a2-minimal-fourth-order-max-sym} and subtracting the zero-mode due to constant configurations that do to contribute to the spectrum we find
{\small \begin{align}
 \left.{\rm tr} \left(  a_2 \left( \Delta_{0^+} \right) \right) \right|_{\varphi}  = \left[ -\frac{2}{135} + \frac{(-6 \alpha_1-9 \alpha_3-\alpha_5+6 b_1+3 b_2+9 b_3) (-14 \alpha_1-21 \alpha_3-5 \alpha_5-12 \alpha_7+6 b_1+3 b_2+9 b_3)}{288 (2
 	\alpha_1+3 \alpha_3+\alpha_5+3 \alpha_7)^2} \right] R^2 \, .
\end{align} }

\subsubsection{$0^-$-sector}

The analysis in this sector is completely equivalent to that of the preceding one, and it yields the following contribution to the second Seeley-DeWitt coefficient
{\small \begin{align}
\left.{\rm tr} \left( a_2 \left( \Delta_{0^-}\right) \right) \right|_{\pi} = \left[ -\frac{769}{69120} - \frac{b_1-b_2}{64 \alpha_2} + \frac{\left( b_1 - b_2 \right)^2}{288 \alpha_2^2} \right] R^2 \, .
\end{align} }
The difference in the numerical coefficient w.r.t. the $0^+$-sector arises because a pure number appears in the Hessian \eqref{eq:hessian-pseudoscalar} as soon as we normalize the principal part of the operator.

\subsubsection{Final result without mass terms}

For the sake of completeness, here we report the final form of the traced second Seeley-DeWitt coefficient in absence of mass terms for the torsion, which reads
\begin{small}
	{\small \begin{align}\label{eq:final-result-without-masses}\nonumber
		{\rm tr} & \left(a_2\right) = R^2 \left\{ -\frac{127553}{69120} + \frac{1}{1152} \Bigg[ -305+  16 b_1^2-4 (8 b_2+9) b_1+16 b_2^2+11520 (2 b_1+b_2)^2+\frac{108 (24 b_1+12
			b_2+24 b_3-1)^2}{(32 b_1+16 b_2-15)^2}+36 b_2 \right.\\\nonumber
		& \left.\left. +1728 (2 b_1+b_2) +\frac{72 (24
			b_1+12 b_2+24 b_3-1)}{32 b_1+16 b_2-15}+\frac{1}{\left(-2
			\alpha_8^2+12 \alpha_8+\alpha_6 (8 \alpha_8+6)+9\right)^2} [  24 \left(96 \alpha_8^4-128
		\alpha_6 \alpha_8^3-480 \alpha_8^3-768 \alpha_6 \alpha_8^2\right.\right.\right.\\\nonumber
		& \left.\left.\left.\left. -504
		\alpha_8^2-384 \alpha_6^2 \alpha_8 -2880 \alpha_6 \alpha_8-1512
		\alpha_8+288 \alpha_6^2+216 \alpha_6  +1152 b_1^2 \left(2 \alpha_6^2+6
		\alpha_6+66 \alpha_8^2+108 \alpha_8+45\right)+144 b_2^2 \left(4 \alpha_6^2+12
		\alpha_6 \right.\right.\right.\right.\right.\\\nonumber
		& \left.\left.\left.\left.\left. +600 \alpha_8^2+864 \alpha_8+333\right) +48 b_2 \left(4 (20 \alpha_8+9) \alpha_6^2-4 \alpha_8 (55 \alpha_8+12)
		\alpha_6+3 \left(-16 \alpha_8^3+\alpha_8^2+132 \alpha_8+54\right)\right)+48 b_1 \left(8 (20 \alpha_8+9)
		\alpha_6^2 \right.\right.\right.\right.\right.\\\nonumber
		& \left.\left. +4 \left(34 \alpha_8^2+192 \alpha_8+81\right) \alpha_6 +3 \left(16
		\alpha_8^3-34 \alpha_8^2+48 \alpha_8+27\right)+12 b_2 \left(4 \alpha_6^2+12
		\alpha_6-3 \left(94 \alpha_8^2+144 \alpha_8+51\right)\right)\right)+405\right)] \\
		& \left.\left. -\frac{72 \sqrt{2}
			\sqrt{(2 \alpha_6+3) (4 \alpha_8+3)}}{\sqrt{2} \sqrt{(2 \alpha_6+3)
				(4\alpha_8+3)}-2 \alpha_8} \left(\frac{12 (2 b_1+b_2-1)}{4 \alpha_8+3}-\frac{72 b_1-72 b_2+16 \alpha_6+15}{2 \alpha_6+3}
		+\frac{4 \sqrt{2} \alpha_8}{\sqrt{(2 \alpha_6+3) (4 \alpha_8+3)}}\right) \right.\right.\\\nonumber
		& \left.\left.  -\frac{72 \sqrt{2} \sqrt{(2 \alpha_6+3) (4 \alpha_8+3)}}{\sqrt{2} \sqrt{(2 \alpha_6+3) (4 \alpha_8+3)}+2 \alpha_8}
		\left(\frac{12 (2 b_1+b_2-1)}{4 \alpha_8+3} -\frac{72 b_1-72 b_2+16 \alpha_6+15}{2 \alpha_6+3}-\frac{4 \sqrt{2} \alpha_8}{\sqrt{(2 \alpha_6+3)
				(4 \alpha_8+3)}}\right) \right.\right.\\\nonumber
		& \left.\left.  -\frac{6 \left(\alpha_6+\frac{3}{2}\right) (8 \alpha_8+6)}{\left(\sqrt{2} \sqrt{(2 \alpha_6+3) (4
				\alpha_8+3)}-2 \alpha_8\right)^2} 
		\left(-\frac{12 (2 b_1+b_2-1)}{4 \alpha_8+3} +\frac{72 b_1-72 b_2+16 \alpha_6+15}{2 \alpha_6+3}-\frac{4 \sqrt{2} \alpha_8}{\sqrt{(2 \alpha_6+3) 
				(4 \alpha_8+3)}}\right)^2 \right.\right.\\\nonumber
		& \left. \left. -\frac{6 \left(\alpha_6+\frac{3}{2}\right) (8 \alpha_8+6)}{\left(\sqrt{2} \sqrt{(2 \alpha_6+3) (4 \alpha_8+3)}+2 \alpha_8\right)^2}
		\left(-\frac{12 (2 b_1+b_2-1)}{4 \alpha_8+3}+\frac{72 b_1-72 b_2+16
			\alpha_6+15}{2 \alpha_6+3}+\frac{4 \sqrt{2} \alpha_8}{\sqrt{(2 \alpha_6+3) (4
				\alpha_8+3)}}\right)^2 \right.\right.\\\nonumber
		& \left.\left. -\frac{6 (24
			b_2 \alpha_6+64 \alpha_8 \alpha_6+24 \alpha_6+60 \alpha_8+48
			b_1 (\alpha_6+6 \alpha_8+6)-36 b_2 (8 \alpha_8+5)+9)^2}{(2 \alpha_6+3)
			(4 \alpha_8+3) \left(-2 \alpha_8^2+12 \alpha_8+\alpha_6 (8 \alpha_8+6)+9\right)} \right.\right.\\\nonumber
		& \left.\left.  +\frac{12 (44 b_1+22 b_2+6 b_3-15) (356 b_1+178 b_2+18 b_3-24
			\alpha_7-113)}{(-56 b_1-28 b_2+6 \alpha_7+17)^2} \right]  \right\} \, .
		\end{align} }
\end{small}



\end{document}